\documentclass[11pt,a4paper]{article}
\usepackage[utf8]{inputenc}
\usepackage[left=3cm,right=3cm,top=3cm,bottom=3cm]{geometry}

\usepackage{amsmath}
\usepackage{amsfonts}
\usepackage{amssymb}
\usepackage{mathtools}
\usepackage{braket}
\usepackage{cite}

\usepackage{graphicx}

\usepackage{color}
\definecolor{nicered}{rgb}{0.7,0.1,0.1}
\definecolor{nicegreen}{rgb}{0.1,0.5,.1}

\usepackage{subcaption}

\usepackage{stackrel}

\usepackage{multirow}

\usepackage{float}

\usepackage{authblk}

\usepackage{comment}

\def\gr#1{{\color{green} #1}}
\def\rd#1{{\color{red} #1}}

\usepackage[colorlinks=true
,urlcolor=magenta
,anchorcolor=black
,citecolor=blue
,filecolor=black
,linkcolor=red
,menucolor=black
,linktocpage=true
,pdfproducer=medialab
]{hyperref}

\title{$S$-wave contribution to rare $D^0 \to \pi^+ \pi^- \ell^+ \ell^-$ decays \\ in the Standard Model and sensitivity to New Physics}

\date{} %\date{July 2023}

\author[a]{Svjetlana~Fajfer}
\author[b]{Eleftheria~Solomonidi}
\author[b]{Luiz~Vale~Silva}

\affil[a]{\it \small Jo\v{z}ef Stefan Institute, Jamova 39, P. O. Box 3000, 1001 Ljubljana, Slovenia,

Faculty of Mathematics and Physics, University of Ljubljana, Jadranska 19, 1000 Ljubljana, Slovenia}

\affil[b]{\it \small Departament de F\'{i}sica Te\`{o}rica, Instituto de F\'{i}sica Corpuscular,

Universitat de Val\`encia -- Consejo Superior de Investigaciones Cient\'{i}ficas,

Parc Cient\'{i}fic, Catedr\'{a}tico Jos\'{e} Beltr\'{a}n 2, E-46980 Paterna, Valencia, Spain}

\begin{document}

\maketitle

Physics of the up-type flavour offers unique possibilities of testing the Standard Model (SM) compared to the down-type flavour sector.
Here, we discuss SM and New Physics (NP) contributions to the rare charm-meson decay $ D^0 \to \pi^+ \pi^- \ell^+ \ell^- $. %Contrarily to analogous rare $B$-meson decays, phase space windows with reduced resonant contributions are limited.
%We exploit a data-driven way to extract non-perturbative quantities, which consists in a step forward in the description of the underlying strong dynamics
In particular, we discuss the effect of including the lightest scalar isoscalar resonance in the SM picture, namely, the $f_0 (500)$, which manifests in a big portion of the allowed phase space.
Other than showing in the total branching ratio at an observable level of about $ 20\% $,
the $f_0 (500)$ resonance manifests as interference terms with the vector resonances, such as at high invariant mass of the leptonic pair in distinct angular observables.
Recent data from LHCb optimize the sensitivity to $P$-wave contributions, that we analyse in view of the inclusion of vector resonances.
%We propose a new definition of the measured observables which is sensitive to the $S$-wave and is straightforward to implement experimentally.
We propose the measurement of alternative observables which are sensitive to the $S$-wave and are straightforward to implement experimentally.
This leads to a new set of null observables, that vanish in the SM due to its gauge and flavour structures.
Finally, we study observables that depend on the SM interference with generic NP contributions from semi-leptonic four-fermion operators in the presence of the $S$-wave.

\section{Introduction}

%%%%%%% Opening %%%%%%%
%Decays mediated by Flavour Changing Neutral Currents (FCNCs) are suppressed in the SM, since they are forbidden at the tree level. Such mechanism, called Glashow-Iliopoulos-Maiani, is much more efficient in the case of charm physics compared to down-type transitions, thus resulting in the dominance of non-perturbative QCD effects that are extremely challenging to describe.

Rare decays played a crucial role in building the Standard Model (SM): it is for instance thanks to $K_L \to \mu^+ \mu^-$ that one gathered indirect information about the existence of the charm-quark before its discovery \cite{Glashow:1970gm}.
Rare charm-meson decays provide complementary information to down-type Flavour Changing Neutral Currents (FCNCs) transitions.
However, given the effectiveness of the Glashow-Iliopoulos-Maiani (GIM) suppression in up-type FCNCs, and the almost diagonal structure of the Cabibbo-Kobayashi-Maskawa (CKM) matrix, this class of transitions is very sensitive to the strong dynamics:
as we will see,
%It is necessary to describe well the SM contribution due to the characteristics of
the available phase space in charm-meson decays
%, which
is entirely populated with ``intermediate'' resonance peaks and their tails, in contrast to analogous bottom-meson decays.
%This is so because the allowed phase space in charm-meson decays is populated with resonant peaks and their tails.
Therefore, for the sake of New Physics (NP) searches in rare charm-meson decays, the SM has to be described sufficiently well;
this is so when the SM acts as a background,
and is also the case when one wants to understand the SM-NP interference in order to set bounds on the NP properties.

%%%%%%% LHCb %%%%%%%
LHCb will largely improve measurements of rare $D$ meson decay channels; for very recent experimental analyses of $ D^0 \to \pi^+ \pi^- \mu^+ \mu^- $ and $ D^0 \to K^+ K^- \mu^+ \mu^- $, see the analysis of Refs.~\cite{LHCb:2021yxk,LHCb_supplementary_material_aps,LHCb_supplementary_material_4},
that extends Refs.~\cite{LHCb:2013gqe,LHCb:2017uns,LHCb:2018qsd,LHCb_supplementary_material_3}.
The total branching fractions are \cite{LHCb:2017uns}:

\begin{equation}\label{eq:full_BRs}
    \mathcal{B} (D^0 \to \pi^+ \pi^- \mu^+ \mu^-) = (9.6 \pm 1.2) \times 10^{-7} \,, \quad \mathcal{B} (D^0 \to K^+ K^- \mu^+ \mu^-) = (1.5 \pm 0.3) \times 10^{-7} \,.
\end{equation}

\noindent
A rich angular analysis is possible, resulting from the high multiplicity of the final state.
This promising experimental programme has to be matched by an increased theoretical precision.
Our ultimate goal here is to provide more robust tests of NP contributions possibly affecting these rare charm-meson decays. For this sake, we reassess the description of the SM contributions.
As it will be discussed in this article,
present data already allows for an enhanced control over the SM background, i.e., contributions of intermediate resonances, and their relative strong phases.
As a result, we will then in particular be able to point out improved observables for NP searches.

%Since rare decays are candidates for discovering BSM effects, once they do not have to compete with large SM contributions, the present analysis unlocks opportunities for BSM physics, in which their couplings (Wilson coefficients) can be probed via observables accessible at colliders \cite{LHCb:2021yxk}.

%(binned and/or unbinned)
%LFU violation
%CP violation

%[comment on future sensitivity: preliminary/too preliminary]
%https://indico.cern.ch/event/760368/contributions/3316100/attachments/1822241/2981045/AContu_RareCharm_TUPFPhop2019.pdf

%%%%%%% Previous theoretical works %%%%%%%
We focus here on the inclusion of intermediate resonances in the description of the decay $D^0 \to \pi^+ \pi^- \ell^+ \ell^-$ ($\ell$ are electrons or muons); we reserve the mode $ D^0 \to K^+ K^- \ell^+ \ell^- $ to future work.\footnote{The lightest resonances coupling more strongly to the kaon pair are $ f_0 (980) $ and $ \phi (1020) $, which manifest at similar energies, the latter being very narrow though; this may produce an interesting interference pattern between the $S$- and $P$-waves in angular observables. A representation of the line-shape of the scalar isoscalar resonance is more difficult to achieve due to its proximity to the kaon pair threshold.}
The strategy adopted is to consider quasi two-body decays,
where the pion pair in the final state originates from strong decays of resonances such as the $ \rho (770)^0 \equiv \rho^0 $,
%The kaon pair originates from $ \phi $.
while the lepton pair originates from EM decays of states such as $ \eta $, $ \eta' (958) \equiv \eta' $, $ \rho^0 $, $ \omega (782) \equiv \omega $, and $ \phi (1020) \equiv \phi $.
The vector resonances are clearly seen in the data collected by LHCb \cite{LHCb:2021yxk,LHCb_supplementary_material_aps,LHCb_supplementary_material_4}.
%[I guess the pseudoscalar contributions can be neglected in a first moment]
For previous theoretical analyses, see for instance Refs.~\cite{Fajfer:1998rz,Burdman:2001tf,Fajfer:2005ke,Bigi:2011em,Cappiello:2012vg,DeBoer:2018pdx}; also, see Refs.~\cite{Feldmann:2017izn,Bharucha:2020eup} in the framework of QCD factorization at low-$q^2 (\ell^+ \ell^-)$ (while as it will be later discussed we avoid this region), where the hadronic uncertainties in this framework are quantitatively assessed, and also for the use of an Operator Product Expansion (OPE) in the very high-$q^2 (\ell^+ \ell^-)$ region (which for different reasons we also avoid, as discussed later).
%4/3*C1+C2 is a feature of quark loops
%in part due to heavy $P$-wave resonances: an OPE is the approach that could describe these heavy resonances altogether
%Bharucha et al.: They refer to a tower of Regge-like resonances; "less effective expansion"; novelty is the treatment a la e+e-; compatible to other approaches, but annihilation is large?; OPE requires large enough bin;  
Other cases of interest in assessing SM contributions in related rare (semi-)leptonic charm-meson decay modes include the ones of Refs.~\cite{Paul:2011ar,deBoer:2015boa,Fajfer:2015mia,Bause:2019vpr,Bharucha:2020eup} (while Ref.~\cite{Sanchez:2022nsh} discusses the mode $ D^+_s \to \pi^+ \ell^+ \ell^- $, not mediated by FCNCs).
See also Ref.~\cite{Gisbert:2020vjx} for a recent theoretical and experimental review.

Beyond the vector and pseudoscalar resonances aforementioned,
further resonances could also lead to an important SM contribution.
We have identified the scalar isoscalar state $f_0 (500) \equiv \sigma$ as a relevant contribution not previously included in past analyses (although pointed out in Ref.~\cite{DeBoer:2018pdx}).
Such a broad state leaves its footprints in the rescattering of pion pairs \cite{Bugg:2006gc,Pelaez:2015qba}; note that the PDG \cite{Workman:2022ynf} mini-review on scalar mesons below $1$~GeV quotes for the $\sigma$ pole position the value $ (449^{+22}_{-16}) - i (275 \pm 12) $~MeV stemming from ``the most advanced dispersive analyses'', which is a precision better than 5\%.
As it will be discussed in this article, although the $S$-wave does not affect some angular observables (in particular those based in $ I_i $, $i = 3, 6, 9$, \cite{DeBoer:2018pdx}), it affects a large set of them (i.e., some observables built from $ I_i $, $ i = 1, 2, 4, 5, 7, 8$), and thus provides novel null tests of NP when the NP interferes with the SM in the presence of the $S$-wave.

%They point out the contribution of $ D^0 \to f_0 (500) \rho (770)^0 $, and that the angular coefficients $ I_{3, 6, 9} $ ($ A_{FB} \propto I_6 $) are unaffected by this contribution. Note that, however, this mode affects the branching ratio.

%%%%%%% Experimental results about the sigma: semi-leptonic charm-meson decays %%%%%%%
We highlight that
the $S$-wave contribution has already been observed in semi-leptonic charm-meson decays.
BESIII \cite{BESIII:2018qmf} has seen an $S$-wave contribution coming from $\sigma$ at the level of $26\%$
%[\rd{cannot be this large; they give this value in their abstract, although their figure does not seem to be compatible with that; it is not clear from the figure because the background is seen therein}]
of the total branching ratio of $D^+ \to \pi^+ \pi^- e^+ \nu_e$. It is worth stressing that this occurs in the absence of interference with the dominant $P$-wave, as is the case for the total branching ratio; also note that this contribution does not manifest as a distinguished peak in the invariant mass of the final pion pair.
Instead, the $S$-wave effect can be better spotted from its interference with the dominant $P$-wave contribution (mainly coming from $\rho^0 \to \pi^+ \pi^-$) in alternative observables: a pronounced asymmetry is thus clearly seen in the differential branching ratio as a function of the angle $\theta_\pi$ describing the orientation of the pion pair.
%as defined in their decay plane
Accordingly, no pronounced asymmetry is seen in $D^0 \to \pi^- \pi^0 e^+ \nu_e$, for which the $S$-wave contribution is absent.
One could expect even more explicit manifestations of the $S$-wave in the differential branching ratio as a function of $\theta_\pi$, and the angle $\phi$ between the decay planes of the lepton and pion pairs, when integrating over carefully chosen slices of the invariant mass of the pion pair, as seen for instance in the analysis of the Cabibbo allowed mode $D^+ \to K^- \pi^+ e^+ \nu_e$ by BaBar \cite{BaBar:2010vmf}, where the $S$-wave contribution, in particular from $ K^\ast_0 (800) \equiv \kappa $ and $K^\ast_0 (1430)$, is at the level of $6\%$; see also Refs.~\cite{FOCUS:2002xsy} and \cite{CLEO:2010enr}.
This shows that some angular observables can be directly used to investigate the $P$- and $S$-wave interference.

%[we reserve the analysis of the inclusion of the $S$-wave in the Cabibbo-allowed rare decay $D^0 \to K^- \pi^+ \mu^+ \mu^-$ \cite{LHCb:2015yuk} to future work: this does not receive contributions from FCNCs, but from bremsstrahlung]
%[Stress the sizable S-wave contribution found in \cite{BESIII:2018qmf}, close to 25\%]

%%%%%%% Experimental results about the sigma: non-leptonic charm-meson decays %%%%%%%
Moreover, although uncertainties are still large, an amplitude analysis of CLEO data \cite{dArgent:2017gzv}
of $ D^0 \to \pi^+ \pi^- \pi^+ \pi^- $ indicates an important contribution of $ D^0 \to \sigma \, \rho^0 $, comparable to the contributions of $ D^0 \to \rho^0 \, \rho^0 $.
%Such $S$-wave term would then impact the analysis of rare decays in the SM. In this respect,
Other topologies affecting rare decays are suggested by the amplitude analyses of multi-hadronic decays $ D^0 \to \pi^+ \pi^- \pi^+ \pi^- $ and $ D^0 \to K^+ K^- \pi^+ \pi^- $  \cite{dArgent:2017gzv,LHCb:2018mzv}, namely, so-called cascade decays in which there is an intermediate $ a_1 (1260)^\pm $ (which affects $D^0 \to \pi^+ \pi^- \ell^+ \ell^-$) or $ K_1 (1270)^\pm $ (which affects $D^0 \to K^+ K^- \ell^+ \ell^-$). Such states would not manifest as peaks in the invariant mass of the lepton or light hadron pairs, since they involve a distinct combination of kinematical variables. In these topologies, the lepton pair results from $\rho^0$ and $\phi$, while the pion and kaon pairs are non-resonant.
Given that the axial vector resonances above are known to a lesser extent than those resonances included in our analysis,
we reserve their analysis for future work.
%[We will reserve the discussion of the SM background resulting from cascade decays for future work.]
%[In the future, we plan including further SM effects, namely, cascade decays.]
%I won't say it peaks in rho, since I don't know other contributions of a0: a0 to rho pi, a0 to omega pi, a0 to phi pi, a0 to gamma pi

\begin{figure}[t]
    \centering
    \includegraphics[scale=0.6]{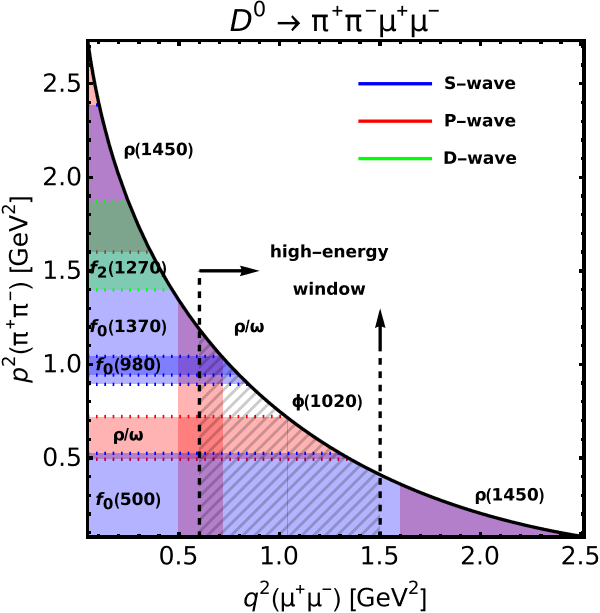}
    \caption{Phase space allowed in the decay $ D^0 \to \pi^+ \pi^- \mu^+ \mu^- $; the invariant mass of the pion (muon) pair is denoted $p^2$ (respectively, $q^2$). Some scalar ({\color{blue}\textbf{blue}}), vectorial ({\color{red}\textbf{red}}) and tensorial ({\color{green}\textbf{green}}) resonant contributions are shown (the very narrow pseudoscalar resonances $ \eta^{(} {'} {}^{)} $, leading to the lepton pair via two-photon exchange, are omitted); the bands correspond to $ (m \pm \Gamma/2)^2 $, with $\Gamma$ taken from Refs.~\cite{Garcia-Martin:2011nna,Pelaez:2015qba,Pelaez:2022qby,Workman:2022ynf}. The ``high-energy window'' referred to in the plot corresponds to $m_{\rho^0}^2 \lesssim q^2 \lesssim 1.5$~GeV${}^2$, for which only $f_0 (500) \equiv \sigma$ gives an important contribution among the $S$-wave contributions, and is indicated by a hashed pattern delimited by dashed vertical lines. Cascade decays are not indicated.}
    \label{fig:DeBoer_Hiller}
\end{figure}

%%%%%%% Other S- and P-waves %%%%%%%
Our study provides the first analysis of the $S$-wave in rare charm-meson decays, and we discuss what can be learnt from this physics case; we focus on the $\sigma$ resonance, which alone impacts a large portion of the allowed phase space, see Fig.~\ref{fig:DeBoer_Hiller} (that extends a figure from Ref.~\cite{DeBoer:2018pdx}).
Considering other scalar isoscalar resonances, let us point out the following:
$f_0 (980)$ is included in the analysis of Ref.~\cite{BESIII:2018qmf}, and is not observed to provide a significant contribution;
$f_0 (1370)$ is a very broad resonance that ``overlaps'' partially with $\rho^0 / \omega \to \ell^+ \ell^-$ in the $ q^2 (\ell^+ \ell^-) $ vs. $ p^2 (\pi^+ \pi^-) $ plane;
$f_0 (1500)$ (of width $\sim 100$~MeV \cite{Workman:2022ynf}) has an important branching ratio into pion pairs of approximately $35\%$ \cite{Workman:2022ynf}, but is restricted to a region that ``overlaps'' little with $\rho^0 / \omega \to \ell^+ \ell^-$;
similarly, $f_0 (1710)$ (of width $\sim 100$~MeV \cite{Workman:2022ynf}) is also restricted to the low-energy window of the lepton pair.
On the other hand,
more is known about the lightest states, which affect a more significant portion of the phase space.
Therefore, we will not include $S$-wave resonances other than the $\sigma$. Instead, we focus on energies $q^2 (\ell^+ \ell^-) \gtrsim m_\rho^2$, reducing the need to include further contributions.
Given the kinematical window we focus on, we do not discuss the Bremsstrahlung contribution (where a soft photon is emitted from $ D^0 \to \pi^+ \pi^- $),
%[\LVS{why J- and A-types are not called Bremsstrahlung?, because they do not peak at $\sim 0$ GeV?: J and A are not soft emissions, but Bremsstrahlung is}]
see Ref.~\cite{Cappiello:2012vg,Low:1958sn} for its description, which is more relevant in the electron-positron than in the muon pair case.\footnote{The differential branching ratio as a function of $p^2 (\pi^+ \pi^-)$ is dominated by $ \mu^+ \mu^- $ resonant contributions (i.e., after integration of the fully differential branching ratio over the variable $q^2 (\mu^+ \mu^-)$), and thus Bremsstrahlung represents a correction that we neglect. This is a very good approximation particularly at low $p^2 (\pi^+ \pi^-)$ \cite{Cappiello:2012vg}.}
For the same reason, $D$-wave resonances are not included.
Moreover, we sum over the lowest lying unflavoured vector resonances, and thus, for instance, $ \rho (1450) $ is not included, further limiting the kinematic window to $q^2 (\ell^+ \ell^-) \lesssim 1.5$~GeV${}^2$.
LHCb \cite{LHCb:2021yxk,LHCb_supplementary_material_aps,LHCb_supplementary_material_4} collected plenty of data in the region delimited by the two above conditions, namely, $m_\rho^2 \lesssim q^2 (\ell^+ \ell^-) \lesssim 1.5$~GeV${}^2$ (no bins simultaneously in both $q^2 (\ell^+ \ell^-)$ and $p^2 (\pi^+ \pi^-)$ are provided in their analysis).
We postpone to future work the discussion of isospin-two contributions to the $S$-wave, which is non-resonant at sufficiently low energies \cite{Workman:2022ynf} and thus in particular its phase motion does not experience a large variation \cite{Durusoy:1973aj}: in practice, it decreases steadily starting from $ 2 \, m_\pi $, and achieves about $ -25 $ degrees at around $1$~GeV. %Note that I know the phase shift, but I don't have a meaningful Omnes representation available

%[\LVS{put second dashed line}]
%<photon^{\ast}|(qqbar)V|eta(')>=<0|(a.polarization^\ast)*(qqbar)V|eta(')>, under C: invariant; under P: invariant (taking the polarization four-vector to flip an overall sign other than moving the Lorentz index), so I would need to consider the momentum times a coefficient; contracting with momentum on both sides, the latter coefficient must be zero (because the current is conserved)
%<photon|(qqbar)A|eta(')>, under C: flips sign, so there is no available Lorentz structure (which are classical)

%\begin{figure}
%    \centering
%    \includegraphics[scale=0.3]{DeBoer_Hiller_3.png}
%    \caption{Modification of Fig.~1 from \cite{DeBoer:2018pdx}: the solid purple horizontal lines indicate the approximate region of $f_0 (500)$ (with parameters taken from \cite{Pelaez:2015qba}); it shows that this state contaminates the high-energy window. The dashed purple horizontal line indicates the contribution of $f_0 (980)$, while the solid yellow horizontal line indicates the approximate region of $f_0 (1370)$.}
%    \label{fig:DeBoer_Hiller}
%\end{figure}

%%%%%%% Distinct modes %%%%%%%
Concerning other rare charm-meson decay modes with pion pairs in the final state,
we note that the channel $D^\pm \to \pi^\pm \pi^0 \ell^+ \ell^-$ is not sensitive to the $S$-wave contributions under discussion and is experimentally more challenging.
The mode $D^0 \to \pi^0 \pi^0 \ell^+ \ell^-$ (which does not receive contributions of the $P$-wave, following Bose-Einstein symmetry) represents an even more significant experimental challenge. These decay modes will thus not be discussed in the following.
Limits on the electronic mode $D^0 \to \pi^+ \pi^- e^+ e^-$ branching ratio are discussed in Refs.~\cite{CLEO:1996jxx,BESIII:2018hqu}.

%%%%%%% B physics %%%%%%%
Before concluding this introduction,
let us point out that
the $S$-wave contribution is relevant also in the bottom sector.\footnote{For the theoretical treatment of $K_{L,S} \to \pi^+ \pi^- \ell^+ \ell^-$ decays, see Ref.~\cite{Sehgal:1992wm,Elwood:1995xv,Pichl:2000ab,Cirigliano:2011ny,Cappiello:2011qc}; see also Ref.~\cite{Bijnens:1994me} for $K_{\ell 4}$ decays.}
For a discussion in the case of $B^0 \to K^+ \pi^- \ell^+ \ell^-$, where the $S$-wave contamination from $B^0 \to K^\ast_0 (\to K^+ \pi^-) \ell^+ \ell^-$ in the reconstruction of the decay chain is at the level of $ \approx 10\% $, see Refs.~\cite{Becirevic:2012dp,Blake:2012mb,Bobeth:2012vn,Doring:2013wka,Das:2014sra,Alguero:2021yus,Descotes-Genon:2023ukb}; note that LHCb has performed measurements of the $S$-wave contribution, e.g., in Refs.~\cite{LHCb:2016ykl,LHCb:2016eyu}.
In the cases of scalar isoscalar states, the $S$-wave has been discussed for $B_{(s)} \to \pi \pi J/\psi$ \cite{Liang:2014tia,Daub:2015xja,Ropertz:2018stk}, which contributes to $B_{(s)} \to \pi \pi \ell^+ \ell^-$;
%$B_{(s)} \to \pi \pi \nu \bar{\nu}$: this is a higher dimensional process
note that the $\sigma$ is expected to provide a sizable contribution, naively as large as $\approx 26\%$, and thus coincident with the result of BESIII \cite{BESIII:2018qmf} in the charm-sector, since $ \mathcal{B} (B^0 \to \rho^0 J/\psi (1S)) \simeq 2.6 \times 10^{-5} $ \cite{LHCb:2014vbo,BaBar:2007yvx}, while $ \mathcal{B} (B^0 \to \sigma J/\psi (1S)) \simeq 0.9 \times 10^{-5} $ \cite{LHCb:2014vbo}.
A process related to the final state with pion pairs
is $B_{(s)} \to K K \ell^+ \ell^-$ \cite{Becirevic:2012dp,Doring:2013wka,Das:2014sra}, due to final-state rescattering \cite{Liang:2014tia,Daub:2015xja,Ropertz:2018stk}.
Important contributions of the $S$-wave are in principle also to be expected in semi-leptonic decays $B^+ \to \pi \pi \ell^+ \nu_\ell$ ($ \ell = e, \mu, \tau $) \cite{Kang:2013jaa,Faller:2013dwa}, and should then be taken into account in future tests of the SM, such as lepton flavour universality; see Ref.~\cite{Leskovec:2022ubd} for a discussion of the extraction of
the $P$-wave contribution from a lattice QCD calculation.
%Apparently, there is an anomaly in an analysis by Belle, which does not seen the S-wave
See Refs.~\cite{Becirevic:2016hea,LeYaouanc:2022dmc,Gubernari:2023rfu,Gustafson:2023lrz} for discussions of the $S$-wave contribution to $ \bar{B} \to D \pi \ell \bar{\nu}_\ell $.

%Messages and comparisons in $B$-physics? (such as $B \to (\pi \pi)_S \ell \ell$ \cite{Daub:2015xja,Ropertz:2018stk} (note that here there is much more phase-space; focusing on high-$q^2_{\mu \mu}$ then one contribution should come from $\sigma$, but I cannot exclude anymore other $f_0$ states, nor $f_2 (1270)$; note $ Br(B \to \rho J/\psi (1S)) = 2.6 \times 10^{-5} $, while $ Br(B \to \sigma J/\psi (1S)) = 0.9 \times 10^{-5} $); why J-type contributions are omitted?, likely because $B \to (\rho_{spec} \to \pi \pi) (J/\psi (1S) \to \ell \ell)$ and $B \to (J/\psi (1S) \to \pi \pi) (\rho_{spec} \to \ell \ell)$ are very different; PDG: $ Br(J/\psi (1S) \to \pi^+ \pi^-) = 1.5 \times 10^{-4} $ and $ Br(J/\psi (1S) \to \mu \mu) = 6.0 \times 10^{-2} $, while $ Br(\rho \to \pi^+ \pi^-) \simeq 1 $ and $ Br(\rho \to \mu \mu) = 5 \times 10^{-5} $; also note CKM suppression in $B \to (\rho_{spec} \to \pi \pi) (\rho \to \ell \ell)$/why would I care about two rhos anyways?).

%%%%%%% Index %%%%%%%
This article is organized as follows: in Sec.~\ref{sec:intermediate_dynamics} we formalize the inclusion of intermediate resonances; then, in Sec.~\ref{sec:observables} we discuss the theoretical expressions of distinct observables; finally, in Sec.~\ref{sec:analysis_data} we present our numerical comparisons with available data; conclusions are provided in Sec.~\ref{ref:conclusions}. In App.~\ref{app:numerical_inputs} we give the expressions of the line-shapes in use, among further useful hadronic information,
and some further comparisons regarding Ref.~\cite{BESIII:2018qmf} are given in App.~\ref{app:sl_decays}.
%and in Appendix~\LVS{XXX} we provide further numerical predictions

\section{Inclusion of intermediate resonances in naive factorization}\label{sec:intermediate_dynamics}

%%% Short distance %%%
To start, we introduce
the Single Cabibbo Suppressed (SCS) effective interaction Hamiltonian density for $ \Delta C = 1 $ up to operators of dimension-six, valid for energy scales $ \mu < \mu_b $ ($ \mu_b $ being the energy scale at which the bottom-quark is integrated out) \cite{Buchalla:1995vs}:

\begin{equation} \label{eq:H_eff}
	\mathcal{H}_{\rm eff} = \frac{G_F}{\sqrt{2}} \left[ \, \sum^2_{i = 1} C_i (\mu) \left( \lambda_d Q_i^d + \lambda_s Q_i^s \right) - \lambda_b \left( C_7 (\mu) Q_7 + C_9 (\mu) Q_9 + C_{10} (\mu) Q_{10} \right) \right] + \mathrm{h.c.}
\end{equation}

\noindent where

\begin{equation}
	\lambda_q = V^\ast_{c q} V_{u q} \,, \qquad q = d, s, b \,.
\end{equation}
The basis of operators is the following:
%[\rd{the signs in $\mathcal{H}_{\rm eff}$ seem correct (I have compared to Buras's notes); I flipped the sign in the definition of $Q_7$ (which was coming from Kagan, Brod, Zupan)}]

\begin{eqnarray}\label{eq:operator_list}
	&& Q_1^d = ( \overline{d} c )_{V-A} ( \overline{u} d )_{V-A} \,, \\
	&& Q_2^d = ( \overline{d}_j c_i )_{V-A} ( \overline{u}_i d_j )_{V-A} \stackrel[]{Fierz}{=} ( \overline{u} c )_{V-A} ( \overline{d} d )_{V-A} \,, \nonumber\\
	&& Q_1^s = (\overline{s} c)_{V-A} (\overline{u} s)_{V-A} \,, \nonumber\\
	&& Q_2^s = (\overline{s}_j c_i)_{V-A} (\overline{u}_i s_j)_{V-A} \stackrel[]{Fierz}{=} (\overline{u} c)_{V-A} (\overline{s} s)_{V-A} \,, \nonumber\\
    && Q_7 = \frac{e}{8 \pi^2} m_c \overline{u} \sigma_{\mu \nu} (\mathbf{1} + \gamma_5) F^{\mu \nu} c \,, \nonumber\\
    && Q_9 = \frac{\alpha_{em}}{2 \pi} ( \overline{u} \gamma_\mu (\mathbf{1} - \gamma_5) c ) ( \overline{\ell} \gamma^\mu \ell ) \,, \nonumber\\
    && Q_{10} = \frac{\alpha_{em}}{2 \pi} ( \overline{u} \gamma_\mu (\mathbf{1} - \gamma_5) c ) ( \overline{\ell} \gamma^\mu \gamma_5 \ell ) \,, \nonumber
\end{eqnarray}
where $ (V - A)_\mu = \gamma_\mu (\mathbf{1} - \gamma_5) $, $i, j$ are colour indices, and $\mu \sim \overline{m}_c (\overline{m}_c)$ is the renormalization scale. The operators $ Q_i^q $, $ q=d,s $ and $ i=1,2 $, are the current-current operators.
Above, we have not kept contributions in $ \lambda_b $ other than the electromagnetic dipole $Q_7$ and the semi-leptonic interactions $ Q_9 $ and $ Q_{10} $, which are kept for the only sake of later convenience.
%(i.e., gluonic and electroweak penguins and the gluonic dipole)
The (short-distance) SM Wilson coefficients $C_7, C_9, C_{10}$, first generated at one-loop via the exchange of EW gauge bosons, are significantly suppressed in the $D$ system \cite{deBoer:2016dcg},\footnote{Because of the GIM mechanism, there are no short-distance contributions to $C_7, C_9, C_{10}$ above the scale $ \mu_b $ at one-loop; $C_7, C_9$ are generated electromagnetically below $ \mu_b $ via single insertions of dimension-6 four-quark operators, while $ C_{10} $ is generated only via double insertions of dimension-6 operators, and thus of higher order in $G_F$ \cite{Cappiello:2012vg}. Such is also the case in di-neutrino decay modes \cite{Bause:2020xzj}.} and furthermore their contributions are accompanied with a CKM suppression;
%give numerical values: I don't see a real need
since $ C_{10} \sim 0 $ in the SM, we will see that some angular observables approximately vanish (i.e., those based in $I_{5,6,7}$). The main SM contribution to an effective $C_9$ comes from long-distance dynamics, as it will be later discussed in this section.
As stressed in Ref.~\cite{DeBoer:2018pdx}, the latter feature is welcomed in the sense that it enhances the sensitivity to NP that contributes to the observables that vanish in the SM, such as having $Q_{10}$ induced by NP which interferes with the large SM long-distance part.
Operators of flipped chirality, i.e., $ Q_7', Q_9', Q_{10}' $, are not displayed, and are virtually absent in the SM, their contributions being relatively suppressed by $m_u/m_c$.
For all purposes, we take $ \lambda_s = - \lambda_d $.

%in the following we assume CP odd phases from the SM to be negligible

\begin{figure}
    \centering
    \includegraphics[scale=0.27]{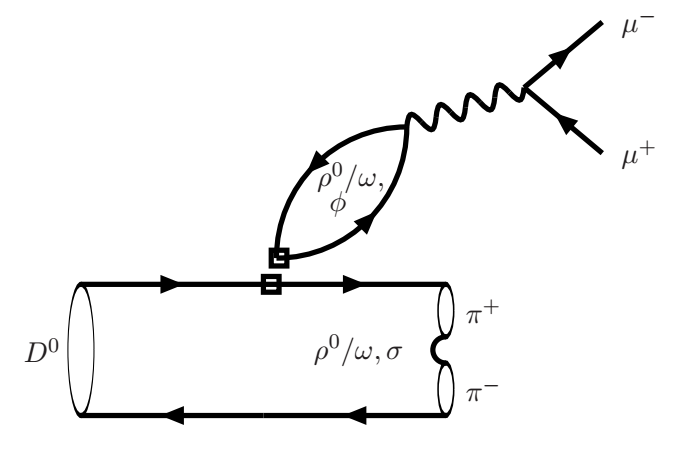} \hspace{9mm}
    \includegraphics[scale=0.23]{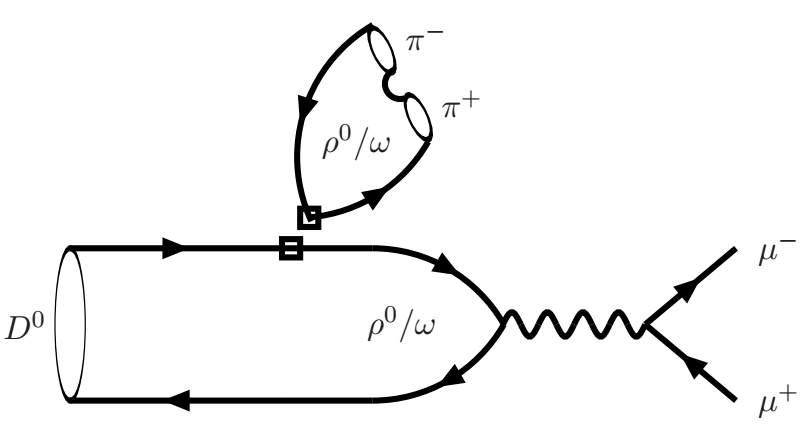} \\
    \vspace{5mm}
    \includegraphics[scale=0.24]{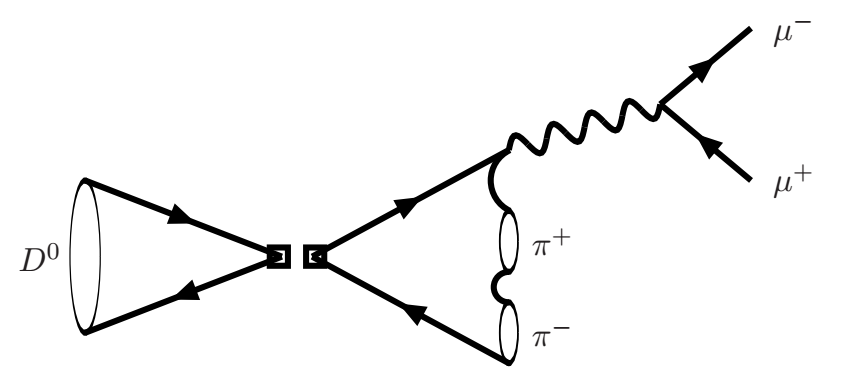} \hspace{9mm}
    \includegraphics[scale=0.27]{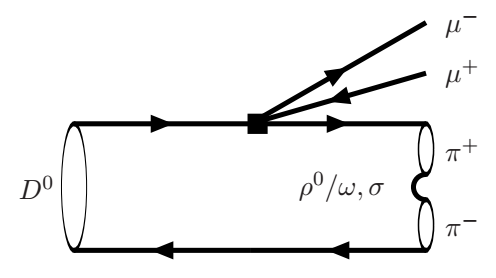}
    \caption{Quasi two-body topologies; the lepton (pion) pair comes from electromagnetic (respectively, strong) decays of the intermediate resonances; (Top, Left) W-type factorization contribution, (Top, Right) J-type factorization contribution, (Bottom, Left) A-type factorization contribution (i.e., annihilation topology); pairs of empty squares represent the two quark colour-neutral bilinears that are factorized. (Bottom, Right) Contributions for which the lepton pair comes from an effective semi-leptonic contact interaction, represented by a solid square.}
    \label{fig:Q2B}
\end{figure}

%%% Introduction: framework for non-perturbative matrix elements and relevant intermediate states %%%
The full decay amplitude of the charm-meson decay is calculated here in the framework of factorization, closely following Ref.~\cite{Cappiello:2012vg}.
We include in our analysis only the quasi two-body topologies with the lowest lying intermediate resonances that are indicated in Fig.~\ref{fig:Q2B}. %\ref{fig:cascade}
Therein,
the lepton pair originates from one vector meson, namely, $\rho^0$, $\omega$, or $\phi$, coupling to a photon (we neglect cases where one isoscalar hadron couples to two photons due to the small resulting effect, as supported by data, see e.g. Ref.~\cite{Dai:2014zta}; similarly, we do not include pseudoscalar resonances in our analysis). The pion pair originates from strong decays of $\rho^0$, $\omega$, or $\sigma$. The latter list does not include the $\phi$ since we assume the Zweig rule to be at play, i.e., we discard the possibility of a light-quark pair rescattering into $s\overline{s}$.
Since the intermediate resonances are electrically neutral, the only operators that contribute in naive factorization are $ Q_2^q $, $q = d, s$.
We employ the next-to-leading order (NLO) value $C_2 = -0.40 $ in the naive dimensional regularization (NDR) scheme at $ m_c $ \cite{Buchalla:1995vs,deBoer:2016dcg}.
%I DON'T KNOW WHAT IS MEANT BY "(NAIVE) FACTORIZATION": IS THERE A SEPARATION OF SHORT AND LONG DISTANCES?

%%% Whole matrix element %%%
We write schematically for the $S$-matrix element of the process:
%[\LVS{momentum transfer?; in and out states are being given in momentum space, while operators in position space}]
%Note that we are not writing \mathcal{H}_{\mathcal{R} \pi\pi} explicitly, nor \mathcal{H}_{D \mathcal{R} \mathcal{V}}, so I won't say the calculation is at tree-level, nor I will call (all) these Hamiltonian densities

\begin{eqnarray}
    && \qquad\qquad \langle \pi^+ \pi^- \ell^+ \ell^- | S |D^0 \rangle \\
    && = \langle \pi^+ \pi^- \ell^+ \ell^- | \int d^4 x \, d^4 w \, d^4 y \, d^4 z \, T \{ \mathcal{H}_{em}^{\rm lept}(z) \, \mathcal{H}_{\mathcal{V}\gamma}(y) \, \mathcal{H}_{\mathcal{R} \pi\pi}(w) \, \mathcal{H}_{D \mathcal{R} \mathcal{V}}(x) \} |D^0 \rangle \nonumber \,,
\end{eqnarray}%I took off the factor (-i)^4 %\sum_{ \mathcal{R} , \mathcal{V}}
with electromagnetic interactions given by \cite{Pich:2018ltt}:\footnote{Partial integration has been used to rewrite $\mathcal{H}_{\mathcal{V}\gamma} \propto F_{\mu \nu} \left(\partial^\mu {\mathcal{V}}^\nu-\partial^\nu {\mathcal{V}}^\mu\right)$ \cite{Cappiello:2012vg}, and we employed the gauge condition $\partial^\mu A_\mu=0$. Moreover, $\mathcal{H}_{em}^{\rm lept}$ consists only of an interaction term, and is not gauge invariant.}

\begin{equation}\label{HVgamma}
    \mathcal{H}_{\mathcal{V}\gamma} = -e \, \left(\frac{f_{\rho^0}}{\sqrt{2}m_{\rho^0}}\left(\rho^0\right)^\mu \,+\frac{1}{3}  \frac{f_{\mathcal{\omega}}}{\sqrt{2}m_{\mathcal{\omega}}} \, \mathcal{\omega}^\mu -\frac{\sqrt{2}}{3}\frac{f_{\mathcal{\phi}}}{\sqrt{2}m_{\mathcal{\phi}}} \, \mathcal{\phi}^\mu \right) \, \square A_\mu \,, \qquad \mathcal{H}_{em}^{\rm lept} = e \, A^\mu \, \overline{\ell} \gamma_\mu \ell \,.
\end{equation}

%--------------------------------------
%\input{Vgamma-DecayConstants}
%--------------------------------------

%[point out other resonances: $f_0 (980)$, $f_0 (1370)$, $f_0 (1500)$, $f_0 (1710)$]
Above, $\mathcal{R}$ is one of the vector or scalar resonances coupling to the pion pair, and $\mathcal{V}$ is the vector resonance coupling electromagnetically to the lepton pair.
The flavour changing interaction $\mathcal{H}_{D \mathcal{R} \mathcal{V}}$ results from insertions of the current-current operators $ Q_2^q $, $q = d, s$, of the weak Hamiltonian density in Eq.~\eqref{eq:H_eff}, while matrix elements of $\mathcal{H}_{\mathcal{R} \pi\pi}$ are discussed in Secs.~\ref{sec:rhopipi} and \ref{sec:sigmapipi} for intermediate vectors and the scalar, respectively.

%\begin{figure}[h]
%    \centering
%    \includegraphics[scale=0.1]{sketch-decays.png}
%    \caption{The chain of vertices entering the decay}    \label{badsketch}
%\end{figure}

%%% Topologies and annihilation %%%
Let us at this point define the specific topologies that show up within factorization given the intermediate states aforementioned.
There are three possible ways to contract the currents, shown graphically in Fig.~\ref{fig:Q2B}:
%(the notation 'W' and 'J' is inspired by \cite{Cappiello:2012vg})

\begin{equation}
    Q_A \equiv -\braket{ \mathcal{R} \mathcal{V}|(\overline{q}q)_{V-A}|0}\braket{0| \overline{u} \gamma^\mu \gamma_5 c |D^0 (p_D)} = \braket{ \mathcal{R} \mathcal{V}|(\overline{q}q)_A (x) |0} i f_D p_D^\mu \, e^{-i \, p_D \cdot x} \,,
\end{equation}
%I will use this phase factor in the following, i.e., the incoming momentum appears with a minus

\begin{equation}\label{eq:W_type_definition}
    Q_W \equiv \left\{
    \begin{array}{l}
    \braket{\mathcal{V}|(\overline{q}q)_V|0}
    \braket{ \mathcal{R} |(\overline{u}c)_{V-A}|D^0} \,, \quad \mathcal{R} = \rho^0, \omega \,, \\
    \\
    - \braket{\mathcal{V}|(\overline{q}q)_V|0} \braket{ \mathcal{R} |(\overline{u}c)_A|D^0} \,, \quad \mathcal{R} = \sigma \,,
    \end{array} \right.
\end{equation}

\begin{equation}
    Q_J \equiv \braket{ \mathcal{R} |(\overline{q}q)_V|0}\braket{\mathcal{V}|(\overline{u}c)_{V-A}|D^0} \,, \quad \mathcal{R} = \rho^0, \omega \,,
\end{equation}
where
%$R$ is the vector or scalar resonance that couples to the pion pair and $V'$ is the vector resonance that couples to the lepton pair
$q=d,s$.
Both quark bilinears are calculated at the same space-time point.
Above, we have already indicated explicitly which currents (whether vector, axial-vector, or both) give non-vanishing contributions, and which resonances are possible.
In particular,
note that
there is no $\sigma$ exchange in the $Q_J$ case, since the (axial)vector $ \braket{ \sigma |(\overline{q}q)_{V (A)} |0} $ matrix element vanishes.
%[\LVS{this statement does not make sense} {\color{blue} why not?} \LVS{you must specify which operator is giving a vanishing matrix element (think about $Q_6$ insertions in our CPV project)}, \LVS{if it is the vector matrix element (e.g., see how charge-conjugation operates), or the axial vector matrix element (e.g., see how parity operates), then I agree it indeed vanishes, but that is not being said explicitly}]
The type of contraction at the origin of $Q_A$, which is the weak annihilation topology,
is proportional to the light quark mass $m_q$, as seen from contracting the axial-vector current $ \overline{q} \gamma_\mu \gamma_5 q $ with the decaying charm-meson four-momentum $p_D^\mu$, and we will thus neglect this contribution compared to the other two, that are non-zero; see e.g. Ref.~\cite{Bauer:1986bm} for a discussion.
%(and Ref.~\cite{Feldmann:2017izn,Bharucha:2020eup}) [it is a different kinematical region, and I don't know exactly what is meant by annihilation in their papers]
%vanishes for all intermediate vector resonances; that is because we assume the Zweig rule to hold, i.e., a $\rho/\omega$ can only couple to another $\rho/\omega$ and not to the $\phi$. The vector form factor relating these vanishes at the limit of equal vector meson masses. {\color{red} The weak annihilation term could exist for the $\sigma$ coupling to $\rho/\omega$ but we consider it to be negligible.} {[\LVS{revision; see Bauer et al.}]}

%%% W-type contributions %%%
We are left with the types of contractions of $ Q_W $ and $ Q_J $, that we shall refer to as `W'- and `J'-type contractions, and to which we now turn and provide further details.
In the case of W-type factorization, we need to evaluate the following vacuum to lepton pair matrix element:
%[\LVS{only the vector part in $\braket{V'|(\overline{q}q)_{V-A}|0}$ contributes}]
%[\LVS{I am erasing an integral since the other bilinear is not being shown; I am omitting integrals in the second line; I added an overall "-1" from the two propagators}]
%would one perform integrations over momentum, the phase factor would lead to momentum conservation factors
%({\color{red} [LVS: comment on axial part of $<V'|(\overline{d}\gamma^\mu d)_L|0>$]})
%[\rd{the phase going with the phi contribution has to be flipped in the numerical studies}: not anymore]
%later, comment on $SU(3)$: relate signs for all resonances

\begin{eqnarray}\label{Wleptonic}
%    && \braket{\ell^+ \ell^- |T\bigl\{\ \int d^4 x \, d^4 y_2 \, d^4z \sum_{V'} \bigl(\lambda_d (\overline{d}(-\gamma^\mu \gamma_5)d)(x)+\lambda_s (\overline{s}(-\gamma^\mu \gamma_5)s)(x)\bigr) \mathcal{H}_{V'\gamma}(y_2)\mathcal{H}_{em}(z) \bigr\}|0} = \nonumber\\
    %&& \braket{\ell^+ \ell^- |T \left\{ \int d^4 x \, d^4 y_2 \, d^4z \sum_{V'} \left( \sum_{q=d,s} \lambda_q (\overline{q} q)_V(x) \right) \mathcal{H}_{V'\gamma}(y_2)\mathcal{H}_{em}(z) \right\} |0} \nonumber\\
    && \braket{\ell^+ \ell^- | \int d^4 y \, d^4 z \, T \left\{ \mathcal{H}_{em}^{\rm lept} (z) \, \mathcal{H}_{\mathcal{V}\gamma} (y) \, \left( \sum_{q=d,s} \lambda_q (\overline{q} q)_V (x) \right) \right\} |0} \nonumber\\ %\sum_{\mathcal{V}=\rho^0,\omega,\phi}
    && \quad = - \sum_{\mathcal{V}=\rho^0,\omega,\phi}\braket{\ell^+ \ell^- | H_{em}^{\rm lept} |\gamma^\ast}\frac{1}{q^2}\braket{\gamma^\ast| H_{\mathcal{V}\gamma} |\mathcal{V}}\frac{1}{P_{\mathcal{V}}(q^2)}\braket{\mathcal{V}| \sum_{q=d,s} \lambda_q (\overline{q}q)_V (x) |0} \nonumber\\
    && \quad\quad = - e^{i q \cdot x} \lambda_d e^2 \, (\overline{u}_\ell\gamma^\mu v_\ell) \left( \frac{c^W_{\rho^0} \, f_{\rho^0}^2}{P_{\rho^0}(q^2)}+\frac{c^W_\omega \, f_\omega^2}{P_{\omega}(q^2)}+\frac{c^W_\phi \, f_\phi^2}{P_{\phi}(q^2)} \right) \,,
\end{eqnarray}
%%[\rd{I didn't cross-check the overall signs}]
where $q^2$ is the invariant mass squared of the lepton pair, and
%$ c^W_{\rho^0} = 1/\sqrt{2} $, $ c^W_\omega = -1/(3\sqrt{2}) $ and $ c^W_\phi = {\color{black}-}\sqrt{2}/3 $
$ c^W_{\rho^0} = 1/2$, $ c^W_\omega = -1/6 $ and $ c^W_\phi = -1/3 $.
%\rd{ previously: $ c^W_{\rho^0} = 1 $, $ c^W_\omega = -1/3 $ and $ c^W_\phi = {\color{black}+}\sqrt{2}/3 $; fitted parameters quoted for these prefactors [LVS: what is the meaning of this last comment?]}
The expressions for the line-shapes will be later discussed in the text (see App.~\ref{app:line_shapes}).\footnote{We reserve the notation $\mathcal{H}$ for the Hamiltonian density, while $H$ denotes the Hamiltonian.}
Note that the $\phi$ contribution comes with the CKM factor $\lambda_s$, where $\lambda_s = -\lambda_d$. For the values of the decay constants, see App. \ref{app:decay_constants}.
%\rd{It might be in the end that $c^W_\phi=-\sqrt{2}/3$, because of our conventions.}
%In the above we have made use of the relations 
%\begin{eqnarray}
%    &\langle \mathcal{\phi}(q,\epsilon)|(\overline{s}s)_V |0\rangle = f_\mathcal{\phi} m_\mathcal{\phi} \epsilon^\ast (q)& \\
%    &\langle \mathcal{\rho}(q,\epsilon)|(\overline{d}d)_V |0\rangle = f_\mathcal{\rho} m_\mathcal{\rho} \epsilon^\ast (q)& \\
%    &\langle \mathcal{\omega}(q,\epsilon)|(\overline{d}d)_V |0\rangle =- f_\mathcal{\omega} m_\mathcal{\omega} \epsilon^\ast (q)&
%\end{eqnarray}

%{\color{blue} From just the quark content we get a minus and from the $\lambda_s$ another minus. $SU(3)$ arguments are not considered. Anyway the sign is absorbed in the phase which is introduced later.}
%and the $\lambda_d$ is factored out, and the relative sign is taken into account in the above expression

%%% J-type contributions %%%
In the J-type factorization, we need to evaluate the following $ D^0 \to \ell^+ \ell^- $ matrix element:

\begin{eqnarray}\label{Jleptonic}
    %&& \braket{\ell^+ \ell^- |T \left\{ \int d^4 x \, d^4 y_2 \, d^4z \sum_{V'=\rho,\omega}\lambda_d(\overline{u}\gamma_\mu c)(x) \mathcal{H}_{V'\gamma}(y_2)\mathcal{H}_{em}(z) \right\} |D^0} \\
    && \braket{\ell^+ \ell^- | \int d^4 y \, d^4 z \, T \left\{ \mathcal{H}_{em}^{\rm lept} (z) \, \mathcal{H}_{\mathcal{V}\gamma} (y) \, \lambda_d (\overline{u}\gamma_\mu c) (x) \right\} |D^0 (p_D) } \\ %\sum_{\mathcal{V}=\rho^0,\omega}
    && \quad = - \sum_{\mathcal{V}=\rho^0,\omega}\braket{\ell^+ \ell^- | H_{em}^{\rm lept} |\gamma^\ast} \frac{1}{q^2}\braket{\gamma^\ast| H_{\mathcal{V}\gamma} |\mathcal{V}}\frac{1}{P_{\mathcal{V}}(q^2)}\braket{\mathcal{V}| \lambda_d (\overline{u}\gamma_\mu c) (x) |D^0 (p_D) } \nonumber\\
    && \quad\quad = - e^{i (q - p_D) \cdot x} \lambda_d e^2 \, (\overline{u}_\ell\gamma_\mu v_\ell) \left( \frac{c^J_{\rho^0} \, f_{\rho^0}}{m_{\rho^0} P_{\rho^0}(q^2)}+\frac{c^J_\omega \, f_\omega}{m_\omega P_{\omega}(q^2)}+\frac{c^J_\phi \, f_\phi}{m_\phi P_{\phi}(q^2)} \right) \nonumber\\
    && \quad\quad\quad \times (A_1(p^2) \, c_1(q^2,p^2)+A_2(p^2) \, c_2(q^2,p^2)+V(p^2) \, c_V(q^2,p^2)+A_0(p^2) c_0(q^2,p^2)) \,, \nonumber
\end{eqnarray}
%%%[\rd{I didn't cross-check the overall signs}]
where $p^2$ is the invariant mass squared of the pion pair, and
%$ c^J_{\rho^0} = 1 $, $ c^J_\omega = +1/3 $ and $ c^J_\phi = 0 $
$ c^J_{\rho^0} = 1/\sqrt{2} $, $ c^J_\omega = +1/(3\sqrt{2}) $ and $ c^J_\phi = 0 $.
Again, the $\phi$ does not contribute due to its quark content (similarly, there is no contribution proportional to $ \lambda_s $). The form factors of $D\to \mathcal{V}$, $ \mathcal{V}=\rho^0,\omega$, are
%the ones of Ref.~\cite{Cappiello:2012vg} and
equal for the two resonances (see App.~\ref{app:form_factors} for details about their parameterizations); the functions $c_i(q^2,p^2)$, $ i=V,0,1,2$, encode the kinematical factors that accompany each form factor \cite{Wirbel:1985ji}.

%%% Comment about relative phases %%%
In the above Eqs.~\eqref{Wleptonic} and \eqref{Jleptonic}, the relative signs and numerical prefactors between $ \rho^0 $ and $ \omega $ can be quickly understood from the quark content of the vector resonances

\begin{equation}
    V_\mu^\phi \equiv \bar{s} \gamma_\mu s \,, \;\;
    V_\mu^\omega \equiv \frac{1}{\sqrt{2}} \left( \bar{u} \gamma_\mu u + \bar{d} \gamma_\mu d \right) \,, \;\;
    V_\mu^{\rho^0} \equiv \frac{1}{\sqrt{2}} \left( \bar{u} \gamma_\mu u - \bar{d} \gamma_\mu d \right) \,, \;\;
\end{equation}
where the quark content of the operators $V_\mu^\mathcal{V}$ is such that they can create or annihilate the vector meson $\mathcal{V}$,
and we enforce the Zweig rule (for corrections, see e.g. Ref.~\cite{Bharucha:2015bzk}). In terms of these operators, the hadronic electromagnetic current can be rewritten as:

\begin{equation}
    (j_{em}^{\rm had})_\mu = Q_s \, V_\mu^\phi + \frac{Q_u + Q_d}{\sqrt{2}} \, V_\mu^\omega + \frac{Q_u - Q_d}{\sqrt{2}} \, V_\mu^{\rho^0} \,,
\end{equation}
where $ Q_u = +2/3 $ and $ Q_d = Q_s = -1/3 $.

%%% Introduction of extra phases %%%
To accommodate further strong dynamics, we will in the following discussion associate a strong phase $\delta_{\{\mathcal{R}, \mathcal{V}\}}$ to each vertex $\braket{\mathcal{R} \mathcal{V}|\mathcal{H}_{D \mathcal{R} \mathcal{V}}|D^0}$; a similar approach is followed by Ref.~\cite{DeBoer:2018pdx}; see also Ref.~\cite{Bharucha:2020eup} (strong phases are extracted from $ e^+ e^- $ data in Refs.~\cite{Lyon:2014hpa,Bharucha:2020eup}).
It will be assumed that these strong phases vary slowly, the faster variations being expected from the line shapes, and one then takes the $\delta_{\{\mathcal{R}, \mathcal{V}\}}$ as constants under the assumption that the main resonances needed for phenomenological applications are included in our analysis. Such strong phases are introduced to represent rescattering effects that take place beyond (naive) factorization.
%(\rd{This leaves us with four arbitrary phases for the couplings of $D^0$ to $\rho^0-\rho^0/\omega$, $\rho^0-\phi$, $\sigma-\rho^0/\omega$ and $\sigma-\phi$ pairs of resonances.})
This leaves us with six arbitrary phases for the couplings of $D^0$ to $\rho^0-\rho^0$, $\rho^0-\omega$, $\rho^0-\phi$, $\sigma-\rho^0$, $\sigma-\omega$ and $\sigma-\phi$ pairs of resonances.
%[\LVS{why there is no omega with all the rhos?}]
We will shortly discuss the phases present in $D^0$ to $\omega-\rho^0$, $\omega-\omega$, $\omega-\phi$
%the relative phase of the $\omega$ with respect to the $\rho^0$ when allowing for further dynamics beyond the previous paragraph
(see the discussion following Eq.~\eqref{brhoom} below).
%We use the same phases for the $\rho$ and $\omega$ assuming isospin symmetry to hold [{\color{red}LVS: include reference}].
In practice, we will see that the presently measured $ d \Gamma / d q^2 $ distribution depends on the phase differences:
%[\LVS{I slightly changed the notation}]

\begin{eqnarray}
    && \Delta_1 \equiv\delta_{ \{ \rho^0/\omega, \rho^0 \} }-\delta_{ \{ \rho^0/\omega, \phi \} } \,, \quad
    \Delta_2 \equiv\delta_{ \{ \rho^0/\omega, \rho^0 \} }-\delta_{ \{ \rho^0/\omega, \omega \} } \,, \\
    && \Delta_3 \equiv \delta_{ \{ \sigma, \rho^0 \} }-\delta_{ \{ \sigma, \phi \} } \,, \quad\quad
    \Delta_4 \equiv \delta_{ \{ \sigma, \rho^0 \} }-\delta_{ \{ \sigma, \omega \} } \,, \nonumber
\end{eqnarray}
since $S$- and $P$-waves do not interfere in $ d \Gamma / d q^2 $; given that the $\omega$ and $\phi$ are narrow resonances, $ \Delta_1 - \Delta_2 $ and $ \Delta_3 - \Delta_4 $ do not play an important role. On the other hand, when discussing angular observables that depend on the $S$- and $P$-waves interference, the following extra phase difference is relevant:

\begin{equation}
    \Delta_{SP} \equiv \delta_{ \{ \sigma, \rho^0 \} }-\delta_{ \{ \rho^0/\omega, \rho^0 \} } \,,
\end{equation}
which completes the list of phase differences in the SM to be discussed below (i.e., out of six phases we have five independent differences among them).

%%%%%%%%%%%%%%%%%%%%%%%%%%%%%%%%%%%%%%%%%%%%%%%%%%%%%%%%%%%%%%%%%%%%%%%%%%%%%%%%%%%%%%%%%%%%%%%%%%%%%%%%%%%%%%%%%%%%%%%%%%%%%%%%%%%%%%%
\subsection{Implementation of the $\pi^+\pi^-$ $P$-wave contribution}\label{sec:rhopipi}

For the coupling of a vector resonance $V$ to the pion pair we use the following expression for the matrix element of $ H_{\mathcal{R} \pi\pi} $ resulting from strong interactions:
%[\LVS{I am erasing the integral over $w$}]

\begin{equation}
    %\int d^4 w \,
    \bra{\pi^+(p_1)\pi^-(p_2)} H_{\mathcal{R} \pi\pi} \ket{V(p, \lambda)}= F_{BW}(p^2) \, b_V \, \epsilon_{V}(p, \lambda) \cdot (p_1-p_2) \,,
    \end{equation}
%{\color{blue} [Ele: I think the rhs corresponds to the momentum space, without the $\int dw$]}
where the phenomenological form factor $F_{BW}$ is the so-called Blatt-Weisskopf barrier factor for a particle of spin-1; see App.~\ref{app:line_shapes} for definitions, and the review on resonances of Ref.~\cite{Workman:2022ynf}. The quantity $b_V$ is assumed not to carry any dynamics, and is extracted from the decay rate of $V\rightarrow \pi^+\pi^-$:
%\footnote{We point out the misprint in the corresponding expression of Ref. \cite{Cappiello:2012vg}.}

\begin{equation}
    \Gamma(V\rightarrow \pi^+\pi^-)=\frac{1}{48\pi}b_V^2 m_V^{-5} \lambda^{3/2}(m_V^2,m_\pi^2,m_\pi^2) \,,
\end{equation}
where $ \lambda (a,b,c)=a^2+b^2+c^2-2(a b +b c + c a) $.
In practice this relation is used only for $V=\rho^0$, for which we take $\mathcal{B}(\rho^0\rightarrow \pi^+\pi^-) = 1$, thus resulting in $ b_{\rho^0} = 5.92 $.

The line-shape of $\rho^0$ is expressed in the Gounaris-Sakurai parameterisation \cite{Gounaris:1968mw}, which implements finite-width corrections (see App.~\ref{app:line_shapes} for details). Following previous literature on $\rho^0/\omega$ contributions to $ e^+ e^- \to \pi^+ \pi^- $, we collect both resonances together by considering the expression:

\begin{equation} \label{brhoom}
    b_{\rho^0/\omega} (p^2) =b_{\rho^0} \left( 1 + a_\omega \, e^{i \phi_\omega} \, \text{RBW}_\omega (p^2) \right) \,,
\end{equation}
where the relativistic Breit-Wigner line-shape $\text{RBW}_\omega (s)$ is given in App.~\ref{app:line_shapes}.
In (naive) factorization, if only the W-type contraction was possible, then $\phi_\omega = 0$; on the contrary, in the J-type contraction, $\phi_\omega = \pi$.
In Eq.~\eqref{brhoom}, both contributions are collected together, and the phase $\phi_\omega$ will also accommodate further hadronic effects beyond (naive) factorization in our study.
%{\color{blue} In practice, the phase $\phi_\omega$ introduced here accounts for the difference between W- and J-type factorization: namely, if only the W-type contraction was possible (as is the case in semileptonic decays, which this parameterization has been inspired from), the $\omega$ term in Eq. \eqref{brhoom} would appear with a plus (and no extra phase); on the contrary, in the J-type contraction it appears with a minus. Since it is more practical to factor out the hadronic part of the amplitude, which depends on the variable $p^2$, the introduction of said phase mitigates the effect of the different signs present in the naive factorization, while it also creates room for accommodating further hadronic effects.} 
In Sec.~\ref{sec:analysis_data}, the parameters $a_\omega$ and $\phi_\omega$ of the coupling of the $\omega$ to two pions are fitted to the experimental differential branching ratio as a function of the invariant mass of the pion pair (a small but non-vanishing value of $a_\omega$ is generated from isospin-breaking effects, mixing the isospin-triplet $\rho$ and the isospin-singlet $\omega$ states). This is different from the implementation of the resonances in the matrix elements of the lepton pair, where the ${\rho^0}$ and $\omega$ contributions are added serially.
%and no relative phase is considered; we adopt such a procedure in lack of statistics around the ${\rho^0}/\omega$ peak in the dilepton mass distribution, as it will be made clearer later.
%%% Matrix element w/o sigma %%%
We then have for the contribution where the pion pair originates from ${\rho^0}/\omega$ resonances:
%%[\LVS{up to a phase factor}]

\begin{eqnarray}\label{fulldpipillrho}
    && \braket{\pi^+ \pi^- \ell^+ \ell^-| S |D^0}^{({\rho^0}/\omega)}= 
    (2 \pi)^4 \, \delta^{(4)} (p + q - p_D) \, \xi_2\frac{b_{{\rho^0}/\omega} (p^2) F_{BW}(p^2)}{P_{\rho^0}(p^2)}(\overline{u}_\ell\gamma_\mu v_\ell) \nonumber\\
    && \; \times \sum_{\mathcal{V}} \Bigg\{ \left[ \frac{c^W_{\mathcal{V}} B_{\mathcal{V}} f_{\mathcal{V}}^2 e^{i\delta_{ \{ {\rho^0/\omega}, \mathcal{V} \} }}}{P_{\mathcal{V}}(q^2)} \left( \frac{2q\cdot(p_1-p_2)}{m_D+\sqrt{p^2}}A_2(q^2)- \left( m_D+\sqrt{p^2} \right) A_1(q^2) \right) \right. \nonumber\\
    && \;\; \left. + \frac{1}{\sqrt{2}} m_{\rho^0} f_{\rho^0} \frac{c^J_{\mathcal{V}} B_{\mathcal{V}} f_{\mathcal{V}} e^{i\delta_{ \{ {\rho^0/\omega}, \mathcal{V} \} }}}{m_{\mathcal{V}}P_{\mathcal{V}}(q^2)} \left( \frac{2q\cdot(p_1-p_2)}{m_D+\sqrt{q^2}}A_2(p^2)- \left( m_D+\sqrt{q^2} \right) A_1(p^2) \right) \right] p_1^\mu \nonumber\\
    && \;\;\; + \left[ \frac{c^W_{\mathcal{V}} B_{\mathcal{V}} f_{\mathcal{V}}^2 e^{i\delta_{ \{ {\rho^0/\omega}, \mathcal{V} \} }}}{P_{\mathcal{V}}(q^2)} \left( \frac{2q\cdot(p_1-p_2)}{m_D+\sqrt{p^2}}A_2(q^2)+ \left( m_D+\sqrt{p^2} \right) A_1(q^2) \right) \right. \nonumber\\
    && \;\;\;\; \left. + \frac{1}{\sqrt{2}} m_{\rho^0} f_{\rho^0} \frac{c^J_{\mathcal{V}} B_{\mathcal{V}} f_{\mathcal{V}} e^{i\delta_{ \{ {\rho^0/\omega}, \mathcal{V} \} }}}{m_{\mathcal{V}} P_{\mathcal{V}}(q^2)} \left( \frac{2q\cdot(p_1-p_2)}{m_D+\sqrt{q^2}}A_2(p^2)+ \left( m_D+\sqrt{q^2} \right) A_1(p^2) \right) \right] p_2^\mu \nonumber\\
    && \;\;\;\;\; + \left[ \frac{c^W_{\mathcal{V}} B_{\mathcal{V}} f_{\mathcal{V}}^2 e^{i\delta_{ \{ {\rho^0/\omega}, \mathcal{V} \} }}}{P_{\mathcal{V}}(q^2)}\frac{-4 i V(q^2)}{m_D+\sqrt{p^2}} \right. \nonumber\\
    && \;\;\;\;\;\; \left. + \frac{1}{\sqrt{2}} m_{\rho^0} f_{\rho^0} \frac{c^J_{\mathcal{V}} B_{\mathcal{V}} f_{\mathcal{V}} e^{i\delta_{ \{ {\rho^0/\omega}, \mathcal{V} \} }}}{m_{\mathcal{V}} P_{\mathcal{V}}(q^2)}\frac{-4 i V(p^2)}{m_D+\sqrt{q^2}} \right] \epsilon^{\mu\nu\lambda{\rho}}p_{1\nu}p_{2\lambda}q_{\rho} \Bigg\} \,,
    %there was an i before the big bracket taken from Cappiello, but they talk about the transition amplitude calligraphic M
\end{eqnarray}
%\color{blue} [Ele: I think this is the $\mathcal{M}$, otherwise it needs a $(2\pi)^4\times \delta^{(4)}(...)$. The numerics are OK.]}
where
%$\mathcal{V}=\rho,\omega,\phi$ and $\mathcal{V}=\rho,\omega$
\begin{equation}
    %\xi_2=C_2\lambda_d e G_F/\sqrt{2}
    \xi_2 = \lambda_d \frac{G_F}{\sqrt{2}} e^2 C_2 (\mu) \,.
\end{equation}
%Constant factors accompanying the vector decay constants are left implicit in the sums over resonances $V,V'$.
The terms coming with $A_1(q^2)$, $A_2(q^2)$, $V(q^2)$ (respectively, $A_1(p^2)$, $A_2(p^2)$, $V(p^2)$) originate from the W-type (J-type) factorization, since the momentum transfer of the $D^0$ form factor is the one of the lepton pair (pion pair).\footnote{In the above we have used the approximation $m_\omega f_\omega \approx m_{\rho^0} f_{\rho^0}$ for the $0\rightarrow V$ term in the J-type factorization, in order to simplify the expression.}
Note that the $A_0$ contribution vanishes because it is accompanied by $q^\mu \overline{\ell}\gamma_\mu \ell=0$ in the W-type factorization, and by $p_1^2-p_2^2$ in the J-type factorization, also vanishing in the case of $\pi^+\pi^-$ final-state mesons.
%When contracted with the $0\to\pi\pi$ current, the $A_0$ contribution vanishes.
In the case of charm-meson decays the J-type contribution gives sizeable effects, as it is manifest from Eq.~\eqref{fulldpipillrho}.\footnote{The analogous J-type contribution in $ B^+ \to K^{(\ast)+} \ell^+ \ell^- $ transitions from current-current operators is $ V^\ast_{ub} V_{us} $-CKM suppressed with respect to the dominant contribution, which goes as $ V^\ast_{cb} V_{cs} $.
}

%%% Normalization %%%
In Eq.~\eqref{fulldpipillrho}, apart from the complex phases that correct the (naive) factorization picture, we have also introduced for the same sake the real and positive parameters $ B_{\rho^0} $, $ B_\omega $ and $B_\phi$ that will be adjusted from data, and are also assumed not to carry any dependence with the energy.
Note that a somewhat similar approach is followed by Ref.~\cite{DeBoer:2018pdx}, which fits the factors controlling the normalizations of the resonances around their respective peaks.

\subsection{Implementation of the $\pi^+\pi^-$ $S$-wave contribution}\label{sec:sigmapipi}

We now consider the effect of the $\sigma=f_0(500)$ resonance.
%following BES \cite{BESIII:2018qmf} who observed said resonance
The $\sigma$ is encoded in the $ w_+ $ and $ r $ form-factors of the $ D^0 \to \pi^+ \pi^- $ matrix element \cite{Wirbel:1985ji,Lee:1992ih,Khodjamirian:2020btr}:

\begin{eqnarray}\label{Dpipiff}
    && \braket{\pi^+(p_1)\pi^-(p_2)|(\overline{u}\gamma^\mu (1-\gamma_5){c}) (x)|D^0(p_D)} = e^{i x \cdot (p - p_D)} \{ i w_+ (p_1+p_2)^\mu \nonumber \\
    && \qquad + i w_-(p_1-p_2)^\mu + h \epsilon^{\mu\alpha\beta\gamma} (p_D)_\alpha (p_1+p_2)_\beta (p_1-p_2)_\gamma + i r q^\mu \} \,.
\end{eqnarray}
The contraction of $ q^\mu $ with the spinorial part of the
leptonic matrix element $ (\overline{u}_\ell \gamma^\mu v_\ell) $ in Eq.~\eqref{Wleptonic} vanishes, and thus
%The form factor of $q^\mu$ can be neglected at the limit of zero lepton masses [understand this better (\LVS{contract with $q_\mu$})].
the effect of the $S$-wave intermediate states appears only in the form factor $w_+$,
%[\LVS{write $ - \langle \pi^+(p_1)\pi^-(p_2)| \mathcal{H}_{R \pi \pi} | \sigma \rangle P^{-1} \langle \sigma | \overline{u}\gamma^\mu \gamma_5{c} (x)|D^0(p_D) \rangle $, $(p_1+p_2)^\mu$ is coming from the second term}]
to which the following $S$-wave term is added:

\begin{equation}\label{eq:form_factor_S_wave}
    w^{S}_{+} (p^2, q^2) = a_S (q^2) \mathcal{A}_S(p^2) \,, \quad a_S (q^2) = a_S (0) / \left( 1-\frac{q^2}{m_A^2} \right) \,.
\end{equation}
Here,
%vector meson dominance
the nearest pole
%i.e., one is replacing the current $\bar{u} \gamma_\mu \gamma_5 c$
is used \cite{BESIII:2018qmf}, for which we have $ m_A = 2.42$~GeV, where $A$ is the axial $D$-meson ($J^P = 1^+$). The quantity $a_S (0)$, assumed to be a constant,\footnote{A dynamical behaviour of $a_S (0)$ could for instance result from the annihilation topology.} represents a magnitude encompassing the strength of the transition $D\rightarrow \sigma$ multiplied by the coupling of $\sigma$ to the pion pair.
We extract it from fitting the experimental data. Following Ref.~\cite{BESIII:2018qmf}, the lineshape $\mathcal{A}_S(p^2)$ is the one of Bugg \cite{Bugg:2006gc}, which is data-driven (and in particular includes small Zweig-violating effects);
%i.e., based on rescattering data: not only, $ J/\psi \to \omega \pi \pi $ should be an important role
%[\LVS{develop; comment on phase-shift}]
its full expression is provided in App.~\ref{app:line_shapes}. The complex phase assigned to the $ \sigma $ is close to the one extracted from $ \pi \pi $ rescattering in the elastic region. We reserve the analysis of alternative line-shapes
%and other $S$-wave contributions
to the future when the quest of higher precision may become more pressing.
%(previously: $\frac{1}{1-\frac{q^2}{m_A^2}}a_s e^{i\phi_s}\mathcal{A}_s(m)$; $\phi_s$ was a phase compared to the $P$-wave amplitude but which we don't have access to with the current observables, and anyway it is absorbed in the phases of the vertices $\sigma-\rho/\omega$ and $\sigma-\phi$).  

With all the above, we incorporate the scalar resonance to our factorization model
%%[\LVS{up to a phase}]
%namely, the hadronic part comes from the scalar resonance and the dilepton comes from an intermediate vector meson:

\begin{eqnarray}
    \langle \pi^+ \pi^- \ell^+ \ell^-| S |D^0 \rangle^{(\sigma)} &=& (2 \pi)^4 \, \delta^{(4)} (p + q - p_D) \, \xi_2 (\overline{u}_\ell\gamma_\mu v_\ell) i \\
    && \qquad \times \sum_{\mathcal{V}}\frac{c^W_{\mathcal{V}} B_{\mathcal{V}}^{(S)}f_{\mathcal{V}}^2 e^{i\delta_{ \{ \sigma, \mathcal{V} \} }}}{P_{\mathcal{V}}(q^2)} a_S (q^2) \mathcal{A}_S(p^2) \,. \nonumber
\end{eqnarray}
The full matrix element is then given by

\begin{equation}\label{eq:S_matrix_decomposition}
    \langle \pi^+ \pi^- \ell^+ \ell^-| S |D^0 \rangle = \braket{\pi^+ \pi^- \ell^+ \ell^-| S |D^0}^{({\rho^0}/\omega)} + \langle \pi^+ \pi^- \ell^+ \ell^-| S |D^0 \rangle^{(\sigma)} \,.
\end{equation}

\subsection{Effective Wilson coefficient}

%%For phenomenological applications it would be useful to encode the resonant matrix elements into an effective Wilson coefficient of the $Q_9$ operator defined in Eq.~\eqref{eq:operator_list}, in which the quark pair carries the chiral, $V-A$ structure and the lepton pair a vector structure.
%\begin{equation}
%    Q_9=(\overline{\ell}\gamma^\mu \ell)(\overline{u}\gamma_\mu (1-\gamma_5) c)
%\end{equation}
%In such a case
%the resulting $C^{\rm eff}_9$ Wilson coefficient
%contain the long-distance contributions mediated by the resonances, and
%would depend on the squared invariant masses $p^2, q^2$ from the long-distance dynamics.

It would be useful to write the previous matrix element in Eq.~\eqref{eq:S_matrix_decomposition} as the matrix element of a semi-leptonic four-fermion operator, with the intermediate resonance at the origin of the lepton pair encoded in an effective Wilson coefficient.
Assuming that the only factorization is the W-type one, as is the case for instance in semi-leptonic non-rare decays, it is easy to match the full hadronic matrix element to that of a $Q_9$ operator, i.e., in which the quark pair carries the chiral $V-A$ structure, and the lepton pair a vector structure, as it would result from the coupling to a single photon.
%%then, by noticing that in Eq.~\eqref{Wleptonic} the matrix element of the electromagnetically-dressed hadronic current that couples to the leptonic pair is rewritten as $\braket{\ell\ell|j^\mu|0}$
As seen from Eqs.~\eqref{eq:W_type_definition} and \eqref{Wleptonic},
the matrix element $ \langle \pi^+ \pi^- | (\bar{u} c)_{V-A} (x) | D^0 \rangle $ for initial and final state mesons has been factorized out from the leptonic matrix element, and we are able to write the latter as $ \langle \ell^+ \ell^- | ( \bar{\ell} \ell )_V (x) | 0 \rangle $ times an effective coefficient that encodes the intermediate resonant dynamics of the lepton pair invariant mass
%({\color{red}up to EW corrections}) %LVS: I am not worried about renormalization, but whether the leptonic operator can lead to a non-leptonic structure %I won't worry about this
%%[\LVS{up to a phase}]

\begin{eqnarray}
    C^{{\rm eff}:W}_9 (\mu; q^2) &=& 8 \pi^2 C_2 (\mu) \left( \frac{c^W_{\rho^0} f_{\rho^0}^2}{P_{\rho^0}(q^2)} B_{\rho^0} e^{i\delta_{ \{ {\rho^0/\omega}, {\rho^0} \} }} \right. \\
    && \;\; + \left. \frac{c^W_\omega f_\omega^2}{P_{\omega}(q^2)} B_\omega e^{i\delta_{ \{ {\rho^0/\omega}, {\omega} \} }}  + \frac{c^W_\phi f_\phi^2}{P_{\phi}(q^2)} B_\phi e^{i\delta_{ \{ {\rho^0/\omega}, {\phi} \} }} \right) \nonumber
\end{eqnarray}
(where we have included the factors beyond (naive) factorization that have been previously discussed), such that the transition $c \to u \ell^+ \ell^-$ is described by
%%[\LVS{sign}]
%following the notation introduced in Eq. \eqref{eq:H_eff} [\LVS{this phrase is too encrypted}].
%assigned at the vertices

\begin{equation}\label{eq:eff_Hamiltonian_C9}
	\mathcal{H}^{c \to u \ell \ell}_{\rm eff} = \frac{G_F}{\sqrt{2}} \lambda_d C_9^{{\rm eff}:W} (\mu; q^2) Q_9 + \mathrm{h.c.}
\end{equation}

Conversely, the matrix element appearing in the J-type contribution is $ \langle \mathcal{V} | (\bar{u} c)_{V-A} | D^0 \rangle $, where $\mathcal{V}$ does not lead to the pion pair, but instead to the lepton pair, so we cannot separate the full matrix element into a hadronic times a leptonic factors calculated at the same space-time point.
%%On the other hand, for the case of the J-type factorization, inspecting Eq. \eqref{Jleptonic} we see that the full matrix element could be recast as
%({\color{red}again, up to EW corrections}) %I won't worry about this
%%\begin{equation}
%%    C^{\rm eff,J}(q^2,p^2)<\ell\ell\pi\pi|(\overline{\ell}\gamma^\mu \ell) (z) (\overline{d}\gamma_\mu(1-\gamma_5) d)|0>\label{Jtype}
%%\end{equation}
%with \begin{equation}
%    C^J_{eff}(q^2,p^2)=e^2 \bigl(\frac{f_\rho}{P_{\rho}(q^2)}-\frac{f_\omega}{3 P_{\omega}(q^2)}\bigr)e^{i\phi_\rho}(A_1(p^2)\cdot c_1(q^2,p^2)+A_2(p^2)\cdot c_2(q^2,p^2)+V(p^2)\cdot c_V(q^2,p^2))
%\end{equation}
%%Eq.~\eqref{Jtype} does not come in the desirable form of a $Q_9$ matrix element, as the initial state is the vacuum instead of $D^0$.
Thus this contribution prevents us from writing, at least straightforwardly, our full amplitude using an effective Wilson coefficient multiplying a semi-leptonic four-fermion operator.

In the following we explore an alternative which would make the use of an approximate effective $C_9$ coefficient viable, if the $\rho^0 /\omega$ were the only resonances creating the pion pair. Starting with the $\rho^0$, we rewrite the J- and W-type contributions in a way that approximates an effective coefficient for $Q_9$. By inspecting Eq.~\eqref{fulldpipillrho}, one condition is that
\begin{equation}
    %%m_\rho f_\rho \frac{f_{\rho}}{m_{\rho}P_{\rho}(q^2)}\frac{1}{m_D+\sqrt{q^2}}F(p^2)\,  \text{  and  } \, \frac{f_\rho^2}{P_{\rho}(q^2)}  \frac{1}{m_D+\sqrt{p^2}}F(q^2) \,, \; F=A_2,V \label{cond1}
    m_{\rho^0} f_{\rho^0} \frac{f_{\rho^0}}{m_{\rho^0}P_{\rho^0}(q^2)}\frac{1}{m_D+\sqrt{q^2}}F(p^2) \simeq \frac{f_{\rho^0}^2}{P_{\rho^0}(q^2)}  \frac{1}{m_D+\sqrt{p^2}}F(q^2) \,, \; F=A_2,V \,, \label{cond1}
\end{equation}

%%as well as 
%%\begin{equation}
%%    m_\rho f_\rho \frac{f_{\omega}}{m_{\omega}P_{\omega}(q^2)}\frac{1}{m_D+\sqrt{q^2}}F(p^2) \, \text{  and  } \, - \frac{f_\omega^2}{P_{\omega}(q^2)}  \frac{1}{m_D+\sqrt{p^2}}F(q^2) \label{cond2}
%%\end{equation}

\noindent
%%should contribute in the same way in the observables;
while a similar discussion holds for the terms that are proportional to the form factor $A_1$.
To achieve our goal, we need firstly to examine if the $m_D+\sqrt{q^2}$ and $m_D+\sqrt{p^2}$ factors can be replaced with $m_D+m_\rho$, as it is usually done in the literature.
Indeed,
%we find that
this narrow-width approximation is good enough.
%for our purposes, seeing that the predictions for the observables do not change significantly with this replacement
What is left of the above conditions in Eq.~\eqref{cond1} comes from the dependencies of the form factors on $q^2$ or $p^2$. Since in our nearest pole parameterisation of the form factors in App.~\ref{app:form_factors} these dependencies go as ${m_{\rm pole}^2}/({m_{\rm pole}^2}-{q^2})$ or ${m_{\rm pole}^2}/({m_{\rm pole}^2}-{p^2})$, and the di-lepton and di-hadron invariant masses are generally much smaller than the pole masses, the two dependencies are soft.

The situation is more complicated for the $\omega$.
Since $ c^W_\omega $ and $ c^J_\omega $ have opposite signs, seemingly the $\omega$ contribution in the leptonic part would disappear in Eq.~\eqref{fulldpipillrho} under the use of the simplifications discussed in the previous paragraph.
However, when considering the original picture before the introduction of $b_{\rho^0 / \omega} (p^2)$ in Eq.~\eqref{brhoom}:

\begin{equation} 
   b_{\rho^0} \left( 1 + a_\omega \,  \text{RBW}_\omega (p^2) \right) c^W_\omega \frac{f_\omega^2}{P_\omega(q^2)}
\end{equation}
from the W-type, and 
\begin{equation} 
   b_{\rho^0} \left( 1 - a_\omega \,  \text{RBW}_\omega (p^2) \right) \frac{1}{\sqrt{2}} c^J_\omega \frac{f_\omega^2}{P_\omega(q^2)}
\end{equation}
from the J-type factorization, we see that an $\omega\to \ell^+ \ell^-$ contribution survives in the form of

\begin{equation}\label{eq:D_to_rho_pipi_omega_ellell}
    b_{\rho^0}  a_\omega \,  \text{RBW}_\omega (p^2) \left( c^W_\omega - \frac{1}{\sqrt{2}} c^J_\omega\right) \frac{f_\omega^2}{P_\omega(q^2)} \,;
\end{equation}
i.e., the contributions $ D^0 \to [\rho^0 \to \pi^+ \pi^-] \omega $ from the W- and the J-type terms largely cancel in naive factorization,
while the surviving $ D^0 \to [\omega \to \pi^+ \pi^-] \omega $ contributions
are suppressed due to the smallness of the factor $a_\omega$ coming from the small coupling of $\omega\to\pi\pi$.
%To introduce the $\omega\to\ell\ell$ contribution back to the simplified expression that groups together W and J- type as in Eq.~\eqref{brhoom}
Finally, the $\omega$ term is introduced in the effective Wilson coefficient with a small parameter $\epsilon_\omega \equiv a_\omega \, \text{RBW}_\omega (p^2) $, where the dependence on $p^2$ is not soft as in the previous paragraph. The presence of a $ p^2 $ dependence represents an impediment for the introduction of an effective $ C_9 $ coefficient, which should apply simultaneously for both $ \rho^0 $ and $ \omega $ decays to a pion pair in presence of both W- and J-type topologies; however, this represents only a small effect, suppressed by $ a_\omega $.
%\LVS{Note that the same cannot be said about the $\omega$ contributions, since as previously seen $ c^W_\omega $ and $ c^J_\omega $ have opposite signs; anyways, the $\omega$ does not give the most important contributions, and we focus on the $\rho$ and $\phi$ contributions.}

Under all of the above simplifications, one is able to define an approximate effective coefficient for $Q_9$ containing $P$-wave contributions as:
\begin{eqnarray}\label{C9mod}
    %%C_{9,P}^{\rm eff}(q^2)=e^2 \bigl(\frac{2 f_\rho^2}{P_{\rho}(q^2)}e^{i\phi_\rho}-\frac{2 f_\omega^2}{3 P_{\omega}(q^2)}e^{i\phi_\rho}{\color{black}+}\frac{\sqrt{2}f_\phi^2}{3P_{\phi}(q^2)}e^{i\phi_\phi}\bigr)
   && C_9^{{\rm eff}:P}(\mu; q^2) = 8 \pi^2 C_2 (\mu) \left[ \left( c^W_{\rho^0}+ \frac{1}{\sqrt{2}} c^J_{\rho^0} \right) \frac{f_{\rho^0}^2}{P_{\rho^0}(q^2)} B_{\rho^0} e^{i\delta_{ \{ {\rho^0/\omega}, {\rho^0} \} }} \right. \\
   && \qquad\qquad\qquad\quad \left. + \left( c^W_\omega- \frac{1}{\sqrt{2}} c^J_\omega\right) \frac{f_\omega^2}{P_\omega(q^2)} B_\omega \epsilon_\omega e^{i\delta_{ \{ {\rho^0/\omega}, {\omega} \} }}+c^W_\phi \frac{f_\phi^2}{P_{\phi}(q^2)} B_\phi e^{i\delta_{ \{ {\rho^0/\omega}, {\phi} \} }} \right] \,, \nonumber
\end{eqnarray}
where the $p^2$ dependence is omitted in $ \epsilon_\omega $, which as previously stressed represents a suppression factor.
%\rd{for simplicity, we ignore this contribution, setting $ \epsilon_\omega \to 0 $}
In contrast, the W- and J-type contributions add up coherently in the case of the $ D^0 \to [\rho^0 \to \pi^+ \pi^-] \rho^0 $ contribution and are unsuppressed.
%[\rd{stress coherent sum of contributions in the rho case, and else!}]
We remind the reader that there is no J-type contribution for the $\phi$, i.e., $c^J_\phi = 0$.
Therefore, we have for the $S$-matrix element of the process:

\begin{equation}\label{eq:C9mod_Pwave}
    \braket{\pi^+ \pi^- \ell^+ \ell^-| S |D^0}^{({\rho^0}/\omega)} \simeq (2 \pi)^4 \, \delta^{(4)} (p + q - p_D) \, C_9^{{\rm eff}:P}(\mu; q^2) \, \langle \pi^+ \pi^- \ell^+ \ell^- | Q_9 | D^0 \rangle^{({\rho^0}/\omega)} \,,
\end{equation}
which should be sufficient for our purposes given the present level of experimental accuracy in the high-energy window of Fig.~\ref{fig:DeBoer_Hiller}.

\begin{comment}
\rd{When considering the approximation Eq.~\eqref{eq:C9mod_Pwave}, the only sizable discrepancy that we find is in the quantity ${d \mathcal{B}}/{dq^2}$, which is overestimated up to $40\%$ in the high-$q^2$ region, see Fig.~\ref{withandwithoutJandWmod}. Therefore the use of the $C_9^{{\rm eff}:P}$ is not ideal, although sufficient to discuss vector resonances with respect to the decay rates, given the present level of experimental accuracy away from the resonance peaks.}
\end{comment}

For the $\sigma$, the discussion is simpler, since there is no J-type contribution:

\begin{equation}
    \langle \pi^+ \pi^- \ell^+ \ell^-| S |D^0 \rangle^{(\sigma)} = (2 \pi)^4 \, \delta^{(4)} (p + q - p_D) \, C^{{\rm eff}:S}_9 (\mu; q^2) \, \langle \pi^+ \pi^- \ell^+ \ell^- | Q_9 | D^0 \rangle^{(\sigma)} \,,
\end{equation}
with the $C^{{\rm eff}:S}_9$ given by:
%being identified with $C^{{\rm eff}:W}_9$ except for having its respective set of phases: 
\begin{eqnarray}\label{C9mod_S}
    C^{{\rm eff}:S}_9 (\mu; q^2) &=& 8 \pi^2 C_2 (\mu) \left( \frac{c^W_{\rho^0} f_{\rho^0}^2}{P_{\rho^0}(q^2)} B_{\rho^0}^{(S)} e^{i\delta_{ \{ {\sigma}, {\rho^0} \} }} \right. \\
    && \;\; \left. +\frac{c^W_\omega f_\omega^2}{P_{\omega}(q^2)} B_\omega^{(S)} e^{i\delta_{ \{ {\sigma}, {\omega} \} }}  + \frac{c^W_\phi f_\phi^2}{P_{\phi}(q^2)} B_\phi^{(S)} e^{i\delta_{ \{ {\sigma}, {\phi} \} }} \right) \,. \nonumber
\end{eqnarray}
%As it will be stressed later in the text,
Due to the cancellation discussed above, around Eq.~\eqref{eq:D_to_rho_pipi_omega_ellell},
the main contribution underlying $ \omega \to \ell^+ \ell^- $ is the one paired with $ \sigma \to \pi^+ \pi^- $.
Were the J-type contraction not considered, this would spoil the assessment from the fits of the size of the contribution $ D^0 \to [\sigma \to \pi^+ \pi^-] [\omega \to \ell^+ \ell^-] $.
%In lack of statistics
Note that $ B_{\rho^0}, B_\omega, B_\phi $ in Eq.~\eqref{C9mod} for the $P$-wave are allowed to be different with respect to Eq.~\eqref{C9mod_S} for the $S$-wave (moreover, an overall relative scale between $P$- and $S$-waves is absorbed into $a_S (0)$).
%\rd{[For simplicity, we set $ B_{\rho^0} = B_\omega $?]}
%For simplicity, we set $ B_\omega $ to zero.

Finally, we have:

\begin{eqnarray}
    \langle \pi^+ \pi^- \ell^+ \ell^-| S |D^0 \rangle &\simeq& (2 \pi)^4 \, \delta^{(4)} (p + q - p_D) \, \left( C_9^{{\rm eff}:P}(\mu; q^2) \, \langle \pi^+ \pi^- \ell^+ \ell^- | Q_9 | D^0 \rangle^{({\rho^0}/\omega)} \right. \nonumber\\
    && \qquad \left. + \, C^{{\rm eff}:S}_9 (\mu; q^2) \, \langle \pi^+ \pi^- \ell^+ \ell^- | Q_9 | D^0 \rangle^{(\sigma)} \right) \,,
\end{eqnarray}
which, due to the J-type contraction, is \textit{not} proportional to:

\begin{equation}
    \langle \pi^+ \pi^- \ell^+ \ell^- | Q_9 | D^0 \rangle \equiv \langle \pi^+ \pi^- \ell^+ \ell^- | Q_9 | D^0 \rangle^{({\rho^0}/\omega)} + \langle \pi^+ \pi^- \ell^+ \ell^- | Q_9 | D^0 \rangle^{(\sigma)} \,.
\end{equation}
%[\LVS{Explain as $(A+B)\times(a+b)$}]
As previously announced,
this prevents us from writing an effective coefficient that would apply simultaneously for both the intermediate $P$- and $S$-waves of the pion pair.

%%Note that even if we were to implement an approximate effective $C_9$ this should be done separately for the P-wave and the S-wave, since the scalar resonance comes only with the W-type factorization, and thus the $C_{9,P}^{\rm eff}$ written from the merging of the W- and J-type contributions is not applicable.

For our numerical results we use the full formulae with W- and J-type factorizations.
Nevertheless, for the sake of greatly simplifying the presentation of formulae in the next section, while keeping a good numerical accuracy,
we employ the notation $ C_9^{{\rm eff}:P} $ and $ C^{{\rm eff}:S}_9 %\equiv C^{{\rm eff}:W}_9 
$ introduced above.
%this should provide good numerical results

%in order to facilitate the discussion on the angular observables we present them in a language of two effective $C_9$ coefficients, one for the $\sigma\rightarrow\pi\pi$ decays and a second approximate one for the $\rho/\omega\rightarrow\pi\pi$ decays,
%keeping in mind that the numerical results do not change substantially for the regions where there are statistically significant data.

%(for example in $I_5$ where we have $\text{Re}(C_9 C_{10}^*)$, when integrated for the P-wave the correct term is $\text{Re}(C_W C_{10}^*)$ and $(C_J C_{10}^*)$ terms, or approximately $C_{9,mod}C_{10}^*$ where $C_{9,mod}$ comes from the W and J combination, but NOT from the S-wave contribution; if we used the experimentally fitted $C_9$ we would have a wrong $C_9^{eff}$ which would naturally include S-wave contributions, thus we would calculate $C_{10}$ wrongly; on the other hand, the $I_5$ integrated for the S-wave will have only the $C_9$ from $W$).

\section{Differential branching ratios and angular observables}\label{sec:observables}

%[\LVS{need to discuss the physics behind at times}]

A set of angular observables can be defined by integrating the differential decay rate of the process over the angular kinematical variables $\theta_\pi, \theta_\ell, \phi $: $ \theta_\ell $ is the angle between the $ \ell^- $-momentum and the $D$-momentum in the di-lepton center of mass frame,
$ \theta_\pi $ is the angle between the $ \pi^+ $-momentum and the negative $D$-momentum in the di-pion center of mass frame, and $ \phi $ is the angle between di-lepton and di-pion decay planes, oriented according to the normal vectors $ \hat{n}_\ell $ and $ \hat{n}_\pi $ of the planes $ ( \ell^- \ell^+ ) $ and $ ( \pi^+ \pi^- ) $ in the $D$ center of mass frame, respectively, from $ \hat{n}_\ell $ to $ \hat{n}_\pi $; with respect to Refs.~\cite{LHCb:2021yxk,LHCb_supplementary_material_aps,LHCb_supplementary_material_4}, our angle $\phi$ differs by $\pi$ (which means that the observables based on $I_4, I_5, I_7, I_8$ flip sign).
The total decay rate can be written as:
\begin{eqnarray}
    \frac{d^5\Gamma}{dq^2 dp^2 d\Omega}=\frac{1}{2\pi}\sum_{i=1}^9 c_i I_i \,,
\end{eqnarray}
where $d \Omega=d \cos\theta_\pi d \cos\theta_\ell d \phi$ and the constants $c_i$ are
\begin{eqnarray}
    && c_1=1 \,, \quad c_2=\cos 2 \theta_\ell \,, \quad c_3=\sin^2\theta_\ell \cos 2\phi \,, \\
    && c_4=\sin 2\theta_\ell \cos\phi \,, \quad c_5=\sin\theta_\ell \cos\phi \,, \quad c_6=\cos\theta_\ell \,, \nonumber\\
    && c_7=\sin\theta_\ell \sin\phi \,, \quad c_8=\sin 2\theta_\ell\sin\phi \,, \quad c_9=\sin^2\theta_\ell \sin2\phi \,. \nonumber
\end{eqnarray}

%%%%%%%%% Introduction of the functions I_i %%%%%%%%%
\noindent
We present the expressions for the coefficients $I_i$ in terms of the long-distance transversity form factors, the effective Wilson coefficients in the SM, distinguishing between the $S$- and the $P$-wave mediated cases, and the local Wilson coefficients introduced by NP. We follow closely the discussion of Refs.~\cite{Cappiello:2012vg,Das:2014sra,DeBoer:2018pdx}.\footnote{We correct Eq.~(A.6) from Appendix~A of Ref.~\cite{Cappiello:2012vg}, considering the conventions for the angles specified above; also, $ \epsilon_{0 1 2 3} = -1 $.
%[\LVS{discuss}] She adopted this convention in deriving our expressions; Gudrun does not need to specify it (she does not provide expressions where this would be needed; she does not have a single paper where all conventions are given at once)
}
Their expressions are as follows (the integrals $ \langle \cdot \rangle_{\pm} $ over $\theta_\pi$ will be defined below):
%[\LVS{dependece of I6 on the lepton mass? -- better adapted to use PL, PR basis, for which one should see that I6 should have a term for which leptons of different chiralities interfere (note chiral flipped cases in the reference by Diganta Das; Im parts of real quantities should be set to zero)}]

\begin{eqnarray}
    && I_1 = \frac{1}{8} \left[ | \mathcal{F}_S|^2 \, \rho^-_{1,S} + \cos^2\theta_\pi |\mathcal{F}_{P} |^2 \, \rho^-_{1,P} + \frac{3}{2} \sin^2 \theta_\pi \left\{ | \mathcal{F}_\Vert |^2 \, \rho^-_{1,P} + | \mathcal{F}_\perp |^2 \, \rho^+_{1,P} \right\} \right] \nonumber\\
    &&\quad +\langle I_1 \rangle_{-} \cos\theta_\pi \stackrel[]{SM}{\to} +\frac{1}{8}\biggl\{ \left[ \cos^2 \theta_\pi| \mathcal{F}_{P} |^2 + \frac{3}{2} \sin^2 \theta_\pi \left\{ | \mathcal{F}_\Vert |^2 + | \mathcal{F}_\perp |^2 \right\} \right] \, | C_{9}^{{\rm eff}: P}|^2 \nonumber\\
     &&\quad\quad +|\mathcal{F}_S|^2 | C_{9}^{{\rm eff}: S} |^2+2\text{Re}\{\mathcal{F}_S \, \mathcal{F}_{P}^\ast \, C_{9}^{{\rm eff}: S} \, (C_{9}^{{\rm eff}: P})^\ast \}\cos\theta_\pi\biggr\}\,,
\end{eqnarray}

\begin{eqnarray}
    && I_2 = -\frac{1}{8} \left[ | \mathcal{F}_S|^2 \, \rho^-_{1,S} + \cos^2\theta_\pi |\mathcal{F}_{P} |^2 \, \rho^-_{1,P} - \frac{1}{2} \sin^2 \theta_\pi \left\{ | \mathcal{F}_\Vert |^2 \, \rho^-_{1,P} + | \mathcal{F}_\perp |^2 \, \rho^+_{1,P} \right\} \right] \nonumber\\
    &&\quad +\langle I_2 \rangle_{-} \cos\theta_\pi \stackrel[]{SM}{\to} -\frac{1}{8}\biggl\{ \left[ \cos^2 \theta_\pi| \mathcal{F}_{P} |^2 - \frac{1}{2} \sin^2 \theta_\pi \left\{ | \mathcal{F}_\Vert |^2 + | \mathcal{F}_\perp |^2 \right\} \right] \, | C_{9}^{{\rm eff}: P}|^2 \nonumber\\
     &&\quad\quad +|\mathcal{F}_S|^2 | C_{9}^{{\rm eff}: S} |^2+2\text{Re}\{\mathcal{F}_S \, \mathcal{F}_{P}^\ast \, C_{9}^{{\rm eff}: S} \, (C_{9}^{{\rm eff}: P})^\ast  \}\cos\theta_\pi\biggr\}\,,
\end{eqnarray}

\begin{equation}
    I_3 = \frac{1}{8} \left[ | \mathcal{F}_\perp |^2 \, \rho^+_{1,P} - | \mathcal{F}_\Vert |^2 \, \rho^-_{1,P} \right] \, \sin^2 \theta_\pi \stackrel[]{SM}{\to} \frac{1}{8} \left[ | \mathcal{F}_\perp |^2 - | \mathcal{F}_\Vert |^2 \right] \, \sin^2 \theta_\pi \, | C_{9}^{{\rm eff}: P}|^2 \,,
\end{equation}

\begin{eqnarray}
    && I_4 =\cos\theta_\pi \sin\theta_\pi \frac{3}{2}\langle I_4 \rangle_{-}+\sin\theta_\pi\frac{2}{\pi}\langle I_4 \rangle_{+} \\
   && \quad \stackrel[]{SM}{\to} -\frac{1}{4} \text{Re} \{ \mathcal{F}_P \mathcal{F}^\ast_\Vert \} \, \cos\theta_\pi\sin \theta_\pi \, | C_{9}^{{\rm eff}: P}|^2 -\frac{1}{4} \text{Re} \{ \mathcal{F}_S \, \mathcal{F}^\ast_\Vert \, C_{9}^{{\rm eff}: S} \,(C_{9}^{{\rm eff}: P})^\ast \} \sin \theta_\pi \,, \nonumber
\end{eqnarray}

\begin{equation}
    I_5 = \cos\theta_\pi \sin\theta_\pi \frac{3}{2}\langle I_5 \rangle_{-}+\sin\theta_\pi\frac{2}{\pi}\langle I_5 \rangle_{+} \stackrel[]{SM}{\to} 0 \,,
\end{equation}

\begin{equation}
    I_6 = - \left[ \text{Re} \left\{ \mathcal{F}_\Vert \mathcal{F}^\ast_\perp \right\} \, \text{Re} \rho^+_2 + \text{Im} \left\{ \mathcal{F}_\Vert \mathcal{F}^\ast_\perp \right\} \, \text{Im} \rho^-_2 \right] \, \sin^2 \theta_\pi \stackrel[]{SM}{\to} 0 \,,
\end{equation}

\begin{equation}
    I_7 = \cos\theta_\pi \sin\theta_\pi \frac{3}{2}\langle I_7 \rangle_{-}+\sin\theta_\pi\frac{2}{\pi}\langle I_7 \rangle_{+} \stackrel[]{SM}{\to} 0  \,,
\end{equation}

\begin{eqnarray}
    && I_8 = \cos\theta_\pi \sin\theta_\pi \frac{3}{2}\langle I_8 \rangle_{-}+\sin\theta_\pi\frac{2}{\pi}\langle I_8 \rangle_{+} 
    \stackrel[]{SM}{\to} - \cos\theta_\pi \sin\theta_\pi \frac{1}{4} \text{Im} \left( \mathcal{F}_P \mathcal{F}^\ast_\perp \right) \, | C_{9}^{{\rm eff}: P}|^2 \nonumber\\
    && \qquad\qquad -\frac{1}{4} \sin\theta_\pi \text{Im} \left\{  \mathcal{F}_S \, \mathcal{F}^\ast_\perp \, C_{9}^{{\rm eff}: S} \, (C_{9}^{{\rm eff}: P})^\ast \right\} \,, \label{I8fact}
\end{eqnarray}

\begin{eqnarray}\label{Isfact}
    && I_9 = \frac{1}{2} \left[ \text{Re} \{ \mathcal{F}_\perp \mathcal{F}^\ast_\Vert \} \, \text{Im} \rho^+_2 + \text{Im} \{ \mathcal{F}_\perp \mathcal{F}^\ast_\Vert \} \, \text{Re} \rho^-_2 \right] \, \sin^2 \theta_\pi \\
    && \quad \stackrel[]{SM}{\to} \frac{\text{Im} \{ \mathcal{F}_\perp \mathcal{F}^\ast_\Vert \}}{4} \sin^2 \theta_\pi \, | C_{9}^{{\rm eff}: P}|^2 \,. \nonumber
\end{eqnarray}

The $0$-transversity form factor is %[\LVS{$ \mathcal{F}_0 $ not being used}]

\begin{equation}
    \mathcal{F}_0 = \mathcal{F}_S + \mathcal{F}_P \, \cos \theta_\pi \,;
\end{equation}
the P-wave form factors can be expressed as:

\begin{eqnarray}
 &&\mathcal{F}_P = -N \frac{b_{\rho^0/\omega} (p^2) F_{BW}(p^2)\sqrt{\beta_\ell(3-\beta_\ell^2)}\lambda_h^{3/4}\lambda_D^{1/4}}{P_{\rho^0}(p^2)} \, \\
 && \qquad\qquad \times \frac{ (m_D + m_{\rho^0})^2 (m_D^2 - p^2 - q^2) A_1(q^2) - \lambda_D A_2(q^2)}{2\sqrt{2}(m_D+m_{\rho^0})  (p^2)^{3/2} } \,, \nonumber\\
 &&\mathcal{F}_\Vert= N \frac{b_{\rho^0/\omega} (p^2) F_{BW}(p^2)\sqrt{\beta_\ell(3-\beta_\ell^2)}\lambda_h^{3/4}\lambda_D^{1/4}}{P_{\rho^0}(p^2)}\, \frac{\sqrt{q^2}(m_D+m_{\rho^0}) A_1(q^2)}{\sqrt{2} p^2 } \,, \\
 &&\mathcal{F}_\perp= -N \frac{b_{\rho^0/\omega} (p^2) F_{BW}(p^2)\beta_\ell^{3/2}\lambda_h^{3/4}\lambda_D  ^{3/4}}{P_{\rho^0}(p^2)}\, \frac{ \sqrt{q^2} V(q^2)}{(m_D+m_{\rho^0})p^2 } \,,
\end{eqnarray}
while for the $S$-wave

\begin{equation}
    \mathcal{F}_S=-N \frac{\sqrt{\beta_\ell(3-\beta_\ell^2)}\lambda_h^{1/4}\lambda_D^{3/4}}{P_{Bugg}(p^2)}\frac{a_S (q^2) }{2\sqrt{2} \sqrt{p^2}} \,,
\end{equation}
where $P_{Bugg}(p^2) = 1/\mathcal{A}_S (p^2)$.
The kinematic factors appearing in these expressions are
$\lambda_h=\lambda(p^2,m_\pi^2,m_\pi^2),\, \lambda_D=\lambda(m_D^2,p^2,q^2),\, \beta_\ell=\sqrt{1-4 m_\ell^2 / q^2}$.
%\footnote{
%\rd{previously:\begin{equation}
%    N=\frac{C_2 G_F \lambda_d}{64\pi^{2.5} e \sqrt{m_D}}
%\end{equation}}
%}
The overall normalization is:

\begin{equation}
    N=\frac{\alpha_{em} G_F \lambda_d}{128 \pi^{7/2}m_D^{3/2}} \,,
\end{equation}
owing to Eq.~\eqref{eq:eff_Hamiltonian_C9}.
%to the parameterization chosen for the effective Hamiltonian and the approximate $C_9$

%%%%%%%%% Wilson coefficients %%%%%%%%%
The Wilson coefficients, effective or not, are encoded in:
%how is the interference produced?: interference of 1+/-gamma5 would depend on lepton mass
\begin{eqnarray}
    && \rho^-_{1,S} = | C_{9}^{{\rm eff}: S} + C_9^{\rm NP} - C'_9 |^2 + | C_{10} - C'_{10} |^2 \,, \label{eq:WCrhos_0}\\
    && \rho^\pm_{1,P} = | C^{{\rm eff}: P}_{9} +C_9^{\rm NP} \pm C'_9 |^2 + | C_{10} \pm C'_{10} |^2 \,, \\
    && \delta \rho = \text{Re} \left[ \left( C^{{\rm eff}: P}_{9} +C_9^{\rm NP} - C'_9 \right) \left( C_{10} - C'_{10} \right)^\ast \right] \,, \\
    && \text{Re} \rho^+_2 = \text{Re} \left[ (C^{{\rm eff}: P}_{9} +C^{\rm NP}_9) \, C_{10}^\ast - C'_9 \, C'^\ast_{10} \right] \,, \\
    && \text{Im} \rho^+_2 = \text{Im} \left[ C'_{10} \, C_{10}^\ast + C'_9 \, ( C^{{\rm eff}: P}_{9} +C^{\rm NP}_9 )^\ast \right] \,, \\
    && \text{Re} \rho^-_2 = \frac{1}{2} \left( | C_{10} |^2 - | C'_{10} |^2 + | C^{{\rm eff}: P}_{9}  +C^{\rm NP}_9|^2 - | C'_9 |^2 \right) \,, \\
    && \text{Im} \rho^-_2 = \text{Im} \left[ C'_{10} \, (C_9^{{\rm eff}:P} +C^{\rm NP}_9)^\ast - C_{10} \, C'^\ast_9 \right] \,,
\label{WCrhos}
\end{eqnarray}
(as seen from the contributing currents in Eq.~\eqref{eq:W_type_definition}, a $\rho^+_{1,S}$ analogously defined does not show up).
%It follows from the discussion in the previous section that
The SM contribution comes from $C_9^{{\rm eff}: S}$ and $C_9^{{\rm eff}: P}$, while NP is at the origin of possibly large Wilson coefficients of the operators $Q'_9, Q_{10}, Q'_{10}$; NP could also contribute to $Q_9$.
%[the Wilson coefficient of $Q'_9$ is suppressed by $m_u / m_c$ $\longrightarrow$ only true in the SM]
%These operators differ in the leptonic part, and have different chirality projectors in the quark-bilinear part;
%all these operators could produce the $\mathcal{F}_0$ contribution.
%One could try to disentangle NP in non-SM operators from the SM contribution.
%
%%%%%%%%% Description %%%%%%%%%
%
Inspecting Eqs.~\eqref{eq:WCrhos_0}-\eqref{WCrhos},
note that: $\text{Im} \rho^\pm_2$ vanish in absence of having simultaneously the presences of $V-A$ and $V+A$ structures of the quark bilinears; these same combinations of Wilson coefficients vanish when no CP-violating phase is present; $\delta \rho$, $\text{Re} \rho^+_2$ and $\text{Im} \rho^-_2$ vanish in absence of having simultaneously the presences of $V$ and $A$ structures of the lepton bilinears.
Since we will focus on the high-$q^2$ energy window of Fig.~\ref{fig:DeBoer_Hiller}, we will not discuss $Q_7$ and $Q'_7$ operators.
Note however that part of the same SM background in the mode $D \to P P + [V' \to \gamma^\ast \to \ell^+ \ell^-]$ also manifests in radiative decays (e.g., $D \to P P + [V' \to \gamma]$, where compared to the semi-leptonic case one has a real photon). These decay modes would provide additional information on the contributions from dipole operators; see e.g. Ref.~\cite{Burdman:1995te,Greub:1996wn,Fajfer:2002bq,Adolph:2020ema,Adolph:2021ncg}. We reserve their analysis to future work.
Performing integration over the di-hadron angle in the following two ways:
\begin{equation}
    \langle I_i \rangle_{-} \equiv \left[ \int^{+1}_{0} d \cos \theta_\pi - \int^{0}_{-1} d \cos \theta_\pi \right] I_i \,, \quad \langle I_i \rangle_{+} \equiv \int^{+1}_{-1} d \cos \theta_\pi I_i \,,
\end{equation}
results in observables that either depend only on the $P$-wave ($\langle I_i\rangle_+$ for $i=3,6,9$ and $\langle I_i\rangle_-$ for $i=4,5,7,8$), receive non-interfering contributions from both the $S$- and the $P$-waves ($\langle I_i\rangle_+$ for $i=1,2$), or depend on the interference of the two waves ($\langle I_i \rangle_-$ for $i=1,2$ and $\langle I_i \rangle_+$ for $i=4,5,7,8$). 
%
%%%%%%%%% Integrated functions I_i %%%%%%%%%
%
Explicitly:
%In particular the expressions for the observables take the following form:
%[\LVS{where is $C_9^{\rm NP}$?}]

\begin{eqnarray}
    && \langle I_1 \rangle_{-} = \frac{1}{4} \text{Re} \left( \mathcal{F}_S \mathcal{F}^\ast_P  ((C_{9}^{{\rm eff}: S} + C_9^{\rm NP} - C_9')(C_{9}^{{\rm eff}: P} + C_9^{\rm NP} - C_9')^\ast+|C_{10}-C_{10}'|^2)   \right) \nonumber\\
    && \quad \stackrel[]{SM}{\to} \frac{1}{4} \text{Re} \left( \mathcal{F}_S \, \mathcal{F}^\ast_P \, C_{9}^{{\rm eff}: S} \, (C_{9}^{{\rm eff}: P})^\ast \right) \,,
\end{eqnarray}

%\begin{equation}
%    \langle I_2 \rangle_{-} = -\frac{1}{4} \text{Re} \left( \mathcal{F}_S \mathcal{F}^\ast_P  ((C_{9}^{{\rm eff}: S} + C_9^{\rm NP} - C_9')(C_{9}^{{\rm eff}: P} + C_9^{\rm NP} - C_9')^\ast+|C_{10}-C_{10}'|^2)   \right) = -\langle I_1 \rangle_{-} \,,
%\end{equation}

\begin{equation}
    \langle I_2 \rangle_{-} = -\langle I_1 \rangle_{-} \,,
\end{equation}

\begin{equation}
    \langle I_3 \rangle_{-} = 0 \,,
\end{equation}

\begin{equation}\label{I4vector}
    \frac{3}{2} \langle I_4 \rangle_{-} = -\frac{1}{4} \text{Re} \left( \mathcal{F}_P \mathcal{F}^\ast_\Vert \right) \, \rho^-_{1,P} \, \stackrel[]{SM}{\to}  -\frac{1}{4} \text{Re} \left( \mathcal{F}_P \mathcal{F}^\ast_\Vert \right) \, | C_{9}^{{\rm eff}: P}|^2 \,,
\end{equation}

\begin{equation}
    \frac{3}{2} \langle I_5 \rangle_{-} = \left[ \text{Re} \left( \mathcal{F}_P \mathcal{F}^\ast_\perp \right) \, \text{Re} \rho^+_2 + \text{Im} \left( \mathcal{F}_P \mathcal{F}^\ast_\perp \right) \, \text{Im} \rho^-_2 \right] \, \stackrel[]{SM}{\to} 0 \,,
\end{equation}

\begin{equation}
    \langle I_6 \rangle_{-} = 0 \,,
\end{equation}

\begin{equation}
    \frac{3}{2} \langle I_7 \rangle_{-} = \text{Im} \left( \mathcal{F}_P \mathcal{F}^\ast_\Vert \right) \, \delta \rho \, \stackrel[]{SM}{\to} 0 \,,
\end{equation}%\bigl(\, \stackrel[]{C_9^{eff}}{\to} 0 \, \bigr)

\begin{equation}\label{I8vector}
    \frac{3}{2} \langle I_8 \rangle_{-} = \frac{1}{2} \left[ \text{Re} \left( \mathcal{F}_P \mathcal{F}^\ast_\perp \right) \, \text{Im} \rho^+_2 - \text{Im} \left( \mathcal{F}_P \mathcal{F}^\ast_\perp \right) \, \text{Re} \rho^-_2 \right] \, \stackrel[]{SM}{\to} - \frac{1}{4} \text{Im} \left( \mathcal{F}_P \mathcal{F}^\ast_\perp \right) \, | C_{9}^{{\rm eff}: P}|^2 \,,
\end{equation}

\begin{equation}
    \langle I_9 \rangle_{-} = 0 \,,
\end{equation}
and (note that $ {d^2\Gamma} / {dq^2 dp^2} = 2 \langle I_1 \rangle_+ - \frac{2}{3} \langle I_2 \rangle_+ $):
%[\LVS{$I_5$, $I_8$}] Ele considered the paper by Jung et al. when doing this exercise; expressions are quadratic in the WCs, you can trace back the associated non-perturbative dynamics from the $^{\ast}$; the relative signs among WCs and their flipped/primed counter-parts correspond to the contribution involving 1 in the quark bilinear, or gamma5

\begin{eqnarray}
    && \langle I_1 \rangle_{+} = \frac{1}{8} \left[ 2| \mathcal{F}_S |^2 \, \rho^-_{1,S} + \frac{2}{3} | \mathcal{F}_P |^2 \, \rho^-_{1,P} + 2 | \mathcal{F}_\Vert |^2 \, \rho^-_{1,P} + 2 | \mathcal{F}_\perp |^2 \, \rho^+_{1,P} \right] \, \label{I1plus}\\
    && \quad \stackrel[]{SM}{\to} +\frac{1}{8}\biggl\{ 2|\mathcal{F}_S|^2 | \, C_{9}^{{\rm eff}: S} |^2 +\left[ \frac{2}{3} |\mathcal{F}_{P} |^2 + 2 ( | \mathcal{F}_\Vert |^2 + | \mathcal{F}_\perp |^2) \right] \, | C_{9}^{{\rm eff}: P}|^2   \biggr\}\,, \nonumber
\end{eqnarray}

\begin{eqnarray}
    && \langle I_2 \rangle_{+} = -\frac{1}{8} \left[ 2| \mathcal{F}_S |^2 \, \rho^-_{1,S} + \frac{2}{3} \left\{ | \mathcal{F}_P |^2 \, \rho^-_{1,P} - | \mathcal{F}_\Vert |^2 \, \rho^-_{1,P} - | \mathcal{F}_\perp |^2 \, \rho^+_{1,P} \right\} \right] \, \label{I2plus}\\
     && \quad \stackrel[]{SM}{\to} -\frac{1}{8}\biggl\{ 2|\mathcal{F}_S|^2 | \, C_{9}^{{\rm eff}: S} |^2 + \frac{2}{3} (|\mathcal{F}_{P} |^2 - | \mathcal{F}_\Vert |^2 - | \mathcal{F}_\perp |^2) \, | C_{9}^{{\rm eff}: P}|^2   \biggr\}\,, \nonumber
\end{eqnarray}

\begin{equation}
    \langle I_3 \rangle_{+} = \frac{1}{6} \left[ | \mathcal{F}_\perp |^2 \, \rho^+_{1,P} - | \mathcal{F}_\Vert |^2 \, \rho^-_{1,P} \right] \stackrel[]{SM}{\to} \frac{1}{6} ( | \mathcal{F}_\perp |^2 \, - | \mathcal{F}_\Vert |^2 ) \, | C_{9}^{{\rm eff}: P}|^2  \,,
\end{equation}

\begin{eqnarray}\label{I4scalar}
    && \frac{2}{\pi} \langle I_4 \rangle_{+} = -\frac{1}{4} \text{Re} \left[ \mathcal{F}_S \mathcal{F}^\ast_\Vert  \left( (C_{9}^{{\rm eff}: S}+C_9^{\rm NP}-C_9')(C_{9}^{{\rm eff}: P}+C_9^{\rm NP}-C_9')^\ast \right. \right. \\
    && \quad \left. \left. +|C_{10}-C_{10}'|^2 \right) \right] \stackrel[]{SM}{\to} -\frac{1}{4} \text{Re} \left[ \mathcal{F}_S \, \mathcal{F}^\ast_\Vert \, C_{9}^{{\rm eff}: S} \, (C_{9}^{{\rm eff}: P})^\ast \right] \,, \nonumber
\end{eqnarray}

\begin{eqnarray}
     && \frac{2}{\pi} \langle I_5 \rangle_{+} = \frac{1}{2} \text{Re} \left[  \mathcal{F}_S \mathcal{F}^\ast_\perp \left( (C_{9}^{{\rm eff}: S}+C_9^{\rm NP}-C_9')(C_{10}+C_{10}')^\ast \right. \right. \\
     && \quad \left. \left. +(C_{9}^{{\rm eff}: P}+C_9^{\rm NP}+C_9')^\ast (C_{10}-C_{10}')  \right) \right] \stackrel[]{SM}{\to} 0\, , \nonumber
\end{eqnarray}

\begin{equation}
    \langle I_6 \rangle_{+} = - \frac{4}{3} \left[ \text{Re} \left( \mathcal{F}_\Vert \mathcal{F}^\ast_\perp \right) \, \text{Re} \rho^+_2 + \text{Im} \left( \mathcal{F}_\Vert \mathcal{F}^\ast_\perp \right) \, \text{Im} \rho^-_2 \right] \stackrel[]{SM}{\to} 0\,,
\end{equation}

\begin{eqnarray}
    && \frac{2}{\pi} \langle I_7 \rangle_{+} = \frac{1}{2}\text{Im} \left[ \mathcal{F}_S \mathcal{F}^\ast_\Vert \left( (C_{9}^{{\rm eff}: S}+C_9^{\rm NP}-C_9')(C_{10}-C_{10}')^\ast \right. \right. \\
    && \quad \left. \left. +(C_{9}^{{\rm eff}: P}+C_9^{\rm NP}-C_9')^\ast (C_{10}-C_{10}')  \right) \right] \stackrel[]{SM}{\to} 0  \,, \nonumber
\end{eqnarray}

\begin{eqnarray}
    && \frac{2}{\pi} \langle I_8 \rangle_{+} = -\frac{1}{4}  \text{Im} \left[  \mathcal{F}_S \mathcal{F}^\ast_\perp \left( (C_{9}^{{\rm eff}: S}+C_9^{\rm NP}-C_9')(C_{9}^{{\rm eff}: P}+C_9^{\rm NP}+C_9')^\ast \right. \right. \\
    && \quad \left. \left. +(C_{10}-C_{10}')(C_{10}+C_{10}')^\ast \right) \right] \stackrel[]{SM}{\to} -\frac{1}{4}  \text{Im} \left[  \mathcal{F}_S \, \mathcal{F}^\ast_\perp \, C_{9}^{{\rm eff}: S} \, (C_{9}^{{\rm eff}: P})^\ast \right] \,, \nonumber
\end{eqnarray}

\begin{equation}
    \langle I_9 \rangle_{+} = \frac{2}{3} \left[ \text{Re} ( \mathcal{F}_\perp \mathcal{F}^\ast_\Vert ) \, \text{Im} \rho^+_2 + \text{Im} ( \mathcal{F}_\perp \mathcal{F}^\ast_\Vert ) \, \text{Re} \rho^-_2 \right] \stackrel[]{SM}{\to} \frac{1}{3} \text{Im} ( \mathcal{F}_\perp \mathcal{F}^\ast_\Vert ) \, | C_{9}^{{\rm eff}: P}|^2 \,.
\end{equation}

\begin{table}[h!]
    \centering
    \renewcommand{\arraystretch}{1.2}
    \begin{tabular}{|lc|c|c|c|c|}
        \hline
        \multicolumn{2}{|c|}{\multirow{2}{*}{$ \int \langle I_i \rangle_{+}^r / \Gamma^r $}} & \multicolumn{2}{c|}{SM: $ C_9^{\rm NP} = C_9' $} & \multicolumn{2}{c|}{NP: $ \tilde{C}_{10} = 0.43 $,} \\
         & & \multicolumn{2}{c|}{$ = C_{10} = C_{10}' = 0 $} & \multicolumn{2}{c|}{$ C_9^{\rm NP} = C_9' = C_{10}' = 0 $} \\
        \cline{3-6}  
        $i$ & $S$-wave & WCs & value (\%) & WCs & value (\%)\\
        \hline
        $ 1^\dagger $ & \gr{$ \circ $} & $ | C_9^{{\rm eff}: S} |^2 $, $ | C_9^{{\rm eff}: P} |^2 $ &  $48$   & SM $ + \, | C_{10} |^2 $ &  $48$ \\
        \hline
        $ 2^\dagger $ & \gr{$ \circ $} & $ | C_9^{{\rm eff}: S} |^2 $, $ | C_9^{{\rm eff}: P} |^2 $ & $-7$  & SM $ + \, | C_{10} |^2 $ &  $-7$ \\
        \hline
        $ 3^\dagger $ &  \rd{$ \times $} & $ | C_9^{{\rm eff}: P} |^2 $ & $-14$ & SM $ + \, | C_{10} |^2 $ & $-14$ \\
        \hline
        $4$ & \gr{$ \checkmark $} & $ C_9^{{\rm eff}: S} \, (C_9^{{\rm eff}: P})^\ast $ & $\pm 2$ & SM $ + \, | C_{10} |^2 $ & $\pm 2$  \\
        \hline
        $5$ & \gr{$ \checkmark $} & -- & $ 0 $ & $ C^{{\rm eff}: S}_9 \, C_{10}^\ast + C_{10} \, (C^{{\rm eff}: P}_9)^\ast $ & $\pm 0.1$  \\
        \hline
        $ 6^\dagger $ &  \rd{$ \times $} & -- & $ 0 $ & $ \text{Re} \left[ C^{{\rm eff}: P}_9 \, C_{10}^\ast \right] $ & $\pm 0.3$ \\
        \hline
        $7$ & \gr{$ \checkmark $} & -- & $ 0 $ & $ C^{{\rm eff}: S}_9 \, C_{10}^\ast + C_{10} \, (C^{{\rm eff}: P}_9)^\ast $ & $\pm 0.4$ \\
        \hline
        $8$ & \gr{$ \checkmark $} & $ C_9^{{\rm eff}: S} \, (C_9^{{\rm eff}: P})^\ast $ & $\pm 1$ & SM $ + \, | C_{10} |^2 $ & $\pm 1$  \\
        \hline
        $ 9^\dagger $ &  \rd{$ \times $} & $ | C_9^{{\rm eff}: P} |^2 $ & $ \sim 0 $ & SM $ + \, | C_{10} |^2 $ & $\sim 0$ \\
        \hline
    \end{tabular}
    
    %%%%%%%%%%%%%%%%%%%%%%%%%%%%%%%%%%%%%%%%%%%%%%%%%%%%%%%%%%%%%%%%%%%%%%%%%%%%%%%%%%%%%%%%%%%%%%%%%%%%%%%%%%%%%%%%%%%%%%%%%%%%%%
    \vspace{3mm}
    \begin{tabular}{|lc|c|c|c|c|}
        \hline
        \multicolumn{2}{|c|}{\multirow{2}{*}{$ \int \langle I_i \rangle_{-}^r / \Gamma^r $}} & \multicolumn{2}{c|}{SM: $ C_9^{\rm NP} = C_9' $} & \multicolumn{2}{c|}{NP: $ \tilde{C}_{10} = 0.43 $,} \\
         & & \multicolumn{2}{c|}{$ = C_{10} = C_{10}' = 0 $} & \multicolumn{2}{c|}{$ C_9^{\rm NP} = C_9' = C_{10}' = 0 $} \\
        \cline{3-6}
        $i$ & $S$-wave & WCs & value (\%)& WCs & value (\%)\\
        \hline
        $1$ & \gr{$ \checkmark $} & $ C_9^{{\rm eff}: S} \, (C_9^{{\rm eff}: P})^\ast $ & $\mp 2$ & SM $ + \, | C_{10} |^2 $ & $\mp 2$  \\
        \hline
        $2$ & \gr{$ \checkmark $} & $ C_9^{{\rm eff}: S} \, (C_9^{{\rm eff}: P})^\ast $ & $\pm 2$  & SM $ + \, | C_{10} |^2 $ & $\pm 2$  \\
        %\hline
        %$I_3$ & $ | C_9^{\rm eff} |^2 $ & 0 & $ | C_9^{\rm eff} |^2 + | C_{10} |^2 $ & 0 \\ %always vanishes
        \hline
        $4^\dagger$ &  \rd{$ \times $} & $ | C_9^{{\rm eff}: P} |^2 $ & $20$  & SM $ + \, | C_{10} |^2 $ &  $20$ \\
        \hline
        $5^\dagger$ &  \rd{$ \times $} & -- & $ 0 $ & $ \text{Re} \left[ C^{{\rm eff}: P}_9 \, C_{10}^\ast \right] $ & $\pm 0.2$ \\
        %\hline
        %$I_6$ & $ 0 $ & 0 & $ \text{Re} \left[ C^{\rm eff}_9 \, C_{10}^\ast \right] $ & 0 \\ %always vanishes
        \hline
        $7^\dagger$ &  \rd{$ \times $} & -- & $ 0 $ & $ \text{Re} \left[ C^{{\rm eff}: P}_9 \, C_{10}^\ast \right] $ & $ \sim 0 $ \\
        \hline
        $8^\dagger$ &  \rd{$ \times $} & $ | C_9^{{\rm eff}: P} |^2 $ & $ \sim 0 $ & SM $ + \, | C_{10} |^2 $ & $\sim 0$ \\
        %\hline
        %$I_9$ & $ | C_9^{\rm eff} |^2 $ & 0 & $ | C_9^{\rm eff} |^2 + | C_{10} |^2 $ & 0 \\ %always vanishes
        \hline
    \end{tabular}
    \caption{%[Not yet including a discussion about J-type contribution: use effective description]
    Summary of the angular observables: the upper table contains $\langle \cdot \rangle_+$ quantities, while the lower one $\langle \cdot \rangle_-$ quantities. In the first column, a tick \gr{$ \checkmark $} indicates an $S$-wave effect through its interference with the $P$-wave, an empty circle \gr{$ \circ $} means that the $S$-wave manifests through an additive term to the $P$-wave instead of an interference term, and a cross  \rd{$ \times $} indicates the absence of any $S$-wave effect. The SM dependencies on the effective Wilson coefficients (WCs) are given in the second column along with a typical value found for the integrated observables in the SM.
    The best fit values of the normalization and relative phases are considered for setting the numerical values given above.
    When two signs are shown, they correspond to different relative phases of the $S$- and $P$-waves ($ \Delta_{SP} $ and $\Delta_{\rho {\rm NP}}$ are taken here to value $ 0\mod \pi/2 $).
    The integration range considered is $ (0.78 \, \text{GeV})^2 < q^2 (\ell^+ \ell^-) < (1.1 \, \text{GeV})^2 $.
    The third column indicates the dependence on the effective SM and on the NP WCs in the presence of a non-vanishing $ \tilde{C}_{10} = V_{ub} V_{cb}^\ast C_{10} $, taken at its current upper bound, along with a typical value for the integrated observables.
    The hermitian conjugate is also understood when the displayed combination of WCs is possibly complex.
    Cases carrying a dagger $\dagger$ indicate quantities already measured by LHCb \cite{LHCb:2021yxk,LHCb_supplementary_material_aps,LHCb_supplementary_material_4}.
    %[\LVS{what is the observable for which numerical values are given?, no overall multiplicative factor I guess}]
    %[\LVS{compare to other tables}]
    %[I am not sure how to quote ranges (perhaps varying $\Delta_1$? it would underestimate the error for angular observables sensitive to the S-wave)]
    %[I think this table should be kept in the main text with $ q^2_{\ell \ell} \gtrsim m_\rho^2 $ and a favorable $\phi_\phi-\phi_\rho$, while binned cases for different strong phase differences should be given in longer tables in the appendices for reference]
    %[ What is the region in $p^2$, $q^2$ being considered??; how ranges are being determined (variation of complex phases?)??; review unit being shown and put in \%; PREPARE longer table with smaller bins}]
    %[ NOTE THAT LHCb's $ \langle S_{2, 5, 6, 7, 8} \rangle $ are consistent with zero, thus setting constraints on the $P$-wave through $\langle I_2 \rangle_{+}$ and $\langle I_7 \rangle_{-}$; LHCb's $ \langle S_{3, 4, 9} \rangle $ displays some deviation from zero, thus setting constraints on the $P$-wave through $ \langle I_3 \rangle_{+} $ and $ \langle I_4 \rangle_{-} $, while $ \langle S_9 \rangle $ vanishes in the SM; what are our predictions for them???}]
    %[ NOTE THAT the $P$- and the $S$-wave cases for 5 and 7 are both sensitive in the same way to $C_{10}$: however, it is possible that due to interferences the $S$-wave comes with a larger pre-factor, particularly in the case of 7; at least, it is sensitive to a different systematic effect}]
    }
    \label{tab:summary_table}
    %[ NOTE THAT LHCb's $ \langle A_{2, 3, 4, 5, 6, 7, 8, 9} \rangle $ are consistent with zero, thus setting constraints on the $P$-wave through $ \text{Im} \left( \mathcal{F}_P \mathcal{F}^\ast_\perp \right) $ and $ \text{Im} ( \mathcal{F}_\perp \mathcal{F}^\ast_\Vert ) $}; these are CP-violating quantities]
    %[ IS IT better to look at $ C_{10}' $??, first decide on size of interference terms}; we won't dedicate much effort to NP]
\end{table}

%%%%%%%%% CP-even, -odd cases %%%%%%%%%
We now define $\overline{I}_i$ as the analogous of $I_i$ for the CP-conjugated process.
The new kinematical conventions are: $ \theta_\ell $ is the angle between the $ \ell^- $-momentum and the $\overline{D}$-momentum in the di-lepton center of mass frame,
$ \theta_\pi $ is the angle between the $ \pi^+ $-momentum and the negative $\overline{D}$-momentum in the di-pion center of mass frame,
%and $ \phi $ is the angle between di-lepton and di-pion decay planes
while following the previous procedure to define the remaining angle $\phi'$, one has $\phi' = \pi - \phi$.
In the comparison of the two processes certain angular observables acquire a sign under CP transformation due to kinematical considerations, $I_i \to \overline{I}_i$ for $ i = 1, 2, 3, 4, 7 $, while $I_j \to -\overline{I}_j$ for $ j = 5, 6, 8, 9 $.
%To facilitate the reader we provide the correspondence of the experimentally provided quantities to the theoretically estimated ones:
LHCb \cite{LHCb:2021yxk,LHCb_supplementary_material_aps,LHCb_supplementary_material_4} provides measurements for the following CP-averaged $S$ and CP-asymmetric $A$ quantities: $ \langle O_i \rangle = \langle I_i \rangle_{f (i)} \pm \langle \overline{I}_i \rangle_{f (i)} $ for $ i = 1, 2, 3, 4, 7 $, and $ \langle O_j \rangle = \langle I_j \rangle_{f (j)} \mp \langle \overline{I}_j \rangle_{f (j)} $ for $ j = 5, 6, 8, 9 $, where $ O \to S $ ($ O \to A $) for the upper (respectively, lower) signs; these measurements by LHCb optimize the sensitivity to $P$-wave effects, namely, $ f (i) = + $ for $ i = 1, 2, 3, 6, 9 $, while $ f (j) = - $ for $ j = 4, 5, 7, 8 $ (see Tab.~\ref{tab:summary_table}).
Since in the current work we neglect CP-odd contributions from the SM, the CP asymmetries of all angular observables vanish in the SM limit.
The CP-averaged quantities are:
%[\LVS{what about $I_1$?}: I1+ and I2+ are related by dGamma/dq2dp2, while I1- and I2- are equal]
%equal to their CP-specific counterparts

\begin{eqnarray}
    & \langle S_2 \rangle {(p^2, q^2)} \equiv \langle I_2 \rangle_{+} \,, \quad & \langle S_3 \rangle {(p^2, q^2)} \equiv \langle I_3 \rangle_{+} \,, \\
    & \langle S_4 \rangle {(p^2, q^2)} \equiv \langle I_4 \rangle_{-} \,, \quad & \langle S_5 \rangle {(p^2, q^2)} \equiv \langle I_5 \rangle_{-} \stackrel[]{SM}{\to} 0 \,, \nonumber\\
    & \langle S_6 \rangle {(p^2, q^2)} \equiv  \langle I_6 \rangle_{+}\stackrel[]{SM}{\to} 0 \,, \quad & \langle S_7 \rangle {(p^2, q^2)} \equiv  \langle I_7 \rangle_{-}\stackrel[]{SM}{\to} 0 \,, \nonumber\\
    & \langle S_8 \rangle {(p^2, q^2)} \equiv \langle I_8 \rangle_{-}  \stackrel[]{SM}{\to} \sim 0 \,, \quad & \langle S_9 \rangle {(p^2, q^2)} \equiv \langle I_9 \rangle_{+} \stackrel[]{SM}{\to} \sim 0 \,. \nonumber
\end{eqnarray}
%\rd{The CP-odd quantities $A_i$ result from combining $ I_i $ and $ \bar{I}_i $ in the opposite way...}
%exchanging the integration over the dipion angle, $ \langle \, \cdot \, \rangle_{\pm} \longleftrightarrow \langle \, \cdot \, \rangle_{\mp} $.
The binned quantities quoted by Refs.~\cite{LHCb:2021yxk,LHCb_supplementary_material_aps,LHCb_supplementary_material_4} are defined as:

\begin{equation} \label{isnormalised}
    \langle O_k \rangle^{[q_{i_1}^2,q_{i_2}^2]} \equiv \frac{1}{\Gamma^{[q_{i_1}^2,q_{i_2}^2]}} \int \langle O_k \rangle^{[q_{i_1}^2,q_{i_2}^2]} \,, \quad O = S, A \,, \quad k = 1, \ldots, 9 \,,
\end{equation}
for a bin $ [q_{i_1}^2,q_{i_2}^2] $,
where the following shortcut notation has been employed:

\begin{equation} \label{intnotation}
    \int f^{[q_{i_1}^2,q_{i_2}^2]} \equiv \int^{q_{i_2}^2}_{q_{i_1}^2} d q^2 \int^{p^2_{\rm max}(q^2)}_{p^2_{\rm min}} d p^2 f (p^2, q^2) \,,
\end{equation}
for any function $f$; the notation $ \Gamma^r $ designates the total width in the $q^2$-bin $ r $.
%Here, $ \Gamma $ is the whole decay width, but for binned quantities the width employed is the one for the same kinematical region.
%
We stress that the observables $\braket{S_8}^r$ and $\braket{S_9}^r$, although vanishing in the SM when employing the approximation $C^{{\rm eff}: P}_9$ for any bin $r$
due to our description of the phases encoded in the transversity form factors $ \mathcal{F}_P $, $ \mathcal{F}_\Vert $, and $ \mathcal{F}_\perp $,
%[\rd{imaginary part among P-wave line-shapes}] [\rd{This suppression could be leveraged if the effect of the $\rho (1450)$ is significant.}] [\rd{I was referring to S7 because it depends on the imaginary part of the interference between the decay amplitudes and there are large strong phase differences between these in B -> J/psi K* decays.}]
obtain non-vanishing values in the original picture (i.e., before the introduction of effective $C_9$ coefficients).
%of W- and J-factorization contractions
Nevertheless, these values remain very small, being suppressed due to the simple
%single pole
parameterisations of the $D \to \mathcal{R}$, $D \to \mathcal{V}$ form factors.
%The estimation could be improved with the implementation of more refined form factors.
%For the observable $I_8$, the effect of vector-vector interference due to the presence of $J$ and $W$ is found to be non-zero but very suppressed, around three orders of magnitude than the $(I_8)_+$. The latter amounts to some sizeable values in a few $q^2$-bins.
Also note that from the above equations $\braket{I_7}_-$ seems to vanish even in the presence of NP. Although this is not the case when the original description is implemented (again, before the effective $C_9$ coefficients were introduced), the calculated values are still very suppressed for the same reason mentioned for $\braket{S_8}^r$ and $\braket{S_9}^r$. On the other hand, as discussed later its $S$-wave sensitive counterpart $\braket{I_7}_+$ yields values comparable to those of the other null-test observables for the same values of NP Wilson coefficients.
%However, $(I_7)_P$ is identically zero in our current hadronic model (even including the J-type factorization), because it is proportional to the difference of some form factors which have equal pole masses. On the contrary, $(I_7)_S$ is sensitive to NP.

%%%%%%%%% Extracting directly WCs %%%%%%%%%
Some relations aiming to isolate the Wilson coefficients with potential phenomenological interest include (see also Ref.~\cite{Feldmann:2017izn}):

\begin{comment}
\begin{eqnarray}
    && | \mathcal{F}_\Vert |^2 \, \rho^-_{1,P} = \frac{3}{2} ( \langle I_1 \rangle_{+} + \langle I_2 \rangle_{+} - 2 \langle I_3 \rangle_{+} ) \,, \\
    && | \mathcal{F}_\perp |^2 \, \rho^+_{1,P} = \frac{3}{2} ( \langle I_1 \rangle_{+} + \langle I_2 \rangle_{+} + 2 \langle I_3 \rangle_{+} ) \,, \\
    && | \mathcal{F}_S |^2 \, \rho^-_{1,S} + \frac{2}{3} | \mathcal{F}_P |^2 \, \rho^-_{1,P} = 2 ( \langle I_1 \rangle_{+} - 3 \langle I_2 \rangle_{+} ) \,.
\end{eqnarray}
Also,
\end{comment}

\begin{eqnarray}
    && \frac{\langle S_8 \rangle_{(p^2, q^2)}}{\langle A_5 \rangle_{(p^2, q^2)}} = \frac{1}{2} \frac{\text{Im} \rho^+_2}{\text{Re} \rho^+_2} \,, \qquad
    \frac{\langle S_9 \rangle_{(p^2, q^2)}}{\langle A_6 \rangle_{(p^2, q^2)}} = -\frac{1}{2} \frac{\text{Im} \rho^+_2}{\text{Re} \rho^+_2} \,, \\
    && \frac{\langle S_5 \rangle_{(p^2, q^2)}}{\langle A_8 \rangle_{(p^2, q^2)}} = -2 \frac{\text{Im} \rho^-_2}{\text{Re} \rho^-_2} \,, \qquad
    \frac{\langle S_6 \rangle_{(p^2, q^2)}}{\langle A_9 \rangle_{(p^2, q^2)}} = 2 \frac{\text{Im} \rho^-_2}{\text{Re} \rho^-_2} \,,
\end{eqnarray}
which are relevant only in the unbinned limit, since $ C_{9}^{{\rm eff}: P}$ carries a dependence on kinematical variables.

\section{Fits and predictions}\label{sec:analysis_data}

%\begin{itemize}
    %\item For tables, give the full results
    %\item Phases could perhaps be fixed from the fit (give numerical values for the best fit point?; discuss later varying phases)
    %\item Give available bounds on NP WCs
    %\item Shift numerical inputs for form factors to Appendix
    %\item Discuss later how to present the tables
%\end{itemize}

We search for footprints of the $S$-wave in three different types of observables. Firstly (I), the ones related to the differential mass distributions, where the effect of the $S$- and $P$-waves is additive.
%the ones that are non-vanishing in the SM also by the P-wave-mediated processes; in these observables, namely the ones related to the differential mass distributions, the effect of the $S$-wave will be additive [\LVS{discuss}].
Secondly (II), we examine the observables that probe the $S$- and $P$-wave interference. Thirdly (III), we look into observables that vanish in the SM, and find some that are sensitive to NP only in the presence of the $S$-wave; we compare these to observables that are sensitive to NP only in the presence of the $P$-wave.
%We also comment on the effect of the inclusion of the J-type factorization in some observables [\LVS{still?}].
Cases (I) and (II) are discussed in Sec.~\ref{sec:SM_fits}; we will in particular extract in this section parameters accounting for normalizations, namely, $ \{ a_\omega, a_S (0) / A_1 (0) , A_1 (0) \, B_{\rho^0}, B_\phi / B_{\rho^0}, B_\omega^{(S)} / B_{\rho^0}^{(S)}, B_\phi^{(S)} / B_{\rho^0}^{(S)} \} $,
%[\rd{not both $A_1 (0)$ and $ B_\rho $}]
and relative strong phases among intermediate resonances, namely, $ \phi_\omega $ and $ \Delta_i $, $ i=1, 3, 4 $.
Due to the suppression factor $\epsilon_\omega$, we do not include $ B_\omega $ nor $ \Delta_2 $ in this list.
The ratio $ B_{\rho^0}^{(S)} / B_{\rho^0} $ is set to the unit, and $a_S (0)$ is adjusted to determine the overall contribution of the $S$-wave.
It is implicitly assumed that NP contamination is negligible in the differential mass distributions.
Case (III) is the subject of Sec.~\ref{sec:SM_NP_fits}.
The three types of observables (I)-(III) are easily identified in Tab.~\ref{tab:summary_table};
the values of the most interesting observables over distinct $q^2$ bins will be discussed in details in the following, and are given in Tabs.~\ref{gammas,i2}, \ref{i2i4i8} and \ref{NPobsevablesPwave}, that deal with cases (I)-(III), respectively.
We stress that we also make comparisons to the LHCb data set that optimizes the sensitivity to the $P$-wave.
We have not included theory uncertainties (e.g., stemming from the use of the factorization approach) in the following discussion beyond the ones attached to the unknown parameters we have fitted for.
%which are difficult to estimate at the present stage

\subsection{SM fits and predictions}\label{sec:SM_fits}

\begin{figure}
    \centering
    \includegraphics[scale=0.45]{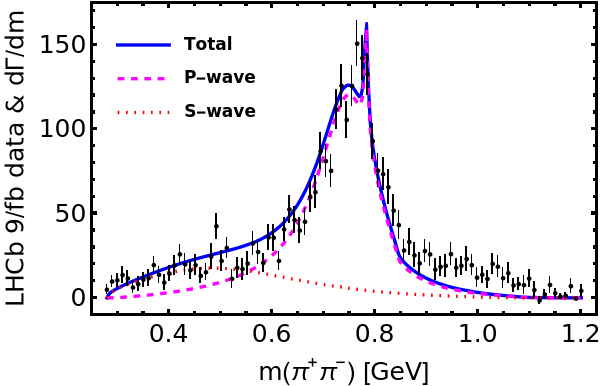} \hspace{3mm}
    \includegraphics[scale=0.44]{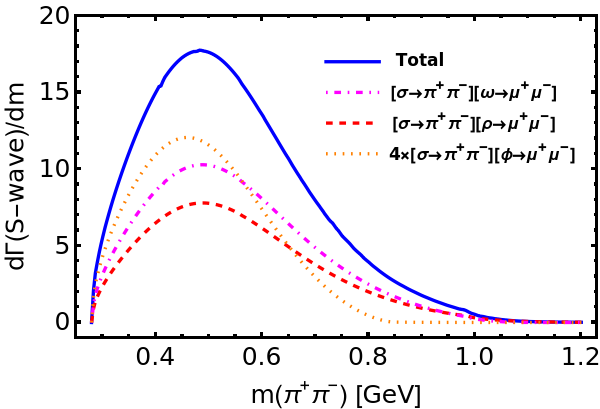}
    \caption{The prediction for the differential decay rate $ d \Gamma / d m $ and LHCb data over the di-hadron invariant mass $ m(\pi^+ \pi^-) \equiv \sqrt{p^2} $ \cite{LHCb:2021yxk,LHCb_supplementary_material_aps,LHCb_supplementary_material_4}. (Left) The contributions from the $S$-wave (dotted red) and the $P$-wave (dashed magenta) add up to the full resonant contribution (solid blue).
    %[\LVS{explain shift that we will have: i.e., curve above the points; not even when we do not perform a combined fit the points go ``on average"/kind of yes, they do; since it is not a marked effect, I won't insist}]
    (Right) Components of the $S$-wave contribution: $ \sigma \rho $ (dashed red), $ \sigma \omega $ (dot-dashed magenta), and $ \sigma \phi $ (dotted orange, multiplied by $4$ for an easier comparison).}
    %normalization
    \label{fig:dgammadp2}
\end{figure}

%%%%%%%%% Sigma in p^2 %%%%%%%%%
The large statistics and fine binning of Refs.~\cite{LHCb:2021yxk,LHCb_supplementary_material_aps,LHCb_supplementary_material_4}
allows for a precision numerical study.
The global fit we perform combines bins of both differential mass distributions as functions of the invariant mass of the lepton ($q^2$) or pion ($p^2$) pairs.
We note that no correlations among bins of $ d \Gamma / d p^2 $ and $ d \Gamma / d q^2 $ have been made available in those references.
We first discuss the features of the $ d \Gamma / d p^2 $ distribution,
which is crucial to establish the $\sigma$ contribution.
Being a very broad resonance, the effect of including the $\sigma$ might be difficult to spot.
However, we do observe a clear contribution in the differential decay rate as a function of $p^2$,
%Indeed, available data about the differential decay rate as a function of $p^2$ already shows the presence of the $S$-wave contribution,
see Fig.~\ref{fig:dgammadp2}.
%as we now discuss
%For a comparison see the plot of $d\Gamma/dp^2$, where the $\sigma$ is sizeable at low $p^2$.
It is clearly seen by eye that including $\sigma$ in the theoretical prediction improves the quality of the fit;
quantitatively, $ \chi^2_{min; \text{w/o } \sigma} - \chi^2_{min} = 10^2 $, clearly favoring its inclusion.\footnote{For this test only, we have reintroduced back to the fit $ B_\omega $ and $ \Delta_2 $, so the improvement comes mainly from the $ d \Gamma / d p^2 $ distribution.}
The $ d \Gamma / d p^2 $ distribution is also used
to probe the small $\omega \to \pi^+ \pi^-$ contribution, together with its relative phase with respect to the $\rho$ contribution.
There is good evidence of the presence of such $ \omega $: $ \chi^2_{min; \text{w/o } \omega} - \chi^2_{min} = 4^2 $, which is also approximately distributed as a $\chi^2$ with a single degree of freedom.
In performing the fits, we have excluded the region $ \pm 70 $~MeV around the mass of the $K_S^0$ to account for the possibility of contamination from $K_S^0 \to \pi^+ \pi^-$.\footnote{This procedure is adopted from Ref.~\cite{BESIII:2018qmf}, which however is a different experiment (and process). In the case of LHCb, $K_S^0$ contributions are not explicitly vetoed. However, vertexing eliminates to a certain degree the aforementioned $K_S^0$ contamination, but there is no quantitative estimate of the resulting efficiency \cite{Mitzel}.}
Also, we have considered data points up to $ 0.9 $~GeV, since beyond this energy virtual kaon pairs (i.e., below their actual threshold)\footnote{Note that this is a source of violation of the Zweig rule, see e.g. Ref.~\cite{Donoghue:1990xh}.} along with other resonances such as $f_0 (980)$ start manifesting more strongly (in the former case, in the dispersive part of the amplitude).
The presence of
%these analytical features and
other resonances, that include beyond the $S$- and $P$-waves also the $D$-wave, together with the isospin-two and Bremsstrahlung contributions, are likely to be at the origin of the poor comparison between our prediction and the data in the high-$p^2$ region, see the left panel of Fig.~\ref{fig:dgammadp2}.
%%Moreover, the large value $ \chi^2_{min} $
%$ \chi^2_{min} / N_{\rm d.o.f.} \simeq 2.4 $ (where $ N_{\rm d.o.f.} \simeq 44 $)
%%that we found is in part driven by the energy region $ 0.8 - 0.9 $~GeV, possibly suggesting that the latter effects not included in our analysis could leak below $ 0.9 $~GeV.
%or yet other effects discussed in the text
%More information is provided in App.~\ref{app:line_shapes}.
The value of $ \chi^2_{min} / N_{\rm d.o.f.} \simeq 2 $ (where $ N_{\rm d.o.f.} \simeq 77 $) has been found, driven mainly by the $ d \Gamma / d p^2 $ data set.

\begin{figure}
    \centering
    \includegraphics[scale=0.45]{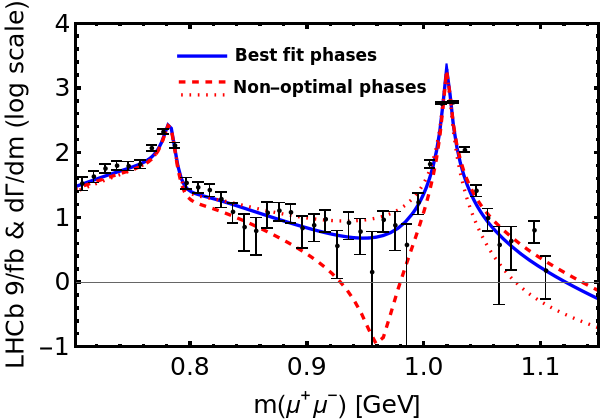} \hspace{3mm}
    \includegraphics[scale=0.45]{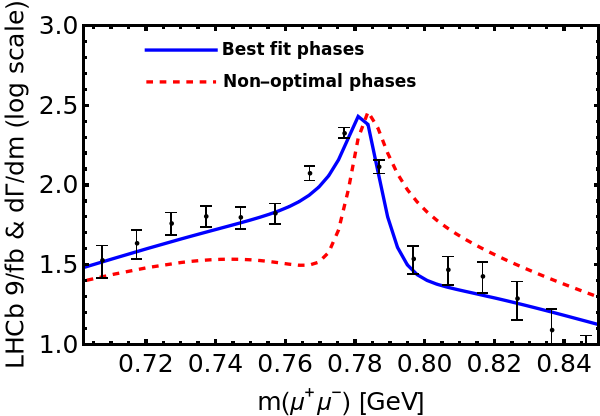}
    \caption{The differential decay rate $ d \Gamma / d m $ and LHCb data over the di-lepton invariant mass $ m(\mu^+ \mu^-) \equiv \sqrt{q^2} $ \cite{LHCb:2021yxk,LHCb_supplementary_material_aps,LHCb_supplementary_material_4}. (Left)
    %The blue band is determined by varying $ \Delta_1 $ around the best-fit point.
    The dashed (dotted) red curve displaying ``non-optimal phases'' corresponds to the optimal $ \Delta_1 $ added with $ \pi / 2 $ ($ -3 \pi / 4 $). (Right)
    %The blue band is determined by varying $ \Delta_4 $ around the best-fit point.
    The dashed red curve displaying ``non-optimal phases'' corresponds to the optimal $ \Delta_4 $ added with $ 5 \pi / 4 $.}
    %[\LVS{normalization}]
    %width of the exp bins: I believe it is clear
    \label{fig:dgammadq2}
\end{figure}

%%%%%%%%% Sigma in q^2 %%%%%%%%%
We now discuss the features of the $ d \Gamma / d q^2 $ distribution.
We fit the data of Refs.~\cite{LHCb:2021yxk,LHCb_supplementary_material_aps,LHCb_supplementary_material_4} in the region $q^2 \geq m_\rho^2$, in order to avoid the many other resonances that we do not address in the present work, shown in Fig.~\ref{fig:DeBoer_Hiller}.
%and Bremsstrahlung effects at very low $q^2$
Fig.~\ref{fig:dgammadq2} displays the result of our fit, which achieves a good qualitative description of the data.
%in which we see that our model is clearly inadequate in the region $q^2 < m_\rho^2$, showing a deficit with respect to the available data therein (which, again, has not been used in the fit for the reason previously pointed out).
%In fact, our fit does not perform well around the $\rho$ peak neither, which could also result from the effects of the aforementioned resonances not included in our analysis;
%\rd{we fix the relative phase of the $\omega$ according to Eq.~XXX, since the statistics around its peak in the present case is lower than in the previous fit, for which the effect of the $\omega$ could be clearly spotted}
Quantitatively, the fit does not perform well at the $\phi$ resonance, underestimating the branching ratio therein; the fit indicates that a broader width of the $\phi$ should be considered, i.e., the predicted values closer to $m_\phi$ tend to be overestimated, while peripheral values away from $m_\phi$ by $ \Gamma^0_\phi = 4.25~\text{MeV} $ \cite{Workman:2022ynf} tend to be underestimated. Accordingly, we observe that a much better fit of the $ d \Gamma / d q^2 $ data is achieved when increasing the width of the $\phi$ by about 60\%, namely, the $ \chi^2_{min} $ drops significantly. %raising theoretical
%(such as the modelling of the complex final-state hadronic environment) Juan says that muon is good to probe the phi even in very dense media
%and experimental questions
This effect should be due to limited momentum resolution at LHCb
(bin migration is found to be negligible in Ref.~\cite{LHCb:2013gqe}), whose effect has not been ``unfolded'', thus broadening the $ \phi $ peak; efficiency variations, instead, are taken into account \cite{Mitzel}.
We fix the $\phi$ width to $ \Gamma^0_\phi $ in our theoretical predictions, and to circumvent the later resolution issue we collect the four bins around the $\phi$ peak into a single bin.
%[\rd{CHECK PAPER by VON HIPPEL AND QUIGG, applied this time for muon pairs (mass similar to pions)} -- my understanding is that since muons are much below the phi mass, this discussion does not apply unchanged]
%[\rd{consider a single bin for the $\phi$!!!}]
%(A somewhat vague possibility is that this is due to broad vector resonances not taken into account in our discussion, such as the $\rho (1450)$ or the $\omega (1420)$, of widths $\sim 290$~MeV and $ \sim 400$~MeV, respectively \cite{Workman:2022ynf}.)
%\rd{For this reason, we considered fitting the value of $ \Gamma^0_\phi $.}
%%\rd{The large value of $ \chi^2_{min} $}
%$ \chi^2_{min} / N_{\rm d.o.f.} \simeq 3.5 $ (where $ N_{\rm d.o.f.} \simeq 34 $)
%%\rd{results also from the $ \phi $ peak region.}
%the two above kinematic regions, i.e., around the $ \rho $ peak, and above the $ \phi $ peak
%%\rd{However, we obtain a good description of the data elsewhere.}
%in the region between the two latter resonances

%%%%%%%%% Further details %%%%%%%%%

From the global fit we find the following value for the overall normalization factor (intervals of about $ 3\sigma $ C.L. are provided in this section):

\begin{equation}
    %0.7 \lesssim A_1 (0) \, B_{\rho^0} \lesssim 0.8 \,,
    0.8 \lesssim A_1 (0) \, B_{\rho^0} \lesssim 1.2 \,,
\end{equation}
%[\rd{DOES IT MAKE SENSE TO GIVE AN INTERPRETATION FOR IT, SINCE LHCb data is given in terms of numbers of events???}]
for the extraction of which we employ also information about
the total branching fraction provided in Eq.~\eqref{eq:full_BRs}.
A value of $ A_1 (0) $ close to $ 0.6 $ as in Ref.~\cite{Melikhov:2000yu} implies $ B_{\rho^0} $
%of around $ 1.3 $
of around $ 1.8 $.
For ratios of normalization factors (or ``fudge factors'') we have:
\begin{eqnarray}
    0.8 \lesssim & B_\phi / B_{\rho^0} & \lesssim 0.9 \,, \\
    0.9 \lesssim & B^{(S)}_\omega / B^{(S)}_{\rho^0} & \lesssim 1.1 \,, \\
    0.05 \lesssim & B^{(S)}_\phi / B^{(S)}_{\rho^0} & \lesssim 0.27 \,.
\end{eqnarray}
%
%(Certain non-perturbative parameters are fitted from data on branching ratios around resonance regions (namely, $a^{\pi \pi}_\phi$, $a^{\pi \pi}_\rho$, $a^{K K}_\rho$ \cite{LHCb:2017uns}, which turn out being of the same order, and much larger than $a^{\pi \pi}_\eta$, $a^{\pi \pi}_{\eta'}$, $a^{K K}_\eta$, determined from a narrow-width approximation).)%I am not comparing to the values Hiller found
%
The $ \phi, \omega \to \mu^+ \mu^- $ resonant branching ratios constrain precisely the parameters $ B_\phi / B_{\rho^0} $ and $ B^{(S)}_\omega / B^{(S)}_{\rho^0} $.
The inclusion of $d \Gamma / d p^2$ data has an important impact in limiting the size of $ B^{(S)}_\phi / B^{(S)}_{\rho^0} $, which reflects differently compared to the other two contributions $ \sigma \omega $ and $ \sigma \rho^0 $, see the right panel of Fig.~\ref{fig:dgammadp2}, due to the different available $ p^2 $ intervals as seen from Fig.~\ref{fig:DeBoer_Hiller}.
It is evident that an important deviation from naive factorization shows up in the extraction of $ B^{(S)}_\phi / B^{(S)}_{\rho^0} $, which lies substantially away from $ 1 $.\footnote{A sizable deviation from factorization is seen in the context of $ B \to K \mu^+ \mu^- $ decays, see e.g. Ref.~\cite{Lyon:2014hpa}.}
It is interesting to point out that the contribution from $ \sigma \phi $ also turns out to be suppressed in the amplitude analysis of $ D^0 \to K^+ K^- \pi^+ \pi^- $ by LHCb \cite{LHCb:2018mzv}.
%\rd{We do not attempt at performing a combined analysis of rare semi-leptonic and purely hadronic four body decays.}
We also extract:

\begin{eqnarray}\label{eq:parameters_from_fit}
    0.001 \lesssim & a_\omega & \lesssim 0.005 \,, \\
    1.1 \, \pi \lesssim & \phi_\omega & \lesssim 1.7 \, \pi \,, \\
    %14~\text{GeV} \lesssim & a_S (0) & \lesssim 22~\text{GeV} \,,
    39~\text{GeV} \lesssim & \frac{a_S (0)}{A_1 (0)} & \lesssim 62~\text{GeV} \,,
\end{eqnarray}
which compare relatively well with $ a_\omega \simeq 0.006 $, $ \phi_\omega \simeq 0.9 \, \pi $ and $ a_S (0) / A_1 (0) \simeq 24~\text{GeV} $ for the analogous semi-leptonic decay $D^+ \to \pi^+ \pi^- e^+ \nu_e$ \cite{BESIII:2018qmf}, see App.~\ref{app:sl_decays} for further discussion.

%%%%%%%%% Relative angles %%%%%%%%%

The fit is also used to extract
the following range for the relative angle $ \Delta_1 = \delta_{ \{ \rho^0/\omega, \rho^0 \} }-\delta_{ \{ \rho^0/\omega, \phi \} } $, see
%the blue band in
the left panel of Fig.~\ref{fig:dgammadq2}: %[\LVS{Pi subtracted}, not anymore]

\begin{equation}
    %0.1 \, \pi \lesssim \Delta_1 \lesssim 0.5 \, \pi \,,
    0.5 \, \pi \lesssim \Delta_1 \lesssim 0.9 \, \pi \,,
    %0.6 \, \pi \lesssim \Delta_3 \lesssim 2.2 \, \pi \,,
\end{equation}
while $ \Delta_3 = \delta_{ \{ \sigma, \rho^0 \} }-\delta_{ \{ \sigma, \phi \} } $ remains unconstrained,
since
%the $ D^0 \to \sigma \phi $ contribution turns out being small, and thus
the contribution from the $\sigma$ plays a less important role in the region between the $\rho^0$ and $\phi$ resonances with respect to the $P$-wave contribution.
As it is clear from the left panel of Fig.~\ref{fig:dgammadq2}, this strong phase has a huge impact in the latter inter-resonant region and the very-high energy region above the $\phi$ resonance, implying modulations of the predicted branching ratios by orders of magnitude in both cases.
%\LVS{It is interesting to point out that the measured values of the branching ratio in the inter-resonant region are approximately the largest ones that could be achieved within our model.}
It is interesting to point out the possible correlation between the inter-resonant and the very-high energy regions due to the $\phi$ line-shape, e.g., a large suppression of the SM prediction in the very-high energy region (making then this region more sensitive to NP contributions) can be correlated to a relatively large branching ratio in the inter-resonant region; a similar effect is seen in Ref.~\cite{DeBoer:2018pdx}.
In the right panel of Fig.~\ref{fig:dgammadq2}, we illustrate the dependence of our prediction on the remaining strong-phase differences.
As it has been discussed around Eq.~\eqref{C9mod}, the contribution of the $\omega \to \ell^+ \ell^-$ paired with the pion pair in a $P$-wave is suppressed;\footnote{We note that allowing for large effects much beyond naive factorization, namely, $ B_\omega \gg B_{\rho^0} $, allows for a good fit of the $ d \Gamma / d q^2 $ data even in the absence of the $S$-wave.} on the other hand, the $\omega \to \ell^+ \ell^-$ can manifest when combined with the pion pair in the $S$-wave.
%which however is only a fraction of the $P$-wave contribution
%Taking into account these effects,
%it then results that the angle $ \Delta_2 = \delta_{ \{ \rho^0/\omega, \rho^0 \} }-\delta_{ \{ \rho^0/\omega, \omega \} } $ is less well constrained compared to $ \Delta_4 = \delta_{ \{ \sigma, \rho^0 \} }-\delta_{ \{ \sigma, \omega \} } $:
We then find for $ \Delta_4 = \delta_{ \{ \sigma, \rho^0 \} }-\delta_{ \{ \sigma, \omega \} } $:

\begin{equation}
    %-0.6 \, \pi \lesssim \Delta_2 \lesssim 0.3 \, \pi \,,
    %0.2 \, \pi \lesssim \Delta_4 \lesssim 0.6 \, \pi \,,
    0.2 \, \pi \lesssim \Delta_4 \lesssim 0.5 \, \pi \,,
\end{equation}
%while $ \Delta_2 = \delta_{ \{ \rho^0/\omega, \rho^0 \} }-\delta_{ \{ \rho^0/\omega, \omega \} } $ remains unconstrained;
see the right panel of Fig.~\ref{fig:dgammadq2}.
%More details are provided in App.~\ref{app:line_shapes}.
It is rather difficult to provide interpretations to the extracted ranges of values for $ \Delta_1 $ and $ \Delta_4 $, or make comparisons to other processes; note that the $ \rho^0 $ and the $ \omega $ or the $ \phi $ are in different isospin irreducible representations, so that the dynamics involved in the rescattering processes with the second resonance (the $ \rho^0/\omega $ in the case of $ \Delta_1 $, and the $ \sigma $ in the case of $ \Delta_4 $) is expected to be substantially different.

%here the question of omega is less important

%%%%%%%%%%%%%%%%%%%%%%%%%%%%%%%%%%%%%%%%%%%%%%%%%%%%%%%%%%%%%%%%%%%%%%%%%%%%%%%%%%%%%%%%%%%%%%%%%%%%%%%%%%%%%%%%%%%%%%%%%%%%%%%%%%%%%%%
\begin{table}
    \centering
    \renewcommand{\arraystretch}{1.4}
    \begin{tabular}{|c|c|c|c|c|c|c|}
    \hline
    \multirow{1}{*}{$q^2$-bin} & \multirow{1}{*}{$\Gamma^r$ (SM)} & $ \frac{\Gamma_\sigma^r}{\Gamma^r} $ (\%) & $\int \langle I_2 \rangle_+^r \times 100$ & $ \frac{\int \langle I_2\rangle _{+,\sigma}^r}{\int \langle I_2\rangle_{+}^r}$ (\%) & $\int \langle I_3 \rangle_+^r \times 100$ & $\int \langle I_4 \rangle_-^r \times 100$ \\
    %\cline{3-7}
    %& & \multicolumn{5}{c|}{ (\%) } \\
    \hline
   $ r^{(\rho: {\rm sup})} $  & $[0.64,0.87]$ &  $[23,43]$    & $[-16,-8.5]$ & $[59,78]$ & $[-7.2,-4.7]$  & $[8.3,13]$\\
   $ r^{(\phi: {\rm inf})} $  & $[1.6,1.9]$ & $[0.3,8]$    & $[-11,-6.2]$  & $[3,45]$ & $[-30,-26]$ & $[36,41]$\\
   $ r^{(\phi: {\rm sup})} $  & $[1.2,1.3]$ &  $[0.8,10]$   & $[-8.7,-4.3]$ &  $[8,53]$ & $[-22,-19]$ & $[26,29]$  \\
       \hline 
    \end{tabular}
    \caption{SM predictions for the non-vanishing observables which only receive $P$-wave contributions ($\int \langle I_3 \rangle_+^r \,, \; \int \langle I_4 \rangle_-^r$), and where the effect of the $S$-wave is additive (i.e., $\Gamma^r = \Gamma^r_{\rho^0/\omega} + \Gamma^r_\sigma$ and $\int \langle I_2 \rangle_+^r = \int \langle I_2 \rangle_{+, \rho^0/\omega}^r + \int \langle I_2 \rangle_{+, \sigma}^r$); a subscript $\sigma$ indicates that only the $S$-wave is kept. The relation $ \Gamma^r = 2 \left( \int \langle I_1 \rangle_+^r - \int \langle I_2 \rangle_+^r / 3 \right) $ holds true. For comparison with LHCb \cite{LHCb:2021yxk,LHCb_supplementary_material_aps,LHCb_supplementary_material_4}, $ \langle S_2 \rangle^r = \int \langle I_2 \rangle_+^r / \Gamma^r $, $ \langle S_3 \rangle^r = \int \langle I_3 \rangle_+^r / \Gamma^r $, and $ \langle S_4 \rangle^r = \int \langle I_4 \rangle_-^r / \Gamma^r $.
    Relevant definitions can be found in Sec.~\ref{sec:observables}, see in particular Eqs.~\eqref{isnormalised} and \eqref{intnotation}.
    The decay rate and the $I_i$'s both need to be multiplied by a common constant factor, $|C_2\lambda_d e G_F/\sqrt{2}|^2\frac{e^2}{m_D} \times 10^{-4} = 2.4 \times 10^{-19}$, with $G_F$, $m_D$ and $\Gamma^r$ in GeV.
    %[\rd{can you provide for inner use unbinned plots of the angular observables as a function of $q^2$?}]
    }
    \label{gammas,i2}
\end{table}
%%%%%%%%%%%%%%%%%%%%%%%%%%%%%%%%%%%%%%%%%%%%%%%%%%%%%%%%%%%%%%%%%%%%%%%%%%%%%%%%%%%%%%%%%%%%%%%%%%%%%%%%%%%%%%%%%%%%%%%%%%%%%%%%%%%%%%%

%%%%%%%%% Binned cases %%%%%%%%%

We now discuss our predictions and the available data for the angular observables. 
Following LHCb \cite{LHCb:2021yxk,LHCb_supplementary_material_aps,LHCb_supplementary_material_4}, we define the ranges:
%[\LVS{I think we should not give predictions for the bins $ r^{({\rm low})}, r^{(\eta)}, r^{(\rho: {\rm inf})}, r^{({\rm high})} $}]

\begin{eqnarray}\label{eq:bin_definitions}
    %%r^{({\rm low})} &\equiv& [ 0.212^2, 0.525^2 ] \; \text{GeV}^2 \,, \nonumber\\
    %%r^{(\eta)} &\equiv& [ 0.525, 0.565 ] \; \text{GeV} \,, \nonumber\\
    %%r^{(\rho: {\rm inf})} &\equiv& \; [ 0.565^2, 0.78^2 ] \; \text{GeV}^2 \,, \nonumber\\
    r^{(\rho: {\rm sup})} &\equiv& \;\; [ 0.78^2, 0.95^2 ] \; \text{GeV}^2 \,, \\
    r^{(\phi: {\rm inf})} &\equiv& \;\; [ 0.95^2, 1.02^2 ] \; \text{GeV}^2 \,, \nonumber\\
    r^{(\phi: {\rm sup})} &\equiv& \;\;\; [ 1.02^2, 1.1^2 ] \; \text{GeV}^2 \,. \nonumber
    %%r^{({\rm high})} &\equiv& \;\;\; [ 1.1, 1.59 ] \; \text{GeV} \nonumber
\end{eqnarray}
Since we focus on the high-energy window of Fig.~\ref{fig:DeBoer_Hiller}, we will discuss predictions for these three bins, while LHCb also provides results for the bins $ [ 0.212^2, 0.525^2 ] \; \text{GeV}^2 $ and $ [ 0.565^2, 0.78^2 ] \; \text{GeV}^2 $; the bin $ [ 0.565^2, 0.78^2 ] \; \text{GeV}^2 $, however, is also used for determining the total branching ratio distribution as a function of $ p^2 $ (the branching ratio outside these four $q^2$-bins is highly suppressed).
%
%\begin{figure}
%    \centering
%    \begin{subfigure}[b]{0.45\textwidth}
%    \includegraphics[scale=0.4]{dgammadp207to09.png}
%%    \end{subfigure}
%    \begin{subfigure}[b]{0.45\textwidth}
%        \includegraphics[scale=0.4]{dgammadp207to09noOmega.png}
%    \end{subfigure}
%    \caption{The differential decay rate (arbitrary numbers) over the dihadron invariant mass $\sqrt{p^2}$ for $\rho$ and $\omega$(left) vs only $\rho$ (right). We use the implementation suggested by BES. (note that such a shape coming from $\omega$ appears in other processes as well) [also, check if I should multiply by $2\sqrt{p^2}$ because it's $dBr/dp^2$ and not $dBr/dp$] }
%    \label{dgammadp2rhovsrhoomega}
%\end{figure}
%
In Tab.~\ref{gammas,i2} we present predicted values for those observables that do not vanish in the SM, in particular in the presence of the $S$-wave, in cases where it does not interfere with the $P$-wave. As seen in this table, the $\sigma$ provides significant contributions, as large as $10 - 40$\% in the binned branching ratios. This fraction is even larger in the case of $\int \langle I_2\rangle _{+,\sigma}^r$, which contributes to the binned branching ratio $ \Gamma^r = 2 \left( \int \langle I_1 \rangle_+^r - \int \langle I_2 \rangle_+^r / 3 \right) $, reaching up to about $50 - 80$\% of $\int \langle I_2\rangle _+^r$. The dominance of the $S$-wave in this observable can be attributed to a suppression of the $P$-wave contribution, due to a cancellation among the transversity form factors as seen from Eq.~\eqref{I2plus} (also manifesting in the case of $\int \langle I_3\rangle _+^r$), which on the other hand are added constructively in the case of $\langle I_1 \rangle_+^r$, cf. Eq.~\eqref{I1plus}.
In performing a comparison of our predictions to LHCb data of the observables $\langle S_2 \rangle^r$, $\langle S_3 \rangle^r$, and $\langle S_4 \rangle^r$
in the three bins $ r^{(\rho: {\rm sup})} $, $ r^{(\phi: {\rm inf})} $, and $ r^{(\phi: {\rm sup})} $
we obtain a $p$-value of $ \mathcal{O} (10) \% $.

As we have seen, our predictions for the angular observables $\langle S_7 \rangle^r$, $\langle S_8 \rangle^r$, and $\langle S_9 \rangle^r$ (approximately) vanish, even in presence of NP; we find however a poor comparison with the hypothesis that they are all zero in the five bins of Eq.~\eqref{eq:bin_definitions}, $ \chi^2 / N_{\rm d.o.f.} \simeq 2.4 $ (where $ N_{\rm d.o.f.} \simeq 15 $), or a $p$-value of $0.2\%$, due to $\langle S_9 \rangle^r$.
%$ \chi^2 / N_{\rm d.o.f.} \simeq 2.1 $ (where $ N_{\rm d.o.f.} \simeq 15 $), or a $p$-value of $0.8\%$ w/o correlations
This may indicate
%an experimental problem,
%in bins for which their measured values deviate significantly from zero
a missing description of the relative strong phases among the transversity form factors $ \mathcal{F}_P $, $ \mathcal{F}_\Vert $, and $ \mathcal{F}_\perp $.
%%[\rd{any particular observable?} it consists of looking for those that deviate from 0, see conclusions]
(Including in this latter test the $\langle S_5 \rangle^r$ and $\langle S_6 \rangle^r$ observables, which also vanish in the SM, we get
$ \chi^2 / N_{\rm d.o.f.} \simeq 2.0 $ (where $ N_{\rm d.o.f.} \simeq 25 $), or a $p$-value of $0.2\%$, which is small also as a consequence of including $\langle S_9 \rangle^r$.)
%$ \chi^2 / N_{\rm d.o.f.} \simeq 1.8 $ (where $ N_{\rm d.o.f.} \simeq 25 $), or a $p$-value of $0.9\%$ w/o correlations
The violation of CP is surely exciting in the context of charm physics, where a sizable level of CP violation has been recently measured by LHCb \cite{LHCb:2019hro,LHCb:2022lry} in hadronic two-body charm-meson decays, see Ref.~\cite{Pich:2023kim} for a theoretical discussion.
On the other hand, the CP asymmetries in rare charm-meson decays are consistent with zero, since in this case we find that $p$-value $= 84\%$.
%$p$-value $= 90\%$ w/o correlations
Note that statistical correlations among bins and across observables are provided by the LHCb analysis; they are small, but have been included.
%%Syst correlations have not been included because syst uncs are sub-leading; for cases where there are non-vanishing preds, theo uncs are likely underestimated
%[some numerical values quantifying the comparison exp vs. pred]
%%[\rd{"Observables measured in disjoint dimuon-mass intervals are statistically uncorrelated.", "Systematic uncertainties between the observables, in each and across different dimuon-mass intervals, are assumed to be fully correlated."}]
Systematic uncertainties are smaller than statistical uncertainties, and are fully correlated (we use the techniques discussed in Ref.~\cite{Charles:2016qtt} to combine both categories of uncertainties in presence of correlations).

In Tab.~\ref{i2i4i8} we provide the values for non-vanishing angular observables that probe the interference of the $S$- and $P$-waves. These observables depend on the relative phase $ \Delta_{SP} = \delta_{ \{ \sigma, \rho^0 \} }-\delta_{ \{ \rho^0/\omega, \rho^0 \} } $ between the $S$- and $P$-waves.
None of the experimentally provided observables from Refs.~\cite{LHCb:2021yxk,LHCb_supplementary_material_aps,LHCb_supplementary_material_4} is sensitive to this phase, hence it is left as a free parameter. A future experimental analysis would probe this phase difference, possibly in combination with the differential distribution over the di-hadron angle, as discussed later in this section.
%\rd{[Should we compare to the BES result for $\phi_s=3.4?$ They are different processes in principle, since here the phase shift can come from FSI.]}
As seen in the table, some sizable values are found, typically smaller but of similar order compared to the ones provided in Tab.~\ref{gammas,i2} that are insensitive to the $S$-wave. 

%Concerning the second category of observables, we find that for the most part the P-wave sensitive  $(\int   dp^2 I_4)_{(-,P)}$ is one order of magnitude larger than $(\int   dp^2 I_4)_{(+,S)}$ [\rd{to do: check dependence, for the vector one, on the relative phase between the $\rho/\omega$ and the $\phi$}].

%%%%%%%%%%%%%%%%%%%%%%%%%%%%%%%%%%%%%%%%%%%%%%%%%%%%%%%%%%%%%%%%%%%%%%%%%%%%%%%%%%%%%%%%%%%%%%%%%%%%%%%%%%%%%%%%%%%%%%%%%%%%%%%%%%%%%%%
%\begin{table}
%    \centering
%    \renewcommand{\arraystretch}{1.2}
%    \begin{tabular}{|c|c|c|c|}
%    \hline
%    $q^2$-bin $r$ &  $\int \langle I_{2}\rangle_{-}^r$ (\%) & $\int \langle I_4\rangle_{+}^r$ (\%) & $\int \langle I_8\rangle_{+}^r$ (\%) \\
%    %\cline{2-4}
%    %& \multicolumn{3}{c|}{$[10^{-2}]$} \\
%    \hline 
   %%$ r^{({\rm low})} $        & 0.50 & -0.21  & 0.15  \\ 
   %%$ r^{(\eta)} $             & 0.22  & -0.11 & 0.09\\ 
   %%$ r^{(\rho: {\rm inf})} $  & -0.57 & 1.75 &  4.84\\
%   $ r^{(\rho: {\rm sup})} $ & -3.63  &  3.49 & 1.18\\ 
%   $ r^{(\phi: {\rm inf})} $  & 8.62   & -9.18& 1.11  \\
%   $ r^{(\phi: {\rm sup})} $  & -2.47 & 2.85 & 3.39\\
   %%$ r^{({\rm high})} $  &      -0.13 & 0.15 & 0.02\\
%       \hline
%    \end{tabular}
%    \caption{OLD SM predictions for the non-vanishing angular observables that probe the interference effect between the $S$- and $P$-waves. The relation $ \int \langle I_1 \rangle_-^r = - \int \langle I_2 \rangle_-^r $ holds true. The same overall multiplicative factor shown in the caption of Tab.~\ref{gammas,i2} applies. [\rd{give at most two digits}]}
%\label{i2i4i8OLD}
% \end{table}

\begin{table}
%%\hspace{-2cm}
\centering
    \renewcommand{\arraystretch}{1.4}
    %%\begin{tabular}{|c|c|c|c|}
    %%\hline
    %%$q^2$-bin $r$ &  $\int \langle I_{2}\rangle_{-}^r$ (\%) & $\int \langle I_4\rangle_{+}^r$ (\%) & $\int \langle I_8\rangle_{+}^r$ (\%) \\
    %\cline{2-4}
    %& \multicolumn{3}{c|}{$[10^{-2}]$} \\
    %%\hline 
   %%$ r^{({\rm low})} $        & 0.50 & -0.21  & 0.15  \\ 
   %%$ r^{(\eta)} $             & 0.22  & -0.11 & 0.09\\ 
   %%$ r^{(\rho: {\rm inf})} $  & -0.57 & 1.75 &  4.84\\
   %%$ r^{(\rho: {\rm sup})} $ & $[-1.9,2.0] c_{SP}+ [-5.0,0.09]s_{SP}$  &  $+[-2.4,1.3]c_{SP}+[-0.08,4.2 ]s_{SP}$ & $[-2.2, 0.03]c_{SP}+[-0.9, 0.8]s_{SP}$\\ 
   %%$ r^{(\phi: {\rm inf})} $  &$[-10.5,9.1]c_{SP}+[-8.2,5.3]s_{SP}$   & $[-11.0,11.3]c_{SP}+[-5.5,9.0]s_{SP} $& $[-3.6, 2.5]c_{SP}+[-4.2, 4.8]s_{SP}$  \\
   %%$ r^{(\phi: {\rm sup})} $  & $[-8.3,7.2]c_{SP}+[-6.7,3.9]s_{SP}$ & $[-7.7,9.0]c_{SP}+[-4.4,7.4]s_{SP}$ & $[-2.7, 1.7]c_{SP}+[-3.3, 3.8]s_{SP}$\\
   %%$ r^{({\rm high})} $  &      -0.13 & 0.15 & 0.02\\
   %%\hline
    %%\end{tabular}
    %%%
    \begin{tabular}{|c|c|}
        \hline
        $q^2$-bin & $\int \langle I_{2} \rangle_{-}^r \times 100$ \\
        \hline
        $ r^{(\rho: {\rm sup})} $ & $[-6.6,-0.8] \, c_{SP}+ [-2.3,-1.1] \, s_{SP}$ \\
        \hline
        $ r^{(\phi: {\rm inf})} $ & $[-7.7,6.1] \, c_{SP}+[-5.3,8.2] \, s_{SP}$ \\ %[-9.1,10.5]
        \hline
        $ r^{(\phi: {\rm sup})} $ & $[-7.1,3.0] \, c_{SP}+[-5.0,5.4] \, s_{SP}$ \\
        \hline
        %%%%%%
        \hline 
        $q^2$-bin & $\int \langle I_4 \rangle_{+}^r \times 100$ \\
        \hline
        $ r^{(\rho: {\rm sup})} $ & $[0.8,5.9] \, c_{SP} + [0.4,1.6] \, s_{SP}$ \\
        \hline
        $ r^{(\phi: {\rm inf})} $ & $[-6.7,8.3] \, c_{SP} + [-8.6,5.4] \, s_{SP} $ \\ %[-11.3,9.6]
        \hline
        $ r^{(\phi: {\rm sup})} $ & $[-3.1,7.6] \, c_{SP} + [-5.9,5.5] \, s_{SP}$ \\
        \hline
        %%%%%%
        \hline  
        $q^2$-bin & $\int \langle I_8 \rangle_{+}^r \times 100$ \\
        \hline
        $ r^{(\rho: {\rm sup})} $ & $[-3.0,-0.2] \, c_{SP} + [-0.4,0.4] \, s_{SP}$ \\
        \hline
        $ r^{(\phi: {\rm inf})} $ & $[-4.6,4.5] \, c_{SP} + [-3.4,4.0] \, s_{SP}$ \\
        \hline
        $ r^{(\phi: {\rm sup})} $ & $[-2.6,3.3] \, c_{SP} + [-1.7,3.3] \, s_{SP}$ \\
        \hline
    \end{tabular}
    \caption{SM predictions for the non-vanishing angular observables that probe the interference between the $S$- and $P$-waves. The parameters appearing stand for $c_{SP} \equiv \cos( \Delta_{SP} )$ and $s_{SP} \equiv \sin( \Delta_{SP} )$. The relation $ \int \langle I_1 \rangle_-^r = - \int \langle I_2 \rangle_-^r $ holds true.
    Relevant definitions can be found in Sec.~\ref{sec:observables}, see in particular Eqs.~\eqref{isnormalised} and \eqref{intnotation}.
    The same overall multiplicative factor shown in the caption of Tab.~\ref{gammas,i2} applies. 
    %[\rd{can you provide for inner use unbinned plots of the angular observables as a function of $q^2$?}]
    }
\label{i2i4i8}
 \end{table}
%%%%%%%%%%%%%%%%%%%%%%%%%%%%%%%%%%%%%%%%%%%%%%%%%%%%%%%%%%%%%%%%%%%%%%%%%%%%%%%%%%%%%%%%%%%%%%%%%%%%%%%%%%%%%%%%%%%%%%%%%%%%%%%%%%%%%%%

%%The ranges provided for each binned angular observable are obtained by varying independently the most uncertain parameters, $ B^{(S)}_\phi / B^{(S)}_{\rho^0} $ and $ a_S (0) / A_1 (0) $ and the phases $\Delta_i$ for $i=1, 3, 4$ within the intervals specified above, while keeping the rest of the parameters at their best fit values.
%\rd{We do not consider variations related to the uncertainties of the normalization factors ($a_S(0)$, $B_\rho,\, B_\omega, \, B_\phi$), since the observables within the defined bins are dominated by the resonances and the corresponding normalization factors largely cancel in the ratios defined in Eq.~\eqref{isnormalised}. [requires further verification: I believe that now one should consider at least the error in the $S$-wave fudge factors; should I say something like uncertainties are a rough estimate? [it is already stated at the head of this section]; comment on parametric uncertainties]}

%%%%%%%%% Sigma and angular variables %%%%%%%%%
%On the other hand,
%\rd{The $\sigma$ is even more prominent when analysing differential decay rates as a function of angular variables. In the observable $I_2$...}

\begin{figure}
    \centering
        \includegraphics[width=0.7\textwidth]{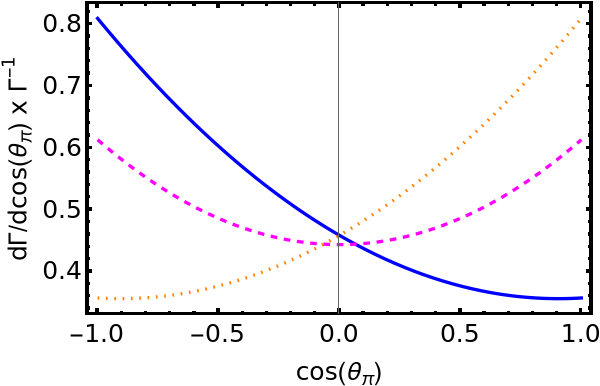}
    \caption{The differential decay rate, after integration of di-lepton energies over the range $r^{(\rho: {\rm sup})} \cup r^{(\phi: {\rm inf})} \cup r^{(\phi: {\rm sup})} = [ 0.78^2, 1.1^2 ] \; \text{GeV}^2 $, as a function of $\cos (\theta_\pi)$.
    %and normalized to the total decay rate in the same bins
    In dashed magenta the observable is shown in the absence of the $S$-wave contribution (rescaled such that $ \int^1_{-1} d \cos (\theta_\pi) d \Gamma / d \cos (\theta_\pi) / \Gamma = 1 $).
    The solid blue and dotted orange lines correspond to extreme cases reached for certain values of the phase difference $ \Delta_{SP} $ between the $S$- and $P$-waves
    that maximize their interference.
    %The decay rate can fall anywhere between these two lines.
    As it is clear from the figure, the interference of the $S$- and $P$-waves can generate a distinguished asymmetry.
    %[{\color{red} need to consider only $q^2$ in the high-energy region defined in the introduction; send to me the numerical information (the angular information and the branching ratio in the same bin), I will generate the graphs}]
    }
    \label{fig:dgammadcostheta}
\end{figure}

Finally,
as announced in the introduction, the $S$-wave can produce distinguished signatures in the differential branching ratio as a function of the angular variables describing the topology of the rare decay.
To illustrate this point, consider: %the relative angle between the pions

\begin{equation}
    \frac{d \Gamma}{d \cos \theta_\pi} = \langle I_{1} \rangle_{+, \rho^0/\omega}^r + \langle I_{2}\rangle_{+, \rho^0/\omega}^r \, (1 - 4 \cos^2 \theta_\pi) - \frac{4}{3} \langle I_{2} \rangle_{+, \sigma}^r -\frac{8}{3} \langle I_{2}\rangle_-^r \, \cos \theta_\pi \,,
\end{equation}
after integration over the $q^2$-bin $r$, where the contributions from the $S$- and $P$-waves alone are indicated in subscript (here, the $\sigma$ and $\rho^0/\omega$ resonances, respectively), and the last term in the right-hand side probes their interference.
As seen in Fig.~\ref{fig:dgammadcostheta}, the presence of the $S$-wave can produce an asymmetry of the distribution with respect to $ \cos \theta_\pi = 0 $.
This provides motivation for binned measurements of the branching ratio as a function of the angular variables.
%[Bins as a function of the angles are not available, nor bins in the di-hadron invariant mass for observables other than the branching ratio --> I cannot clearly indicate a benefit from the latter]

%%%%%%%%%%%%%%%%%%%%%%%%%%%%%%%%%%%%%%%%%%%%%%%%%%%%%%%%%%%%%%%%%%%%%%%%%%%%%%%%%%%%%%%%%%%%%%%%%%%%%%%%%%%%%%%%%%%%%%%%%%%%%%%%%%%%%%%
\subsection{Semi-leptonic operators from generic NP}\label{sec:SM_NP_fits}

%[WE focus on SM times NP]

%We have that: $ |\tilde{C}_9| < 1.2 $ and $ |\tilde{C}_{10}| < 0.51 $ @ 95\% C.L., where $ |\tilde{C}| = |V_{ub} V_{cb}^\ast C| $, see Refs.~\cite{Fajfer:2015mia,Fajfer:2023efd} (``The bounds for $ \tilde{C}_i $ apply also to the chirally flipped coefficients $ \tilde{C}_j' $.'').

We want to know the impact of having dimension-6 operators that can mediate the transition $ c \to u \ell^+ \ell^- $ at the quark level due to interactions mediated by heavy NP.
We focus on vector and axial-vector structures.
Present bounds @ 95\% C.L. are \cite{Fajfer:2015mia}:

\begin{equation}\label{eq:present_bounds}
    |\tilde{C}^{\rm NP}_9|, |\tilde{C}'_9| < 1.2 \,, \quad |\tilde{C}_{10}^{(')}| < 0.43 \,,
\end{equation}
where $ |\tilde{C}| = |V_{ub} V_{cb}^\ast C| $, and
the former bound results from the $ D^+ \to \pi^+ \mu^+ \mu^- $ branching ratio \cite{LHCb:2020car}, while the second from the $ D^0 \to \mu^+ \mu^- $ branching ratio \cite{LHCb:2022jaa}.
Slightly better bounds are found from collider searches for contact interactions manifesting in $ p p \to \mu^+ \mu^- $ \cite{Fuentes-Martin:2020lea}.
%[\rd{need to check work by Admir \cite{Fuentes-Martin:2020lea}: he claims a better sensitivity to NP from direct collider bounds}]
In view of these constraints, it is justified to assume that NP does not affect the previous discussion about the differential branching ratio as a function of the invariant masses of pion and lepton pairs. However, NP could still affect the differential branching ratio in the low and very-high di-lepton invariant mass regions \cite{DeBoer:2018pdx}. It can also affect distinct angular observables as we now discuss.

As seen from the expressions provided in Sec.~\ref{sec:observables}, there are distinct observables that depend on these Wilson coefficients.
In Tab.~\ref{NPobsevablesPwave} we provide predictions for those observables sensitive to the SM-NP interference in presence of a non-vanishing $C_{10}$ Wilson coefficient (its SM value is very suppressed, as discussed around Eq.~\eqref{eq:H_eff}).
The cases $\braket{I_5}_-$ and $\braket{I_6}_+$ are sensitive to the SM-NP interference through the $P$-wave, while $\braket{I_7}_-$ approximately vanishes.
These observables, which isolate the NP interference with the SM $P$-wave, are given as functions of the phase difference

\begin{equation}
    \Delta_{\rho {\rm NP}} \equiv \delta_{ \{ \rho^0/\omega, \rho^0 \} }-\delta_{ Q_{10} } \,,
\end{equation}
where $ \delta_{ Q_{10} } $ allows for a possible strong phase when considering insertions of the $Q_{10}$ operator (beyond the one from the pion pair line-shape).
Predictions are shown in Tab.~\ref{i2i4i8}.

On the other hand,
the cases $\braket{I_5}_+$ and $\braket{I_7}_+$ are sensitive to the SM-NP interference in the presence of the $S$-wave.
These observables
%depending on the NP interference with the $S$-wave
depend on the above phase $ \Delta_{\rho {\rm NP}} $ together with $ \Delta_{SP} $.
The latter phase difference can be probed based on the observables whose predictions are given in Tab.~\ref{i2i4i8}, and the observable shown in Fig.~\ref{fig:dgammadcostheta}.
%\begin{equation}
%    \Delta_{\sigma NP} \equiv \delta_{ \{ \sigma, \rho^0 \} }-\delta_{ Q_{10} }
%\end{equation}
%\LVS{Experimental information on the S-P interference, SM-dominated observables presented in Table \ref{i2i4i8} would help in a better determination of these S-wave related null tests.}
Given the dependence on both phase differences, we do not give explicitly the expressions for the related angular observables.
By varying these phases,
we stress that we find values of the angular observables comparable to the ones found for the analogous $P$-wave null tests in Tab.~\ref{i2i4i8}.
%at the 1-2 percent level

Given the bounds shown above in Eq.~\eqref{eq:present_bounds}, detecting NP requires sub-percentage precision in the measurement of the angular observables. Having reached such precision, some bins of the angular observables sensitive to the $S$-wave provide additional complementary information to favor or disfavor an observation of a possible NP manifestation based on the $P$-wave cases.
In the future, a global fit could extract all relevant phases, together with possible NP contributions.
It is possible that a clever strategy could circumvent the need to extract at least some of the strong phases affecting the angular observables.

\begin{table}[]
    \centering
    \renewcommand{\arraystretch}{1.4}
   \begin{tabular}{|c|c|}
        \hline
        $q^2$-bin & $\int \langle I_{5} \rangle_{-}^r \times 100$ \\
        \hline
        $ r^{(\rho: {\rm sup})} $ & $[0.49,0.83] \, c_{\rho {\rm NP}}+ [-1.5,-1.3] \, s_{\rho {\rm NP}}$ \\
        \hline
        $ r^{(\phi: {\rm inf})} $ & $[-0.36,0.50] \, c_{\rho {\rm NP}}+[-0.83,-0.60] \, s_{\rho {\rm NP}}$ \\
        \hline
        $ r^{(\phi: {\rm sup})} $ & $[0.31,0.66] \, c_{\rho {\rm NP}}+[-0.09,0.49] \, s_{\rho {\rm NP}}$ \\ %[-1,-0.7]
        \hline
        %%%%%%
        \hline
        $q^2$-bin & $\int \langle I_6 \rangle_{+}^r \times 100$ \\
        \hline
        $ r^{(\rho: {\rm sup})} $ & $[0.7,1.2] \, c_{\rho {\rm NP}} + [-2.1,-1.7] \, s_{\rho {\rm NP}}$ \\
        \hline
        $ r^{(\phi: {\rm inf})} $ & $[-0.57,0.78] \, c_{\rho {\rm NP}} + [-1.3,-1.0] \, s_{\rho {\rm NP}} $ \\
        \hline
        $ r^{(\phi: {\rm sup})} $ & $[0.5,1.1] \, c_{\rho {\rm NP}} + [-0.14,0.78] \, s_{\rho {\rm NP}}$ \\
        \hline
    \end{tabular}
    \caption{Observables that vanish in the SM, arising from the interference of the $P$-wave and NP, here calculated for $C_9'=C_{10}'=C_9^{\rm NP}=0$ and non-zero $C_{10}$. The parameters appearing stand for $c_{\rho {\rm NP}}=\cos( \Delta_{\rho {\rm NP}} )$ and $s_{\rho {\rm NP}}=\sin( \Delta_{\rho {\rm NP}} )$. The other $P$-wave dependent observable $\braket{I_7}_-$ approximately vanishes.
    The NP does not interfere with the SM in the decay rate, and can thus be neglected.
    Relevant definitions can be found in Sec.~\ref{sec:observables}, see in particular Eqs.~\eqref{isnormalised} and \eqref{intnotation}.
    The same overall multiplicative factor shown in the caption of Tab.~\ref{gammas,i2} applies; additionally, there is an extra $ \tilde{C}_{10}$ that multiplies the observables.
    %[\rd{can you provide for inner use unbinned plots of the angular observables as a function of $q^2$?}]
    }
    %there is a normalization problem, the result must be $ \propto C_2 \times C_{10}^\ast $, while here it is instead $ \propto |C_2|^2 \times C_{10}^\ast $: Ele extracted the numerics things properly
    \label{NPobsevablesPwave}
\end{table}
%%%%%%%%%%%%%%%%%%%%%%%%%%%%%%%%%%%%%%%%%%%%%%%%%%%%%%%%%%%%%%%%%%%%%%%%%%%%%%%%%%%%%%%%%%%%%%%%%%%%%%%%%%%%%%%%%%%%%%%%%%%%%%%%%%%%%%%

%%%%%%%%%%%%%%%%%%%%%%%%%%%%%%%%%%%%%%%%%%%%%%%%%%%%%%%%%%%%%%%%%%%%%%%%%%%%%%%%%%%%%%%%%%%%%%%%%%%%%%%%%%%%%%%%%%%%%%%%%%%%%%%%%%%%%%%

%%%%%%%%%%%%%%%%%%%%%%%%%%%%%%%%%%%%%%%%%%%%%%%%%%%%%%%%%%%%%%%%%%%%%%%%%%%%%%%%%%%%%%%%%%%%%%%%%%%%%%%%%%%%%%%%%%%%%%%%%%%%%%%%%%%%%%%

%%%%%%%%%%%%%%%%%%%%%%%%%%%%%%%%%%%%%%%%%%%%%%%%%%%%%%%%%%%%%%%%%%%%%%%%%%%%%%%%%%%%%%%%%%%%%%%%%%%%%%%%%%%%%%%%%%%%%%%%%%%%%%%%%%%%%%%
\section{Conclusions}\label{ref:conclusions}

Recent experimental data by LHCb open up the opportunity for precision physics with rare charm-meson decays, a task that can be assisted by complementary information coming from experiments such as BESIII, and by Belle~II in different rare decay modes. For this sake, better theoretical predictions are needed, in particular the description of resonances,
without which it will not be possible to disentangle non-SM contributions from the large SM background;
better theoretical predictions of the SM are also needed in order to describe possible interference terms with non-SM contributions.
We employ a factorization model for the inclusion of intermediate hadronic states contributing to $ D^0 \to \pi^+ \pi^- \ell^+ \ell^- $ in the SM, and discuss in details different contributing topologies. Within this framework, the novelty of this work concerns the inclusion of the lightest scalar isoscalar state, which is a very broad resonance manifesting in long-distance pion pair interactions and impacts a large portion of the allowed phase space, see Fig.~\ref{fig:DeBoer_Hiller}.
We highlight that $ D^0 \to \pi^+ \pi^- \ell^+ \ell^- $ data already show the clear emergence of such $S$-wave effects, see Fig.~\ref{fig:dgammadp2}.
Moreover, current data also allows the study of the strong phases among intermediate resonances, see Fig.~\ref{fig:dgammadq2}.

%It cannot be stressed enough, however, that this requires a careful analysis of the SM background.
The decay $ D^0 \to \pi^+ \pi^- \ell^+ \ell^- $ offers the possibility to define a rich set of angular observables.
We then discuss angular observables that are sensitive to both the $S$- and $P$-waves.
Predictions are given in Tabs.~\ref{gammas,i2} and \ref{i2i4i8}.
We have been able to understand the overall pattern of the measured angular observables $ \langle S_i \rangle^r $, $i=2, \ldots, 8$, in distinct $q^2$-bins $r$.
To further improve our understanding of SM contributions,
we suggest experimentalists measure additional observables to further test and better characterise the contributions of the $S$-wave, such as following the strategy illustrated in Fig.~\ref{fig:dgammadcostheta}.

Such additional observables have also an interest other than improving non-perturbative aspects of the SM description.
Indeed, the search for NP consists of one of the main motivations for looking into this category of rare decay processes.
If any deviation is seen while performing a null test of the SM, a comprehensive analysis will be needed to verify and characterize it.
We emphasize the potential for complementary tests of NP via its interference with the SM in the presence of the $S$-wave, which provide distinct null tests of the SM, as seen from Tab.~\ref{NPobsevablesPwave}.

In order to improve the description of the differential branching ratio,
in particular the one as a function of the pion pair invariant mass,
future theoretical directions include incorporating other $S$- and $P$-wave resonances and the $D$-wave following a similar theoretical framework, isospin-two contributions, and the addition of cascade decays.
%\rd{Better look into
%the $\phi$ region
%implementation of strong phases [what else to include?].}
More studies will be needed to understand the set of angular observables measured by LHCb in more details, since with our simple factorization model some tension appears in the description of the angular observable
$ \langle S_9 \rangle^r $.
%in bins close to the $\phi$ mass
It would also be interesting to extend our analysis to include $ D^0 \to K^+ K^- \mu^+ \mu^- $ and radiative decay modes.
%%%%%%%%%%%%%%%%%%%%%%%%%%%%%%%%%%%%%%%%%%%%%%%%%%%%%%%%%%%%%%%%%%%%%%%%%%%%%%%%%%%%%%%%%%%%%%%%%%%%%%%%%%%%%%%%%%%%%%%%%%%%%%%%%%%%%%%
\subsection*{Acknowledgements}

We thank Fernando~Abudin\'{e}n, Damir~Becirevic, Thomas~Blake, Luka~Leskovec, Lei~Li, Serena~Maccolini, Dominik~Mitzel, Juan~Nieves, Eulogio~Oset, Tommaso~Pajero, Antonio~Pich, Luka~Santelj and Shulei~Zhang for helpful communication.
%L.~V.~S. is grateful to Sebastian~J\"{a}ger for early discussions.
E.~S. is grateful to Gudrun Hiller and the rest of the participants for helpful discussions during the ``First CharmInDor mini workshop''.

\vspace{3mm}
\noindent
This work has been supported by MCIN/AEI/10.13039/501100011033, grants PID2020-114473GB-I00 and PRE2018-085325, and by Generalitat Valenciana, grant PROMETEO/2021/071.
This project has received funding from the European Union’s Horizon 2020 research and innovation programme under the Marie Sklodowska-Curie grant agreement No 101031558.
S.~F. acknowledges the financial support from the Slovenian Research Agency (research core funding No. P1-0035).
L.~V.~S. is grateful for the hospitality of the Jo\v{z}ef Stefan Institute and the CERN-TH group where part of this research was executed.
%
%\LVS{Erasmus funding for Ele's stay in Ljubljana?}

%%%%%%%%%%%%%%%%%%%%%%%%%%%%%%%%%%%%%%%%%%%%%%%%%%%%%%%%%%%%%%%%%%%%%%%%%%%%%%%%%%%%%%%%%%%%%%%%%%%%%%%%%%%%%%%%%%%%%%%%%%%%%%%%%%%%%%%
\appendix

\section{Hadronic inputs}\label{app:numerical_inputs}

\subsection{Decay constants}\label{app:decay_constants}

We have from Ref.~\cite{Bharucha:2015bzk}:

\begin{equation}
    \langle \phi | \bar{s} \gamma_\mu s | 0 \rangle = \epsilon_\mu^\ast m_\phi f_\phi \,, \;\;
    \hat{c}^q_\omega \langle \omega | \bar{q} \gamma_\mu q | 0 \rangle = \epsilon_\mu^\ast m_\omega f^{(q)}_\omega \,, \;\;
    \hat{c}^q_{\rho^0} \langle \rho^0 | \bar{q} \gamma_\mu q | 0 \rangle = \epsilon_\mu^\ast m_{\rho^0} f^{(q)}_{\rho^0} \,,
\end{equation}
with $ \hat{c}^u_{\rho^0} = -\hat{c}^d_{\rho^0} = \hat{c}^u_\omega = \hat{c}^d_\omega = \sqrt{2} $.
%[\LVS{parallel decay constant?; not needed here; Feldmann et al. quote it}]%Perhaps there is a decay constant with sigmamunu, which might be called perpendicular?; I am not sure how to compare with Feldmann et al., but they have a single Lorentz structure (i.e., no sigmamunu); in the equations above, both parallel and perpendicular should be contained; I don't know if there is a relation between parallel above, and a rho polarized exclusively in the parallel direction
%Indeed, it is about sigmamunu, see hep-lat/0301020; the perpendicular seems required in LCSR, I reckon in our naive factorization approach it does not show up
We consider a single decay constant for both matrix elements of $u$- and $d$-quark bilinears, i.e., $ f^{(q)}_\omega \to f_\omega $ and $ f^{(q)}_{\rho^0} \to f_{\rho^0} $, which is good enough for our purposes.
The decay constants are then:

\begin{equation}
    f_{\rho^0} = 216 (3)~\text{MeV} \,, \quad f_\omega = 197 (8)~\text{MeV} \,, \quad f_\phi = 233 (4)~\text{MeV} \,.
\end{equation}
(Mixing effects $ \omega - \rho^0 $ and $ \omega - \phi $ have been included, but are small.)
\subsection{Form factors}\label{app:form_factors}

For the $D\to \mathcal{V}$ form factors, for both $\mathcal{V}=\rho^0, \omega$, we use the nearest pole approximation introduced in Ref.~\cite{Wirbel:1985ji}, which has the general form
\begin{equation}
    F(q^2)=F(0) / \left( 1-\frac{q^2}{m_\text{pole}^2} \right) \,.
\end{equation}
The pole masses implemented are $2.42$ GeV ($J^P = 1^+$) for $F = A_1$ and $A_2$, and $2.01$ GeV ($J^P = 1^-$) for $F = V$.
%and $2.42$ GeV for the $D\to\sigma$ form factor
%\rd{Check again these values! Are they up to date? Cappiello et al. quote Wirbel}
%%\rd{[Cappiello et al. quote wrongly $\bar{s}c$ poles instead of $\bar{d}c$]}
We define:

\begin{equation}
    r_V = \frac{V (0)}{A_1 (0)} \,, \quad r_2 = \frac{A_2 (0)}{A_1 (0)} \,,
\end{equation}
for which Ref.~\cite{BESIII:2018qmf} gives $r_V = 1.695 \pm 0.083 \pm 0.051$ and $r_2 = 0.845 \pm 0.056 \pm 0.039$ (with a correlation of $\rho_{r_V,r_2} = -0.206$), where the first (second) uncertainty is statistical (respectively, systematic).
%%[\rd{This value of $r_V$ leads to a poor comparison with $d \Gamma / d \cos \theta_e$ for $D^0 \to \pi^- \pi^0 e^+ \nu_e$ of \cite{BESIII:2018qmf}: a better value is $\sim r_V / 2$}]

%[\LVS{mention Lattice}] There is none for resonances

%There are searches of the decay modes: $ D^+ \to \phi e^+ \nu_e $ \cite{BESIII:2015kin}. WE DO NOT NEED D to phi FFs, because we are enforcing Zweig

%\section{Inputs for short-distance physics/Available experimental data}

%Some contributions leading to $D^0 \to P^+ P^- \ell^+ \ell^- $, $ P = \pi, K $, $ \ell = e, \mu $, are: $ D^0 \to f_0 (500) \ell^+ \ell^- $, $ D^0 \to \rho (770)^0 \ell^+ \ell^- $, $ D^0 \to \phi (1020) \ell^+ \ell^- $.

%Some recent references giving parameterizations of form factors are:

%\begin{itemize}
    %\item BESIII \cite{BESIII:2015kin}, form factors in the decay $ D^+ \to \omega e^+ \nu_e $, where subsequently $\omega \to 3 \pi$
    %\item BESIII \cite{BESIII:2018qmf} (see also CLEO \cite{CLEO:2011ab}), form factors in the decays $ D^0 \to \rho^- e^+ \nu_e $ and $ D^+ \to \rho^0 e^+ \nu_e $
%\end{itemize}

%There are furthermore observations of the decay modes: $ D^+ \to f_0 (500) e^+ \nu_e $ \cite{BESIII:2018qmf}

%$ D^+ \to \overline{K}_1 (1270)^0 e^+ \nu_e $ \cite{BESIII:2019eao}, $ D^0 \to \overline{K}_1 (1270)^- e^+ \nu_e $ \cite{BESIII:2021uqr}

%%%%%%%%%%%%%%%%%%%%%%%%%%%%%%%%%%%%%%%%%%%%%%%%%%%%%%%%%%%%%%%%%%%%%%%%%%%%%%%%%%%%%%%%%%%%%%%%%%%%%%%%%%%%%%%%%%%%%%%%%%%%%%%%%%%%%%%
\subsection{Line shapes}\label{app:line_shapes}

We reproduce the line shape of $f_0 (500)$ \cite{Bugg:2006gc}:
%(needless to say, a Breit-Wigner is precarious for the $f_0 (500)$)

\begin{equation}
     \mathcal{A}_S (s) = \left[ M^2 - s - g_1^2 (s) \frac{s - s_A}{M^2 - s_A} z (s) - i \, M \, \Gamma_{tot} (s) \right]^{-1} \,,
\end{equation}

\begin{equation}
    \Gamma_{tot} (s) = \sum^4_{i=1} \Gamma_i (s) \,,
\end{equation}

\begin{equation}
    M \, \Gamma_1 (s) = g_1^2 (s) \frac{s - s_A}{M^2 - s_A} \rho_1 (s) \,,
\end{equation}

\begin{equation}
    \rho_1 (s) = \sqrt{ 1 - 4 \, m_\pi^2 / s } \,,
\end{equation}

\begin{equation}
    g_1^2 (s) = M \, \left( b_1 + b_2 \, s \right) \, \exp [ - (s - M^2) / A ] \,,
\end{equation}

\begin{equation}
    z (s) = j_1 (s) - j_1 (M^2) \,,
\end{equation}

\begin{equation}
    j_1 (s) = \frac{1}{\pi} \left[ 2 + \rho_1 (s) \, \log \left( \frac{1 - \rho_1 (s)}{1 + \rho_1 (s)} \right) \right] \,,
\end{equation}
%\ell n

\begin{eqnarray}
    && M \, \Gamma_2 (s) = 0.6 \, g_1^2 (s) \, ( s / M^2 ) \\
    && \qquad\;\;\; \times \exp \left[ - \alpha (s - 4 \, m_K^2) \Theta (s - 4 \, m_K^2) - \alpha' (4 \, m_K^2 - s) \Theta (4 \, m_K^2 - s) \right] \rho_2 (s) \,,\nonumber
\end{eqnarray}

\begin{equation}
    \rho_2 (s) = \sqrt{ 1 - 4 \, m_K^2 / s } \, \Theta (s - 4 \, m_K^2) + i \, \sqrt{ 4 \, m_K^2 / s - 1 } \, \Theta (4 \, m_K^2 - s) \,,
\end{equation}

\begin{eqnarray}
    && M \, \Gamma_3 (s) = 0.2 \, g_1^2 (s) \, ( s / M^2 ) \\
    && \qquad\;\;\; \times \exp \left[ - \alpha (s - 4 \, m_\eta^2) \Theta (s - 4 \, m_\eta^2) - \alpha' (4 \, m_\eta^2 - s) \Theta (4 \, m_\eta^2 - s) \right] \rho_3 (s) \,,\nonumber
\end{eqnarray}

\begin{equation}
    \rho_3 (s) = \sqrt{ 1 - 4 \, m_\eta^2 / s } \, \Theta (s - 4 \, m_\eta^2) + i \, \sqrt{ 4 \, m_\eta^2 / s - 1 } \, \Theta (4 \, m_\eta^2 - s) \,,
\end{equation}

\begin{equation}
    M \, \Gamma_4 (s) = M \, g_{4 \pi} \, \rho_{4 \pi} (s) / \rho_{4 \pi} (M^2) \, \Theta (s - 16 \, m_\pi^2) \,,
\end{equation}

\begin{equation}
    \rho_{4 \pi} (s) = 1.0 / \left[ 1 + \exp (7.082 - 2.845 \, s / \text{GeV}^2) \right] \,,
\end{equation}

\begin{equation}
    s_A \simeq 0.41 \, m_\pi^2 \,, \quad \alpha = 1.3 \, \text{GeV}^{-2} \,, \quad \alpha' = 2.1 \, \text{GeV}^{-2} \,,
\end{equation}
and (solution (iii) of Ref.~\cite{Bugg:2006gc}): $ M = 0.953 $~GeV, $ b_1 = 1.302 $~GeV, $ b_2 = 0.340 $/GeV, $ A = 2.426 $~GeV$^2$, $ g_{4 \pi} = 0.011 $~GeV.

For the line shape of the $\rho (770)^0$ in $ \pi^+ \pi^- $ decays, we adopt the Gounaris-Sakurai parameterization \cite{Gounaris:1968mw}: %[\LVS{where does the numerator appear? --> the part proportional to d is small, the mass^2 in the numerator is taken into consideration}]

\begin{equation}
    P_{\rho^0}(s)=m_{\rho^0}^2 - s + f (s) - i \, m_{\rho^0} \, \Gamma_{\rho^0} (s) \,,
\end{equation}

%{\color{blue}
%\begin{equation}
%    \text{GS}_{\rho^0} (s) = \frac{m_{\rho^0}^2 + d \, m_{\rho^0} \, \Gamma_{\rho^0}^0}{m_{\rho^0}^2 - s + f (s) - i \, m_{\rho^0} \, \Gamma_{\rho^0} (s)} \,,
%\end{equation}}

\begin{equation}
    f (s) = \Gamma_{\rho^0}^0 \frac{m_{\rho^0}^2}{k_{\rho^0}^3} \left\{ k (s)^2 \left[ h (s) - h (m_{\rho^0}^2) \right] + k_{\rho^0}^2 \, (m_{\rho^0}^2 - s) \, h' (m_{\rho^0}^2) \right\} \,,
\end{equation}

\begin{equation}
    \Gamma_{\rho^0} (s) = \Gamma_{\rho^0}^0 \left( \frac{k (s)}{k_{\rho^0}} \right)^3 \frac{m_{\rho^0}}{\sqrt{s}} \,,
\end{equation}

%\begin{equation}
%    d = \frac{3}{\pi} \frac{m_\pi^2}{k_{\rho^0}^2} \ell n \left( \frac{m_{\rho^0} + 2 \, k_{\rho^0}}{2 \, m_\pi} \right) + \frac{m_{\rho^0}}{2 \, \pi \, k_{\rho^0}} - \frac{m_\pi^2 \, m_{\rho^0}}{\pi \, k_{\rho^0}^3} \,,
%\end{equation}

\begin{equation}
    h (s) = \frac{2}{\pi} \frac{k (s)}{\sqrt{s}} \log \left( \frac{\sqrt{s} + 2 \, k (s)}{2 \, m_\pi} \right) \,,
\end{equation}
%\ell n

\begin{equation}
    k (s) = \left( \frac{1}{4} s - m_\pi^2 \right)^{1/2} \,, \quad k_{\rho^0} = \left( \frac{1}{4} m_{\rho^0}^2 - m_\pi^2 \right)^{1/2} \,.
\end{equation}
We also have \cite{BESIII:2018qmf,CMD-2:2001ski,Shulei}:

%{\color{blue} Remove?
%\begin{equation}
%    \mathcal{A}_{{\rho^0}/\omega} (s) = \text{GS}_{\rho^0} (s) \times \left( 1 + a_\omega \, e^{i \phi_\omega} \, \text{RBW}_\omega (s) \right) \times F_1 (s) \,,
%\end{equation}
%}

\begin{equation}
    \text{RBW}_\omega (s) = \frac{s}{m_\omega^2 - s - i \, m_\omega \, \Gamma_\omega^0} \,,
\end{equation}
and

%[\LVS{where does $ \frac{p^\ast}{p^\ast_0} $ appear?}] --> part of it goes to the phase space, part to the normalization, see alphaGS

%{\color{blue} Remove?
%\begin{equation}
%    F_1 (s) = \frac{p^\ast}{p^\ast_0} \frac{B (p^\ast)}{B (p^\ast_0)} \,,
%\end{equation}
%}

\begin{equation}
    F_{BW} (p^2) = B(p^*) / B(p_0^*) \,,
\end{equation}

\begin{equation}
    B (p^\ast) = \frac{1}{\sqrt{1 + r^2_{BW} \, (p^\ast)^2}} \,.
    \label{blattweiss}
\end{equation}
The value of $ r_{BW} $ is taken to be $ 3.0 / $GeV (i.e., the inverse of a non-perturbative scale) \cite{BESIII:2018qmf}.
%$ (3.0 \pm 0.3)/ $GeV
%IT may then not make sense to include such a factor for muon pair emission
The function $p^\ast (p^2) = \sqrt{\lambda (p^2,m_\pi^2,m_\pi^2)} / ( 2 \sqrt{p^2} )$, and $ p^\ast_0 = p^\ast (m^2_{\rho^0}) $.
The $\phi$ and the $\omega$ line-shapes, when the latter decays to the lepton pair, are just Breit-Wigner line-shapes:

\begin{equation}
    P_\phi (s) = m_\phi^2 - s - i \, m_\phi \, \Gamma_\phi^0 \,, \quad P_\omega (s) = m_\omega^2 - s - i \, m_\omega \, \Gamma_\omega^0 \,,
\end{equation}

The masses and widths are \cite{Workman:2022ynf}:

\begin{eqnarray}
    && m_{\rho^0} = 775.3~\text{MeV} \,, \;\; \Gamma^0_{\rho^0} = 147.4~\text{MeV} \,, \\
    && m_\omega = 782.7~\text{MeV} \,, \;\; \Gamma^0_\omega = 8.7~\text{MeV} \,, \\
    && m_\phi = 1019.46~\text{MeV} \,, \;\; \Gamma^0_\phi = 4.25~\text{MeV} \,.
\end{eqnarray}
%\rd{As discussed in the main text, $ \Gamma^0_\phi $ is left as a free parameter in the present study.}

%%%%%%%%%%%%%%%%%%%%%%%%%%%%%%%%%%%%%%%%%%%%%%%%%%%%%%%%%%%%%%%%%%%%%%%%%%%%%%%%%%%%%%%%%%%%%%%%%%%%%%%%%%%%%%%%%%%%%%%%%%%%%%%%%%%%%%%
\section{Further comments on semi-leptonic decays}\label{app:sl_decays}

To reproduce the values in Table~I of \cite{BESIII:2018qmf} relative to $D^+$ decays,
%[has to understand better $D^0$ decays: related by isospin, see \cite{CLEO:2011ab}]
we find: $a_S (0) = 8.6 \pm 0.4$~GeV, $ a_\omega = 0.006 \pm 0.001 $, and $ A_1 (0) = 0.36 $. The strong phases extracted in their analysis are $ \phi_S = 3.4044 \pm 0.0738 $, which is somewhat the analogous of $\Delta_{SP}$ defined in the main text, and $ \phi_\omega = 2.93 \pm 0.17 $ \cite{Shulei}; this latter angle is consistent with $\pi$ from the isospin decomposition of the $(\bar{d} d)_V$ current, that generates the states $\rho^0$ and $\omega$.
%
%\footnote{We note that the relative phase found from $ e^+ e^- \to \pi^+ \pi^- $ (sensitive, therefore, only to $P$-wave contributions) is $ 0.22 \pm 0.06 $ \cite{CMD-2:2001ski}, compatible with \cite{CMD-3:2023alj}, while the relative phase expected from the isospin limit for $ V^0 \to \gamma \to \ell^+ \ell^- $ (i.e., $\sum e_q (\bar{q} q)_V$) is $ 0 $ \cite{Fajfer:2005ke}.}%This is a question of rho(I=1)/omega(I=0) mixing, which is an isospin breaking effect since they have different isospin quantum numbers; some people (Maltman+) however talk about a direct omega->pipi effect (Susan Gardner+ refer to model-dependence when talking about the separation between direct vs. mixing-induced effect); \cite{Fajfer:2005ke} discusses V^0 -> gamma -> ell ell
%
Instead,
we employ in this work the values extracted from a fit to the data of Refs.~\cite{LHCb:2021yxk,LHCb_supplementary_material_aps,LHCb_supplementary_material_4}, see Fig.~\ref{fig:dgammadp2}.
In doing so, we obtain the values quoted in Eq.~\eqref{eq:parameters_from_fit},
%$a_S (0) = 10.4$~GeV ($ a_\omega = 0.004 $), about 20\% (40\%) larger (smaller) than the central value shown above
which in the case of $a_S (0)/A_1 (0)$ is
about 2 times larger than the value shown above.
The comparison, however, is not straightforward, since the $\sigma$ contributes in three dynamical ways when combined with the $ \rho^0, \omega, \phi $ that lead to the lepton pair.
%[\LVS{why does the contribution seem smaller then? (note that the rho carries the same form factors, but there we did not have a J-type contribution, i.e., $25\% \times 2.2 / 2 = 15\%$; update: $25\% \times 2.2 / 2$, but now the normalization factors are treated separately, the S-wave phi is suppressed, and the P-wave phi is the main contribution, and thus the comparison is not that simple anymore)}]
%and $ \phi_\omega = 4.7 \sim 3 \, \pi / 2 $ [\LVS{revisit}]
Note that the resonance that decays into pion pairs originates from both $u$- (in the W-type topology) and $d$-quark pairs (in the J-type topology), which differs from the situation depicted above for $\phi_\omega$.
%(In both cases, it is implicitly assumed the absence of sizable contributions from a NP sector.)
Likely, the extraction of the phase $\phi_\omega$ from data is contaminated by the presence of further resonances discussed in the main text that we do not include in our analysis, and the presence of further intermediate hadrons (i.e., vector mesons that lead to the lepton pair) in the full charm-meson decay process.
%It seems thus that the latter phase depends strongly on the process from which it is extracted, e.g., whether there are further hadrons in the final state compared to the semi-leptonic transition analysed by \cite{BESIII:2018qmf}.: I cannot compare the two like that, one is ddbar, the other a combination of uubar and ddbar
%I won't say "whether there are other waves contributing", i.e., for those we do not take into account, because in principle up to about 0.9~GeV we considered all contributions/difficult to say, because of the tension 0.8-0.9~GeV

%%To reproduce the values in Table~I of Ref.~\cite{BESIII:2018qmf} relative to $D^+$ decays,
%%we find (see in App.~\ref{app:line_shapes} the normalization set by Eq.~\eqref{eq:alpha_GS}):
%%\begin{equation}
%%    A_1 (0) = 0.36
%%\end{equation}
%%A different value of $ A_1 (0) $ translates into a rescaling of $ B_{\rho^0} $, whose value extracted from a fit is given in the main text.
%%[\rd{the value of $A_1 (0) = 0.36$ does NOT compare well with \cite{CLEO:2011ab}, which uses a different normalization and line-shape for the $\rho^0$ resonance: $ A_1 (0) \simeq \sqrt{\pi} \, A_1 (0) |_{\text{CLEO}} \simeq 1.0 $; there is a dimensional problem in Eq.~(1) of \cite{CLEO:2011ab} and in Eq.~(2) of \cite{BESIII:2015kin}, so it is difficult to me to conclude that there is really a problem; De Boer \& Hiller give $A_1 (0) = 0.59$}]

(For comparison with Ref.~\cite{BaBar:2010vmf}, there is an overall normalization factor, adapted for the line shape in use here:
%[\LVS{where does it appear?}]-->it does not appear in the text, it is used to compare to BaBar

\begin{equation}\label{eq:alpha_GS}
    \alpha_{GS} = \sqrt{ \frac{3 \, \pi \, \mathcal{B}_{\rho^0}}{p^\ast_0 \, \Gamma_{\rho^0}^0} } \frac{\Gamma_{\rho^0}^0}{m_{\rho^0}} \,, \quad \mathcal{B}_{\rho^0} = 1 \,. )
\end{equation}

%\input{Estimating the phases from the angular observables.tex}

%\section{Definition of 'calligraphic'/helicity vector form factors}

%\input{omegarho}

%\input{isospinargument}

%%\input{Amplitude_analyses.tex}

%\input{LHCb_extracts.tex}

%\input{angularObservablesModified}

%\input{BESIII_extracts.tex}

%\input{BABAR_extracts.tex}

%\input{CMD2_extracts.tex}

\bibliography{mybib}{}
\bibliographystyle{unsrturl}

\end{document}